\documentclass[bibyear]{aa} 

\usepackage{hyperref}
\hypersetup{
    colorlinks=true,
    citecolor = blue,
    linkcolor=blue
            }
\usepackage{graphicx}
\usepackage{txfonts}
\usepackage{rotating}

\newcommand{\kms}{km\,s$^{-1}$}
\newcommand{\ms}{m\,s$^{-1}$}
\newcommand{\Halpha}{H$\alpha$}

\begin{document}

\title{Surface activity of a Rossby sequence of cool Hyades stars\thanks{Based on data acquired with the Potsdam Echelle Polarimetric and Spectroscopic Instrument (PEPSI) using the Large Binocular Telescope (LBT). The LBT is an international collaboration among institutions in the United States, Italy, and Germany.}}

\author{K.~G.~Strassmeier\inst{1,2}, I.~Ilyin\inst{1}, M.~Steffen\inst{1}, and S.~A.~Barnes\inst{1}}

\institute{Leibniz-Institute for Astrophysics Potsdam (AIP),
     An der Sternwarte 16, D-14482 Potsdam, Germany\\
     \email{kstrassmeier@aip.de}
\and
Institute for Physics and Astronomy, University of Potsdam, D-14476 Potsdam, Germany}

   \date{Received xxx x, 2025; accepted xxx x, 2025}

  \abstract
   {}
   {The Hyades cluster is key for the study of rotational, activity, and chemical evolution of solar-like low-mass stars. Here we present quantitative surface-activity information for a sequence of 21 Hyades dwarf stars with effective temperatures 6160~K to 3780~K (all cooler than the red edge of the Li dip), rotation periods 5\,d to 16\,d and normalized Rossby numbers ($\mathrm{Ro_n}$) between 0.14 to 0.54 with respect to the Sun (Ro(Sun)=1).}
   {High-resolution Stokes-V spectra and least-squares deconvolution of thousands of spectral lines per spectrum are employed for measuring longitudinal surface magnetic field. From Stokes-I spectra we obtain velocities, lithium abundances, metallicity, and chromospheric \ion{Ca}{ii} infrared-triplet (IRT) fluxes.}
   {The average metallicity, +0.186$\pm$0.045 (rms), for our stars with $T_{\rm eff}\geq 4200$\,K is in good agreement with the metallicity in the recent literature. Lithium abundances A(Li) range from 95-times solar (A(Li)$\approx +3.0$) on the warm end of the sample to $1/25$ solar (A(Li)$\approx -0.4$) on the cool end. We confirm the tight relation of A(Li) with $T_{\rm eff}$ and extend it to K--M stars with even smaller Li abundances than previously measurable. A formal relation with rotational period and velocity in the sense higher Li abundance for faster rotators is present. Targets rotating faster than $v\sin i$ of 6~\kms\ ($P_{\rm rot} \approx 8$\,d) appear Li saturated at A(Li)$\approx$3.0\,dex. \ion{Ca}{ii} IRT fluxes for our sample indicate (logarithmic) chromospheric radiative losses $R^\prime_{\rm IRT}$ in the range $-4.0$ to $-4.9$ in units of the bolometric flux. These radiative losses also show a relation with $T_{\rm eff}$, $P_{\rm rot}$, and $v\sin i$, but opposite to A(Li) in an inverse sense with higher radiative losses for the slower=cooler rotators. Longitudinal magnetic-field strengths were measured in the range zero to $-100$\,G and $+150$\,G with phase-averaged, disk-integrated, unsigned values $\langle |B|\rangle$ of 15.4$\pm$3.6(rms)\,G for targets warmer than $\approx$5000\,K and 91$\pm$61(rms)\,G for targets cooler than this. These unsigned field strengths relate to $P_{\rm rot}$, $v\sin i$, and $\mathrm{Ro_n}$, but in a dual-slope fashion. The short-period bonafide single M-target RSP\,348 was found to be a double-lined spectroscopic binary of dM3e+dM5e classification.}
   {We conclude that the Rossby-number dependencies of the surface activity tracers A(Li), $R^\prime_{\rm IRT}$, and $\langle |B|\rangle$ on our Hyades dwarf sequence primarily originate from convective motions, expressed by its turnover time, and only to a smaller and sometimes inverse extent from surface rotation and its related extra mixing.}


   \maketitle

\section{Introduction}

Stellar angular momentum loss and its associated spin-down due to magnetic braking are of key interest for understanding the evolution of our Sun in detail and our Solar System in general. Stars in open clusters always played an important role in finding the relation of rotation and age simply because of their determinable ensemble age but also because of their homogeneous formation history expressed, for example, by their unified metallicities. A star's rotation period as a function of mass and age is the observable global parameter for a quantitative spin-down description (Kraft \cite{k67}, Skumanich \cite{sku}, Noyes et al. \cite{noy:har}, and many thereafter). It enabled the formulation of gyrochronology (Barnes \cite{b07}), the determination of stellar age from a rotation period, with a (re)emphasis of stellar evolution and in particular of convective turnover timescales (Barnes \& Kim \cite{bar:kim}, Corsaro et al. \cite{cor:bon}). For a complete spin-down picture, however, it became obvious that the surface magnetic field morphology plays a more dominant role (e.g., Mestel \cite{mest}, Garraffo et al. \cite{gar:dra}) because it is only the open field that can funnel charged particles off the star and generates mass and angular-momentum loss while the closed field only traps such particles but not adds to the wind (Parker \cite{p58}, but see also Mestel \cite{mest}, Vidotto et al. \cite{vid14}, van~Saders et al. \cite{vansad}, Jardine et al. \cite{jar:vid}). Observational evidence for a magnetic morphology shift in older solar-type stars was presented from spectropolarimetry of bright field stars (e.g., Metcalfe et al. \cite{met1}, \cite{met2}). Magnetic field related questions such as how this field is anchored to the convection zone or what role the small-scale field and its hemispheric current helicity play remain largely unanswered (see, e.g., Hale \cite{hale}, Fan \cite{fan}, R\"udiger \& K\"uker \cite{rue:kue}).

Measurements of surface magnetic fields of cool cluster stars are difficult to obtain because of the large distances of cluster stars vis-a-vis nearby field stars, and are therefore still rare (Folsom et al. \cite{fol}, Wanderley et al. \cite{wander}). The difficulty lies in the intrinsically weak Zeeman signals and the spatially unresolvable coexistence of local bipolarity. The latter ensures that the longitudinal component of a surface magnetic field, when averaged over the visible hemisphere of a star, approaches to near zero if the contributions from positive polarity regions match the ones from negative polarity. Breaking this symmetry requires a time series of spectra. Note that the polarization degree due to the Zeeman-effect from the kilo-Gauss magnetic field of sunspots is typically about 10\%, and only 0.01\% for electron-scattering processes (e.g., Solanki \cite{sol93}). Detecting a signal from photospheric solar-like stellar magnetic fields is therefore best done from circular polarization (CP) but requires high-resolution spectra with high signal-to-noise ratio (S/N) over a large wavelength range. Target brightness is thus a basic selection criterium. But only with the advent of spectrum denoising methods (Donati et al. \cite{don:sem}, Carroll et al. \cite{carr07}, Martinez Gonzales et al. \cite{mar:ase},  Kochukhov et al. \cite{koclsd}, Tkachenko et al. \cite{tka}) was it possible to measure Gauss-level magnetic field strengths also on solar-like cool stars, as compared to the many kG-strong fields of Ap stars. Folsom et al. (\cite{fol}) were the first to apply even Zeeman Doppler Imaging (ZDI; Semel \cite{semel}) to cluster stars, in particular also to the Hyades and younger stellar associations. Two of the five Hyades targets observed by Folsom et al. (\cite{fol}) were binaries (Mel25-151 and Mel25-43) and one a possible binary (Mel25-179). The two remaining stars, RSP\,68 and RSP\,198, were confirmed singles based on the absence of RV variations, both stars are also in the sample of the present paper.

One current working assumption is that only the open large-scale field in its lowest energy state contributes to the mass and angular momentum loss (Jardine et al. \cite{jar13}). If so, a statistical relation should exist between magnetic field morphology and rotation period in the sense of slower rotation in the case of  more open field lines. See et al. (\cite{see:jar}) used a sample of 22 solar-mass field stars from the literature that had a magnetic surface map and used them to determined the amount of open flux. The age span of their field-star sample was between 24\,Myr and 5\,Gyr but with very uncertain ages for the individual targets and consequently with significant inhomogeneities of the stellar  properties. They conclude that the spin-down of main sequence solar-mass stars is likely dominated by the dipolar component of the magnetic field. In another study of six solar analogues between 100 and 600\,Myr of age, Ros\'en et al. (\cite{ros}) found that the magnetic energy in higher-order spherical harmonics was larger than in the lower-order harmonics, indicating a more complex magnetic field morphology with fewer open fields. Folsom et al. (\cite{fol}) found a relation of decreased magnetic-field strength with age, albeit with large scatter, and a power law decrease with Rossby number (ratio of rotation to convection). This dependency was confirmed for 292 M-dwarfs observed as part of the CARMENES planet-catch survey (Reiners et al. \cite{reiners}). The study of Kochukhov et al. (\cite{koch}) from Zeeman line broadening and intensification, for a sample of 15 solar-like field stars, suggested that the age and Rossby-number dependency is solely due to the magnetic filling factor while all stars exhibit roughly the same local field strengths. Saar \& Linsky (\cite{saa:lin}) had already found such a trend of increasing field strength with later spectral type together with a trend of increasing filling factor with magnetic activity.

The primary goal of the present paper is to determine stellar magnetic properties for a significant sample of stars of equal age and metallicity, but different mass, in order to quantitatively explicate a possible morphology change with rotation period or Rossby number. We employ new polarimetric observations of Hyades stars with masses between 1.2 and 0.6 solar masses and rotation periods between five and 16 days, respectively. Our sample thus includes stars with Rossby numbers in the range 0.1--0.4, while of same age. (This corresponds to $0.013 < \mathrm{Ro_n} < 0.54$, where Ro$_n$ is the corresponding value on the scale where Ro$_n,\odot$ = 1.) From this, we hope to learn more about the role of the magnetic field topology for rotational evolution and eventually be able to implement these magnetic topologies, energies, and fluxes as boundary conditions for next-generation global dynamo simulations (e.g., Schrinner \cite{schr}) or, for example, provide constraints for identifying solar-like magnetic cycles (e.g., Lehmann et al. \cite{lehm}). We rely on magnetic-field measurements for 21 Hyades cluster members from Stokes-V data taken with PEPSI at the 11.8\,m LBT. PEPSI's spectral resolution is twice as high as for previously published field determinations of Hyades stars. Moreover, the light-gathering power of the LBT enabled an unprecedented peak S/N of up to 1180 per pixel for RSP\,233 on the bright end and still 140 for RSP\,348 on the faint end. These new observations are described in Sect.~\ref{S2}. The target sample is introduced in Sect.~\ref{S3}, while Sect.~\ref{S4} presents a redetermination of its relevant stellar parameters including radius, gravity, metallicity, Rossby number, and rotational velocity among others. Our activity measurements are presented in Sect.~\ref{S5} and include lithium abundances, \ion{Ca}{ii} IRT emission-line fluxes, and magnetic fields along with its assumptions and data set-up in form of least-squares deconvolved (LSD) line profiles. Section~\ref{S6} is a discussion of the various stellar-activity tracers versus temperature and rotation and Sect.~\ref{S7} summarizes our findings. Our second paper aims at ZDI of above targets, then providing the morphology of the surface magnetic field.

\begin{table*}[!tbh]
\caption{Relevant stellar parameters for the present Hyades sample.}\label{T1}
\begin{flushleft}
\begin{tabular}{lllllllllllllll}
\hline\hline
\noalign{\smallskip}
RSP$^1$   & $P_{\rm rot}$ & $D$ & $L$ & $M_\star$ & $R_\star$ & $T_{\rm eff}$ & RV & $v\sin i$ & $\xi_{\rm t}$ & $\zeta_{\rm t}$ & $i$  & $\tau$ & $\mathrm{Ro}$ & $\mathrm{Ro_n}$\\
      &(d)           & (pc)& ($L_\odot$)& (M$_\odot$) & (R$_\odot$) & (K)& (\ms) & \multicolumn{3}{c}{(\kms)} & ($^\circ$) & (d) & (--) & (--) \\
\noalign{\smallskip}\hline\noalign{\smallskip}
344   &5.14 & 45.147 & 1.63 & 1.20 & 1.155 &6075 &39984 &8.6   & 1.00  &  4.5 & 49 & 18.8  & 0.273 & 0.365  \\
233   &5.45 & 45.413 & 1.71 & 1.21 & 1.178 &6088 &38754 &9.8   & 1.00  &  4.5 & 64 & 18.2  & 0.299 & 0.400  \\
340   &5.87 & 46.993 & 1.90 & 1.22 & 1.213 &6158 &39603 &8.8   & 1.00  &  4.5 & 57  &  14.6  & 0.402 & 0.538 \\
137   &7.47 & 47.225 & 1.32 & 1.15 & 1.081 &5954 &37942 &6.2   & 0.95  &  4.3 & 58  & 24.4  & 0.306 & 0.409\\
177   &8.38 & 45.467 & 0.98 & 1.10 & 0.995 &5769 &37940 &5.4   & 0.90  &  3.7 & 64  & 31.7  & 0.264 & 0.353 \\
440   &8.93 & 46.149 & 1.16 & 1.14 & 1.049 &5856 &40815 &5.4   & 0.95  &  4.0 & 65  & 28.5  & 0.313 & 0.419 \\
227   &8.96 & 46.749 & 0.70 & 1.05 & 0.899 &5578 &38340 &4.9   & 0.90  &  2.9 & 75  & 38.4  & 0.233 & 0.312\\
134   &9.10 & 45.550 & 0.31 & 0.93 & 0.712 &5125 &37294 &3.7   & 0.85  &  1.9 & 69  & 53.4  & 0.170 & 0.227\\
439   &9.16 & 52.093 & 0.30 & 0.92 & 0.717 &5035 &40566 &3.8   & 0.85  &  1.8 & 74  & 56.4  & 0.162 & 0.217 \\
225   &9.39 & 46.627 & 0.62 & 1.03 & 0.866 &5517 &39158 &3.9   & 0.90  &  2.6 & 57  & 40.4  & 0.232 & 0.310 \\
95     &9.82 & 40.735 & 0.34 & 0.98 & 0.750 &5090 &36731 &3.8   & 0.85  &  1.9 & 80 & 54.6  & 0.180 & 0.241 \\
429   &9.82 & 48.507 & 0.33 & 0.95 & 0.742 &5103 &40578 &3.8   & 0.85  &  1.9 & 84  & 54.1  & 0.182 & 0.243 \\
198   &10.26& 50.824 & 0.45 & 1.00 & 0.796 &5320 &38470 &2.7   & 0.85  &  2.2 & 43   & 47.0  & 0.218 & 0.292 \\
68     &10.57& 45.443 & 0.29 & 0.85 & 0.747 &4916 &31328 &2.7   & 0.85  &  1.6 & 49   & 60.5  & 0.175 & 0.234 \\
587   &11.79& 50.718 & 0.16 & 0.83 & 0.613 &4640 &42334 &2.0   & 0.70  &  1.6 & 50  & 70.5  & 0.167 & 0.223 \\
571   &12.22& 49.206 & 0.10 & 0.77 & 0.537 &4446 &41902 &2.2   & 0.70  &  1.6 & 82  & 80.4  & 0.152 & 0.203\\
216   &12.64& 44.958 & 0.25 & 0.88 & 0.699 &4896 &39308 &2.5   & 0.85  &  1.6 & 63  & 61.1  & 0.207 & 0.277 \\
409   &12.90& 45.721 & 0.065& 0.67 & 0.486 &4191 &40887 &1.7   & 0.70  &  1.6 & 63  & 97.8  & 0.132 & 0.177\\
133   &13.51& 47.174 & 0.048& 0.63 & 0.438 &4089 &37916 &1.5   & 0.70  &  1.6 & 66  & 106   & 0.131 & 0.175\\
542   &14.44& 40.638 & 0.037& 0.62 & 0.406 &3994 &42379 &1.4   & 0.70  &  1.6 & 80 & 118.5 & 0.122 & 0.163 \\
556   &15.99& 48.159 & 0.019& 0.59 & 0.327 &3778 &42634 &1.0$^2$   & 0.70  &  1.6 & 75  & 157   & 0.102 & 0.136 \\
\noalign{\smallskip}
\hline
\end{tabular}
\tablefoot{$^1$For other names see Table~\ref{T1-App} in the Appendix.  $^2$Best-fit $v\sin i$ value is given but we note that it is below the empirical lower limit of $\approx$1.3\,\kms. $D$ distance from \emph{Gaia} DR3 parallax (Gaia collab. \cite{DR3}) with related luminosity $L$, mass $M$, and radius $R$, all in solar units. $T_{\rm eff}$ effective temperature from photometry. RV is the radial velocity from our spectra. $v\sin i$ is the projected rotational velocity from our spectra (errors are typically $\pm$0.3\,\kms). The parameters $\xi_{\rm t}$, $\zeta_{\rm t}$, and $i$ denote the microturbulence, macroturbulence, and the inclination of the rotational axis. $\tau$ is the convective turnover time from Barnes \& Kim (\cite{bar:kim}), $\mathrm{Ro}$ is the Rossby number $P_{\rm rot}/\tau$, and $\mathrm{Ro_n}$ the normalized Rossby number ($\mathrm{Ro_{n,\odot}}=1$).}
\end{flushleft}
\end{table*}

\section{Observations}\label{S2}

Spectra were obtained with PEPSI at the 2$\times$8.4\,m LBT in southern Arizona. We employed both polarimeters in the pair of symmetric straight-through Gregorian foci of the two LBT mirrors, dubbed SX and DX. Two pairs of octagonal 200$\mu$m fibers per polarimeter feed the ordinary and extraordinary polarized beams via a five-slice image slicer per fiber into the spectrograph. It produces four spectra per \'echelle order which are recorded in a single exposure with an average spectral resolution of $R=\lambda/\Delta\lambda$=130\,000 ($\approx$0.046\,\AA ). This resolution is usually sampled by 4.2 pixels on the CCD. However, as of 2020, we employed a binning mode of 2$\times$2 pixels, two pixels in cross-dispersion and two pixels in dispersion direction, which then samples the spectral resolution with 2.1 (super)pixels but with four times more flux. Both telescope sides were used with a fiber diaphragm on the sky with a projected diameter of 1.5\arcsec . The Foster prism, the atmospheric dispersion corrector (ADC), two fiber heads, and two fiber viewing cameras are rotated as a single unit with respect to the parallactic axis on sky. The quarter-wave retarder is inserted into the optical beam in front of the Foster prism for the CP measurements (per polarimeter), and retracted for linear polarization (LP)  measurements if LP is desired. The spectrograph and the polarimeters were described in detail in Strassmeier et al. (\cite{pepsi}, \cite{spie-austin}).

Observations of the Hyades commenced in three observing runs. The first one between UT Dec.\,2--18, 2020 (dubbed S20). Full nights were available only after UT\,Dec.~5. The second run was between UT Dec.\,28, 2021 and Jan.\,13, 2022 (S21) and the third run between UT Nov.\,18 to Dec.\,7, 2022 (S22). Each target was observed once per clear night. Unfortunately, snow storms caused a loss of four consecutive nights on Dec. 10--13, 2020 and again on Dec.\,28--Jan.\,3, 2022, with consequent impacts on the rotational phase distribution for 10 of the 16 targets in the first run, leaving us with only three successful targets from the second run, respectively. All of these targets were re-observed in the third run together with three additional targets. The total number of Hyads with phase resolution is thus 22, of which 21 are single stars. Ten targets have two-epoch data sets but one epoch always with only partial phase coverage. Overall phase coverage per target is summarized in Table~\ref{T1-App}. The wavelength settings for all observations were with cross disperser (CD)~III covering 4800--5441~\AA\ and CD~V covering 6278--7419~\AA\ simultaneously. All targets have additionally one integration with CD\,VI in Stokes~I, covering 7419--9067\,\AA\ for the \ion{Ca}{ii} infrared triplet. Its S/N per pixel ranges between 970 for RSP\,344 to 371 for RSP\,542. We note that two targets (RSP\,133 and 556) had to be observed in 2024. 

We always used the two 8.4\,m LBT mirrors in binocular mode, that is like a single 11.8\,m telescope, which then requires two consecutive exposures for the circular Stokes component. Left-hand and right-hand CP spectra were obtained with retarder angles of 45\degr\ and 135\degr , respectively, and with the beam-splitting Foster prism position angle set to 0\degr . The position angle of the Foster prism itself had been calibrated with a standard visual binary at the beginning of each run and was verified occasionally throughout the runs. Exposure times were set according to target brightness and were 2$\times$10\,min for RSP\,340, 233, 344, 137, 440, 177, and 225, 2$\times$15min for RSP\,95, 198, 429, 216, and 587, 2$\times$20min for RSP\,227, 134, 439, 571, 409, and 542, 2$\times$25min for RSP\,133, and 2$\times$30\,min for RSP\,348 and 556. Stokes~I is the sum of both CP exposures. Quantile 95\%\ S/N per pixel for Stokes-I is up to 1180 in CD~V and up to 890 in CD~III for RSP\,233 ($V$=7.3\,mag), and as low as 140 in CD~V and 44 in CD~III for RSP\,348 ($V$=14.2\,mag). S/N of Stokes-V spectra is $\approx$60\%\ of S/N of Stokes~I depending on wavelength. 

At this point we recall that our S/N values are purely based on photon noise. While systematic errors are minimized in our reduction process, for example, by a very rigorous treatment of scattered light, such errors will remain in the data. A direct comparison of the derived stellar parameters from the three instruments HARPS, PEPSI, and ESPRESSO showed sufficient general agreement but with small differences that could be important for certain science cases (Adibekyan et al. \cite{adi}). 

Data reduction was performed with the software package SDS4PEPSI (``Spectroscopic Data Systems for PEPSI'') based on the original code of Ilyin (\cite{4A}), and described in some detail in Strassmeier et al. (\cite{sun}, \cite{pepsi}). The specific steps of image processing include bias subtraction and variance estimation of the source images, super-master flat field correction for the CCD spatial noise, scattered light subtraction, definition of \'echelle orders, wavelength solution for the ThAr images, optimal extraction of image slicers and cosmic spikes elimination, normalization to the master flat field spectrum to remove CCD fringes and the blaze function, a global two-dimensional fit to the continuum, and the rectification of all spectral orders into a one-dimensional spectrum.

\begin{figure*}
{\bf a.}\hspace{60mm}{\bf b.}\hspace{60mm}{\bf c.}\\
   \includegraphics[width=60mm,clip]{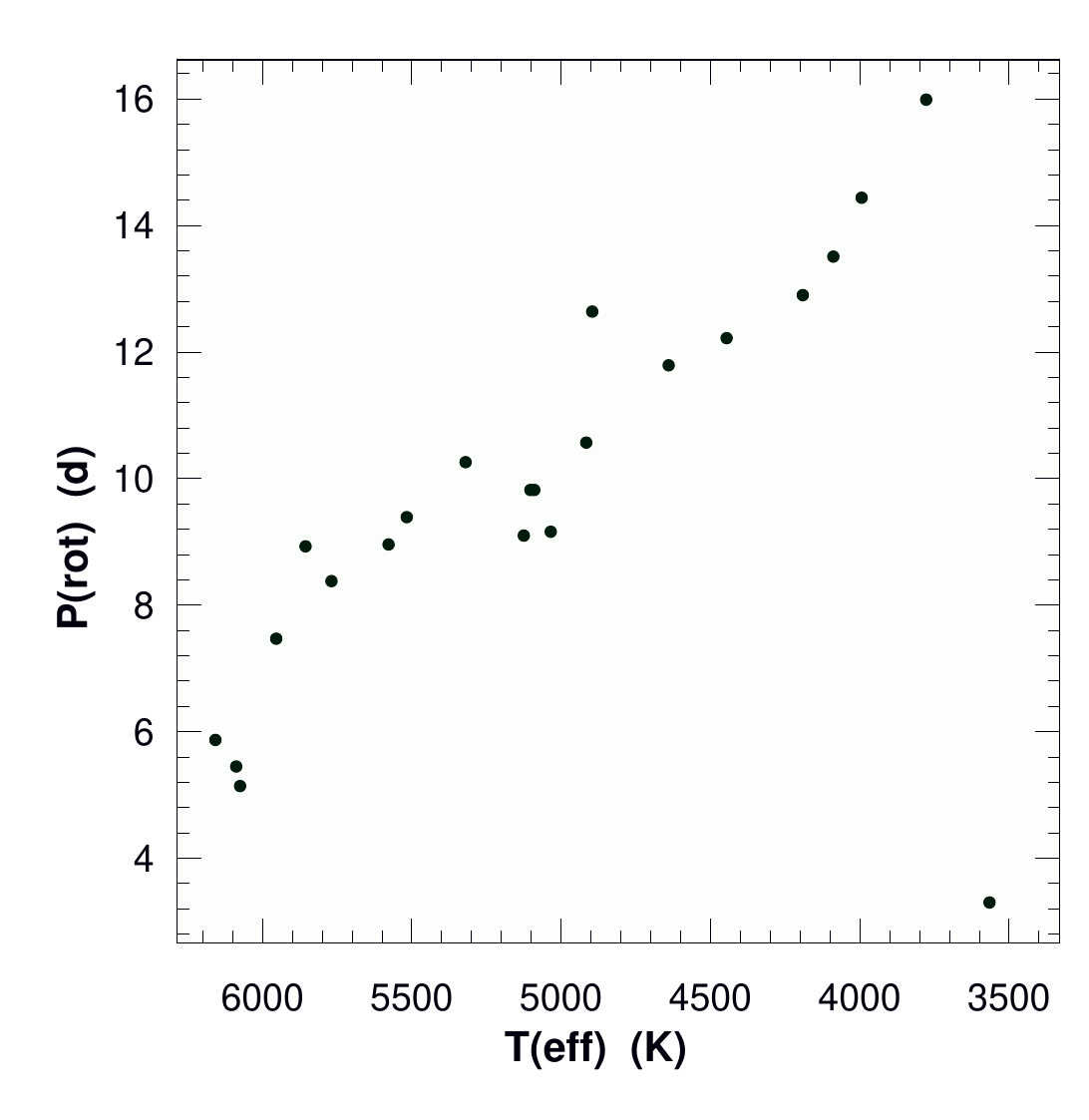}
   \includegraphics[width=60mm,clip]{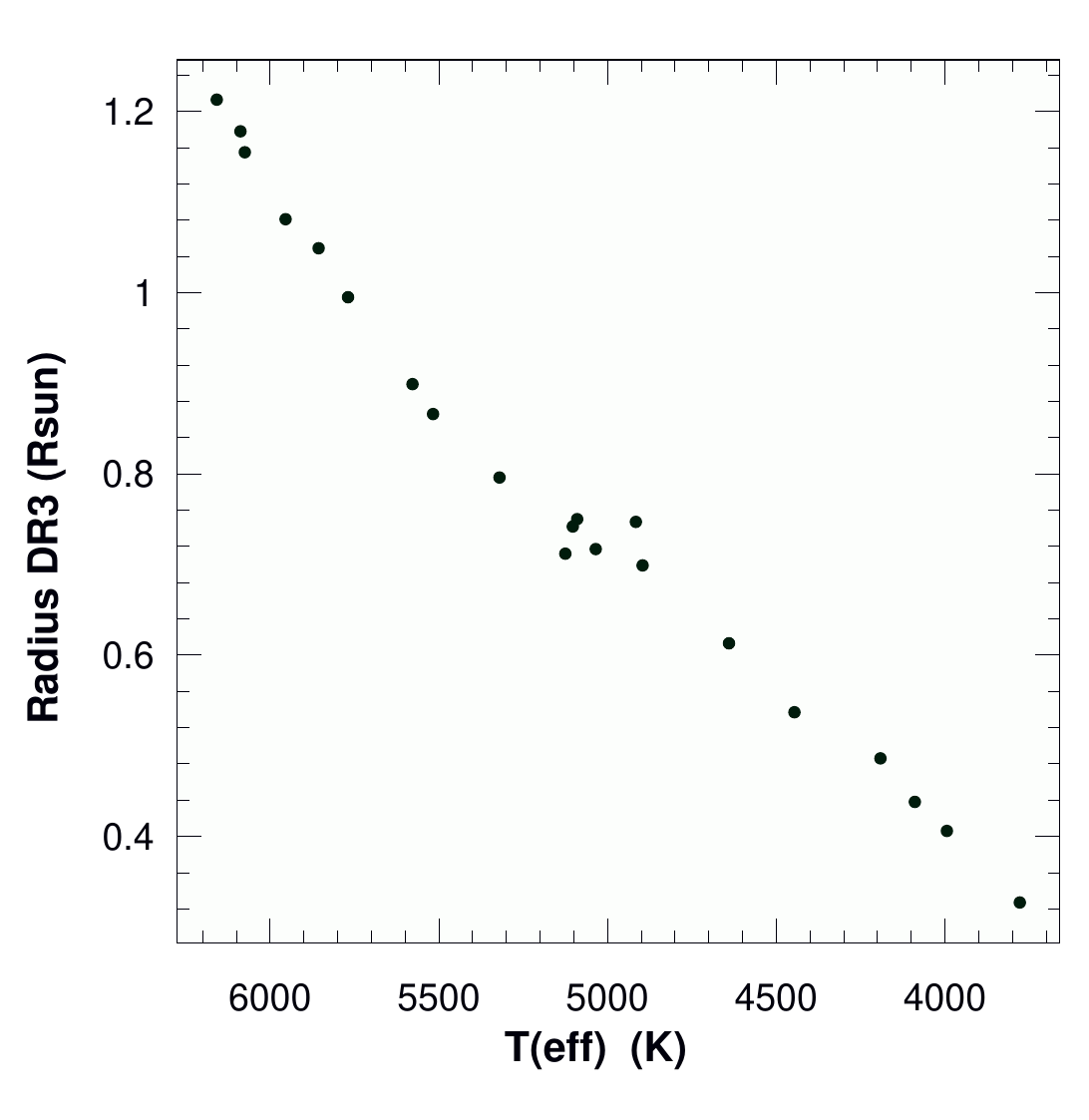}
   \includegraphics[width=60mm,clip]{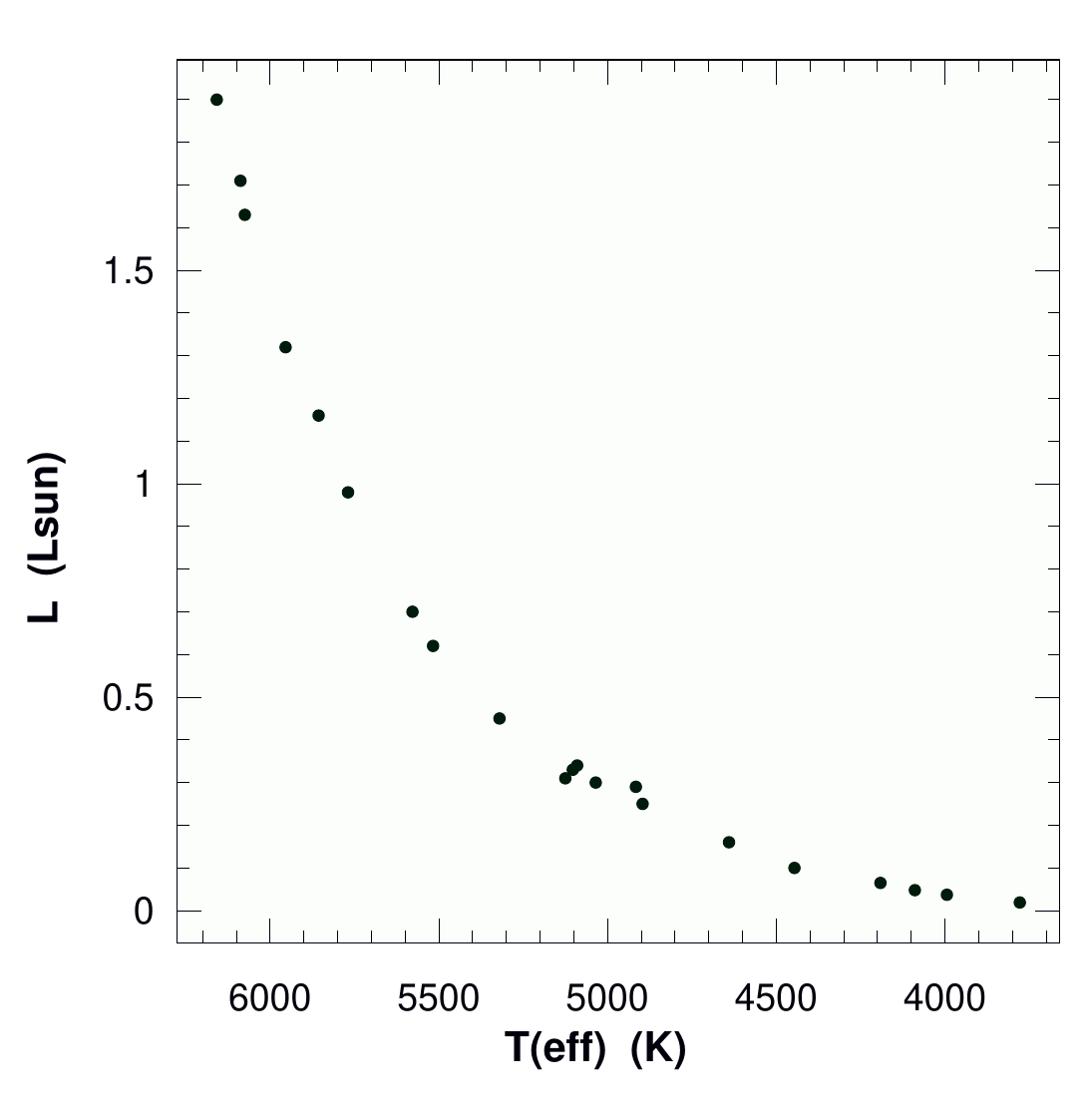}
   \caption{Temperature distributions for the Hyads in our sample. $a.$ Versus rotation period. The target in the lower right corner is the cool M-dwarf binary RSP\,348 (not included in the other panels). $b.$ Versus radius as obtained from the \emph{Gaia} DR3 parallax and $c.$ versus luminosity $L$. }
         \label{F1}
\end{figure*}

\section{Target sample}\label{S3}

The Hyades constitute a relatively young cluster in the age regime $\approx$700\,Myr (with an often-cited value of 625$\pm$50\,Myr, Perryman et al. \cite{perry}), and is the closest open cluster to the Sun (47.5\,pc, Gaia collaboration \cite{gaia-DR2}). Table~\ref{T1-App} summarizes our target sample from an observational point of view and Table~\ref{T1} details its relevant stellar parameters. Figure~\ref{F1} plots it as a function of effective temperature for the quantities rotation period (panel a), radius (panel b), and luminosity (panel c). We note that all our stars are on the slow-rotator sequence in the $P_{\rm rot}$-$T_{\rm eff}$ plane (Barnes~\cite{b03}). We follow the nomenclature of Douglas et al. (\cite{doug19}) and list the target identifications according to their entry in the R\"oser et al. (\cite{rsp}) catalog (RSP), the \emph{Hipparcos} catalog (HIP; van Leeuwen \cite{hip}), and the Two Micron All Sky Survey catalog (2MASS; Skrutski et al. \cite{2mass}). The brightness entry is the apparent $V$ magnitude from the homogeneous recalibration of the Hyades by Joner et al. (\cite{joner}), if available, otherwise from the US Naval Observatory CCD Astrograph Catalog (UCAC4; Zacharias et al. \cite{UCAC}) or the all-sky census from R\"oser et al. (\cite{rsp}). The rotation periods are all of photometric origin, and are either from \emph{Kepler} K2 (Douglas et al. \cite{doug19}, \cite{doug16}) or from ground-based monitoring at Lowell Observatory (Radick et al. \cite{rad87}, Radick \cite{rad95}), from SuperWASP (Delorme et al. \cite{delo}), or from the All-Sky Automated Survey ASAS (Kundert et al. \cite{kund}; privately communicated to Douglas et al. \cite{doug14}). One period is from time-series spectropolarimetry (RSP\,198; Folsom et al. \cite{fol}). Six of our targets have period determinations from two independent sources but are not grossly contradictory. Effective temperatures ($T_{\rm eff}$) are from Douglas et al. (\cite{doug19}) and are also based on photometry. A starting value for the projected rotational velocity $v\sin i$ was taken from Paulson et al. (\cite{paul}) or, if unavailable, from Mermilliod et al. (\cite{mmu}).  Individual $v\sin i$ values are revised in the present work.

All our sample targets are confirmed single-star Hyades members according to four criteria. We follow again the notations by Douglas et al. (\cite{doug19}) and verify single-star compliance with four yes/no (Y/N) flags in Table~\ref{T1-App} in column ``Single?''. The four flags stand for: 1) radial velocities (RV) are with a rms of less than 2\,\kms\ of the system velocity, 2) photometry with a deviation in the color-magnitude diagram of less than 0.375\,mag, 3) binarity unconfirmed, 4) confirmed \emph{Gaia} DR2 RV with an error less than 2\,\kms . We note that only the cool M-dwarf RSP\,348 ($r'$=13.6\,mag, $V$=14.2\,mag) had incomplete information but was, prior to our observations, not proposed nor confirmed as a binary. Our data in App.~\ref{App-A} (Fig.~\ref{F-A1}) show it to be a double-lined spectroscopic binary (SB2) and the target was therefore removed from the magnetic sample.

RSP\,68 is listed with a M1-3V companion candidate of $\approx$0.3\,M$_\odot$ detected from speckle interferometry (Patience et al. \cite{pat:ghe}) and AO imaging (Guenther et al. \cite{gue}). However, no significant RV variations were found for it by Folsom et al. (\cite{fol}) and we conclude that it is either a background target or a very long period companion. We thus treat RSP\,68 as an effectively single star. Similar may be the case for RSP\,133. During a period in 2022 with seeing below or equal to 0.5\arcsec, we noted a faint companion on our guider images, approximately 1.3\arcsec\ away and $V\approx 17$\,mag. However, Kopytova et al. (\cite{kop}) found no companion outside of 2.79\,AU of RSP\,133 in their lucky imaging survey.

\section{Census of stellar parameters}\label{S4}

\subsection{Temperature, gravity, luminosity, radius}

We adopt effective temperatures from Douglas et al. (\cite{doug19}) based on Tycho-2 $BV$ (Hog et al.~\cite{tycho}), \emph{Gaia} $G$-band, and 2-MASS $JK$ photometry as the most homogeneous set of measured temperatures. Spectroscopic temperatures are available for only subsets of our targets (Paulson et al. \cite{paul}, Dutra-Ferreira et al. \cite{dutra}, Cummings et al. \cite{cumm}). The differences between the spectroscopic and the photometric temperatures are typically less than 100\,K per star but the deviations are generally larger for the cooler stars with $T_{\rm eff} < 5000$\,K. We note that in the spectroscopic analysis by Schuler et al. (\cite{sch06}) they used photometric temperatures from $B-V$ from Allende Prieto \& Lambert (\cite{all:lam}) and transformed to $V-K$. Jeffery et al.~(\cite{jef:tay}) presented model-determined temperatures for 88 single Hyads from eight sets of stellar models and compared them with photometric temperatures. It shows basically good agreement but some systematic differences for stars cooler than 5500\,K remain. 

Luminosity, $L$, was calculated based on the \emph{Gaia} DR3 distance together with the apparent $V$ magnitude (mostly from Joner et al. \cite{joner}) and the photometric $T_{\rm eff}$ from Douglas et al. (\cite{doug19}). From $L$ and $T_{\rm eff}$, we derive the radius, $R_\star$, and the logarithmic gravitational surface acceleration, $\log g$, from the relations $L \propto R^2 T_{\rm eff}^4$ and $g \propto M / R^2$, respectively. Because the latter values are all very close to $\log g$=4.5 (4.3--4.7), and because its choice is not critical for our differential Stokes analysis, we adopt model atmospheres fixed to $\log g$=4.50 for all targets except the two M dwarfs RSP\,542 and RSP\,556 for which we adopt $\log g$=5.0. Masses ($M_\star$) were estimated originally by Douglas et al. (\cite{doug19}) based on the \emph{Gaia} DR2 parallax with above $T_{\rm eff}$'s and are given for orientation.

\subsection{Metallicity}\label{feh}

For stellar metallicity, in particular [Fe/H], we compare to the cluster average value of [Fe/H] = +0.18$\pm$0.03~dex from Dutra-Ferreira et al. (\cite{dutra}), that is, a logarithmic abundance of $\mathrm{A(Fe)}) = 7.62\pm0.02$. It is based on 3D model atmospheres, two optimized line lists (one for NLTE line formation and the other for the cluster giants), and consistently applied to high-quality HARPS (giants) and UVES (dwarfs) spectra at resolutions of 110,000 and 60,000, respectively. Three giants and 14 dwarfs were analyzed. The metallicity reported earlier by, for example, Paulson et al. (\cite{paul}) is about 2$\sigma$ smaller, +0.13$\pm$0.01~dex, but judged compatible with the Dutra-Ferreira et al. (\cite{dutra}) value. Metallicities obtained from isochrone fittings show a tendency to even higher values, order +0.25 (Brandner et al. \cite{bra:cal} and references therein). While the best-fit isochrone (710\,Myr, [Fe/H]=+0.25) provides a good fit to the upper and lower main sequence, it still underestimates the luminosity of stars in the mass range between $\approx$0.3--0.85 M$_\odot$.

The metallicities from our lithium fits in Sect.~\ref{S_Li} are mostly constrained by the Fe\,{\sc i} 6707.43\,\AA\ line blend and, with decreasing temperature, by an increasing amount of TiO. For the very cool stars with $T_{\rm eff} < 4500$\,K the atomic transitions do not provide reliable metallicities from Fe anymore. Our iron abundances for these targets in Table~\ref{T2} are thus unrealistic, as one can already see by the $\chi^2$ fit values. A complete chemical abundance analysis including molecular contributors is beyond the scope of this paper. We note that the [Fe/H] errors in Table~\ref{T2} are internal fitting errors and rather small. Their absolute (external) errors are likely a factor ten larger mostly due to the expected continuum-setting uncertainties. The unknown continuum suppression by molecular lines appears also the reason why the fits adopt increasingly lower metallicities the cooler the target (some even with subsolar [Fe/H]). It is an artifact from the continuum suppression likely due to an overestimation of some of the TiO molecular line opacities. Our sample's averaged relative iron abundance is +0.186$\pm$0.045 (rms) when excluding the TiO-dominated stars with $T_{\rm eff} < 4500$\,K. This value agrees very well with the canonical metallicity of +0.18 from Dutra-Ferreira et al. (\cite{dutra}) .

\subsection{Rossby number}\label{ross}

The Rossby number, $\mathrm{Ro}$, has become an accepted independent variable against which various activity indicators for cool stars are assessed (e.g., Noyes et al. \cite{noy:har}, Durney \& Latour \cite{dur:lat}). It is also a parameter in dynamo models and compares timescales between the Coriolis force and the advection. In the present paper, we calculate semi-empirical values as $\mathrm{Ro} = P/\tau_c$, where $P$ is the rotation period of the star, and $\tau_c$ is the convective turnover timescale. We use the convective turnover timescales listed in Table\,1 of Barnes \& Kim (\cite{bar:kim}), interpolating as needed from the stellar effective temperatures, specifically the global one, representing an average over the entire convection zone. While it is true that different tabulations of $\tau_c$ list differing values, there is largely only a scaling difference between them, so that the power law relationships that characterize variables against $\mathrm{Ro}$ are unaffected. In order to remove the bulk of the differences in $\mathrm{Ro}$ that is incurred from the choice of a  particular source for the convective turnover timescales, we further normalize the values of $\mathrm{Ro}$ by the respective solar value ($\mathrm{Ro}_\odot = 0.747$ using $P_{\odot} = 26.09$\,d and $\tau_{c\odot} = 34.9$\,d). It provides values of $\mathrm{Ro_n} = \mathrm{Ro}/\mathrm{Ro_{\odot}}$. This has the effect of translating the numerical value of the solar $\mathrm{Ro}$ from 0.75 to exactly 1.00, facilitating inter-comparisons between our work and others that might use alternative convective turnover timescales (e.g., Mathur et al. \cite{Math2025}). These values are listed in Table~\ref{T1}, and used in the figures in this paper.

\subsection{Projected rotational velocity, inclination}

Another set of stellar parameters concerns the line broadening, mainly the rotational velocity, $v\sin i$, the micro turbulence, $\xi_{\rm t}$, and the (radial-tangential) macro turbulence, $\zeta_{\rm t}$. We adopt canonical values for $\xi_{\rm t}$ from Dutra-Ferreira et al. (\cite{dutra}) (their Table~5) and for $\zeta_{\rm t}$ from Gray (\cite{gray}) (his Table~B.1) but redetermine $v\sin i$ from our $R$=130\,000 spectra. Involved is here the inclination, $i$, of the stellar rotational axis with respect to the line of sight. It remains a weakly constrained parameter. Only for RSP\,198 and RSP\,68 measured $i$ values from ZDI are available from Folsom et al. (\cite{fol}) of $53^{+18}_{-11}$\degr\ and $61^{+29}_{-14}$\degr, respectively. We measure $v\sin i$ by means of an autocorrelation function (ACF) technique employing the large wavelength ranges available from CD3 and CD5. An ACF is computed from phase-combined spectra only; these combined spectra are build from all individual (phase-resolved) spectra by averaging them with equal weight to a single spectrum per target of very high S/N of typically 2000:1 per pixel. We note that the lithium analysis in Sect.~\ref{S_Li} also uses these combined spectra and lists the combined S/N values in Table~\ref{T3}. Wavelength ranges for the autocorrelation mask exclude regions with telluric contamination and strong lines like, for example, \Halpha\ in the red CD or the Mg\,{\sc i} triplet in the blue CD. The width of the ACF is then deconvolved from the combined instrumental profile and adopted macroturbulence listed in Table~\ref{T1} with a two-piece linear fit to its FWHM versus $T_{\rm eff}$ (one fit for $T_{\rm eff}\leq 4000$\,K, one for $T_{\rm eff}>4000$\,K). This approach basically follows the technique used by Fekel (\cite{fek97}). A formal error is computed from the squared sums of the ACF width of different wavelength regions plus the O-Cs from the FWHM vs. $T_{\rm eff}$ fits but is unrealistically small. Instead, we estimate the real $v\sin i$ error to be approximately $\pm$0.3~\kms\ from repeated applications to different wavelength chunks. 

With the above $v\sin i$ values, we estimate the most likely inclination $i$ from the rotation period and the \emph{Gaia} DR3-based stellar radius in Table~\ref{T1}. Its equal likeliness range is the result from the errors of the parallax and effective temperature, and from $v\sin i$ and the period. The inclination is thus a heuristically determined value and not a measurement. The typical uncertainty range is 10-15$\degr$, basically driven by the error of $T_{\rm eff}$ and $v\sin i$, which are typically $\pm$50~K and $\pm$0.3~\kms, respectively. We note that for one target (RSP\,556) the $v\sin i$ measurement is below our empiric spectroscopic detection limit and accordingly more uncertain. The most likely value for $i$ is given in Table~\ref{T1}. 

\begin{figure}
   \includegraphics[width=87mm,clip]{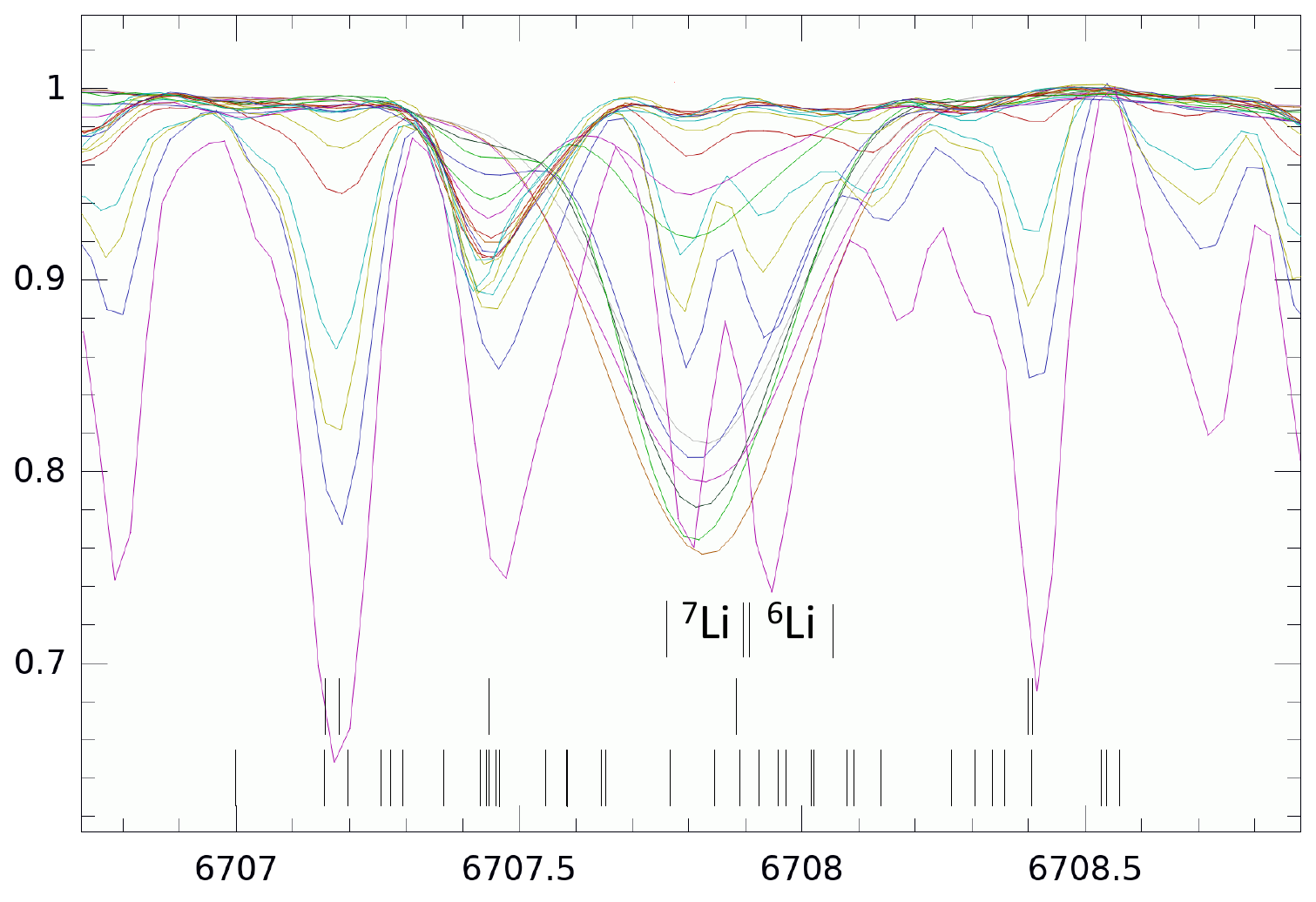}
   \caption{Comparison of the lithium 6707.8-\AA\ region for our sample stars. Shown are the phase-combined spectra. Short vertical dashes indicate the wavelengths of blending lines: Bottom row of dashes is the line list from Strassmeier \& Steffen (\cite{xiboo}), middle row are the strongest CaH and TiO lines from the solar umbral atlas, and the upper four dashes indicate the two $^7$Li and $^6$Li doublets, respectively. The x-axis is wavelength in \AA ngstroem. }
         \label{F2}
\end{figure}

\subsection{Radial velocity}

Radial velocities (RV) in this paper were determined in the course of building the LSD line profiles from Stokes~I and V spectra for the magnetic-field measurements and were zeroed by the standard Th-Ar calibration. Our spectra are on average 445$\pm$278(rms)~\ms\ higher than the \emph{Gaia} DR3 velocities (Katz et al.~\cite{katz}). The 445\,\ms\ difference is just a zero-point difference and can be corrected for if desired. PEPSI's annual RV stability is estimated to be around 6--10~\ms. While relative RVs during one night can be of even higher precision, the above large rms of 278~\ms\ indicates that it is the rotational modulation of the surface activity and the time coverage of our sample that set the external RV accuracy. The RVs in Table~\ref{T1} are phase averages from all spectra available. All targets exhibit modulation of the RVs due to spots and other surface inhomogeneities in the range of above rms value. 

\begin{figure}
   \includegraphics[width=70mm,clip]{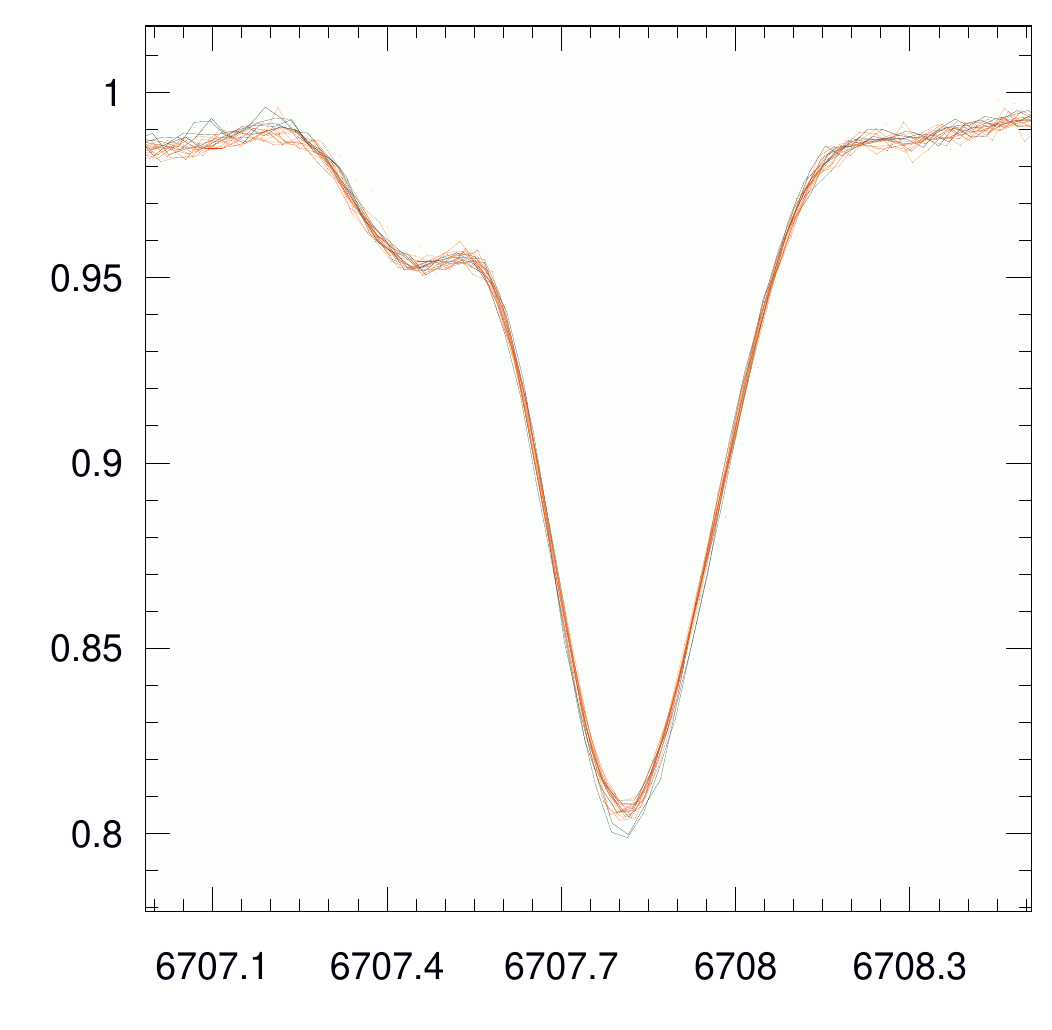}
   \caption{Nineteen individual Li\,{\sc i} 6707.8-\AA\ line profiles of RSP\,177 ($T_{\rm eff}$=5770\,K). It indicates the expected range of line-core changes over time. The black profiles are from 2020 (season S20), the red profiles from 2022 (season S22), according to Table~\ref{T1-App}. The maximum line-core variability is in this case 0.8\%. The x-axis is wavelength in \AA ngstroem. }
         \label{F3}
\end{figure}

\section{Surface activity measurements}\label{S5}

\subsection{Lithium abundance}\label{S_Li}

We employed Turbospectrum (Plez~\cite{turbo}) and its fitting program TurboMPfit (Steffen et al. \cite{mpfit}) for a direct line synthesis fit of the wavelength region 6706--6709\,\AA . Our fits are based on a predetermined set of 1D LTE spectra synthesized from MARCS (Gustafsson et al. \cite{marcs}) model atmospheres for different effective temperatures $T_{\rm eff}$, gravities $\log g$, relative metallicities [Fe/H], microturbulence velocities $\xi_{\rm t}$, lithium abundances A(Li), and  isotope ratios $^6$Li/$^7$Li. TurboMPfit compares then the observed spectrum with the library spectra to find the best match based on a chi-square minimization. For numerical details, we refer to the descriptions in Strassmeier et al. (\cite{vpnep}). All abundances in this paper were determined from the phase-averaged spectra with exceptionally high S/N, varying from 628 (RSP\,556) to 3296 (RSP\,177). Figure~\ref{F2} shows a zoom into the Li region for all stars. The detailed fits are shown in the Appendix in Fig.~\ref{F_App1}. Figure~\ref{F3} is an example (RSP\,177, $T_{\rm eff}$=5770\,K) for the expected changes in the line core due to rotational modulation and/or cyclic variability, if present. Table~\ref{T2} provides the total lithium abundance A(Li) according to Eq.~\ref{eq-ALi} and the $^6$Li/$^7$Li isotopic ratio whenever detected:
\begin{equation}
\mathrm{A(Li)} = \log_{10} { \left( { \frac{N_{\rm{Li}}}{N_{\rm{H}}}} \right) } + 12 \ . \label{eq-ALi}
\end{equation}

\begin{table}[!tbh]
\caption{LTE lithium and iron abundances.}\label{T2}
\begin{flushleft}
\begin{tabular}{lllll}
\hline\hline
\noalign{\smallskip}
RSP   & A(Li)  & $^6$Li/$^7$Li & [Fe/H]  & $\chi^2$ \\
         & (H=12) & (\%\,)        & (solar) & fit      \\
\noalign{\smallskip}\hline\noalign{\smallskip}
 344 & 2.747$\pm$0.001  &\dots & 0.149$\pm$0.007 & 91 \\
 233 & 2.870$\pm$0.001  &\dots & 0.140$\pm$0.007 & 152 \\
 340 & 3.012$\pm$0.001  &\dots & 0.153$\pm$0.009 & 126 \\
 137 & 2.669$\pm$0.001  &\dots & 0.128$\pm$0.005 & 55 \\
 177 & 2.403$\pm$0.001  &\dots & 0.185$\pm$0.003 & 92 \\
 440 & 2.609$\pm$0.001  &\dots & 0.167$\pm$0.005 & 63 \\
 227 & 1.737$\pm$0.001  &\dots & 0.212$\pm$0.003 & 71 \\
 134 & 0.422$\pm$0.010  &14.9  & 0.178$\pm$0.004 & 164 \\
 439 & 0.352$\pm$0.012  &15.7  & 0.216$\pm$0.004 & 114 \\
 225 & 1.458$\pm$0.002  &\dots & 0.183$\pm$0.003 & 190 \\
  95 & 0.214$\pm$0.010  &14.7  & 0.211$\pm$0.003 & 270 \\
 429 & 0.317$\pm$0.011  &10.5  & 0.175$\pm$0.003 & 195 \\
 198 & 0.509$\pm$0.008  &\dots & 0.178$\pm$0.002 & 578 \\
  68 & 0.008$\pm$0.010  &\dots & 0.209$\pm$0.002 & 756 \\
 587 & 0.005$\pm$0.008  &\dots & 0.319$\pm$0.004 & 876 \\
 571 & --0.086$\pm$0.006&\dots & 0.350$\pm$0.008 & 4154 \\
 216 & 0.053$\pm$0.010  & 15.5 & 0.217$\pm$0.002 & 787 \\
 409 & --0.30$\pm$0.02  &\dots & 0.025$\pm$0.1  & 28709 \\
 133 & --0.32$\pm$0.02  &\dots & --0.11$\pm$0.1 & 23236 \\
 542 & --0.42$\pm$0.03  &\dots & --0.32$\pm$0.1 & 42728 \\
 556 & --1.00$\pm$0.03  &\dots & --0.32$\pm$0.1 & 35586 \\
\noalign{\smallskip}
\hline
\end{tabular}
\tablefoot{Errors for A(Li) and [Fe/H] are internal (fitting) errors. Isotope ratios, $^6$Li/$^7$Li, are only listed if the best formal fit requires an isotopic ratio $>3$\% while $T_{\rm eff}>4500$\,K. The $\chi^2$ values of the fit refer to a 44-pixel range in wavelength space centered at Li\,{\sc i}. }
\end{flushleft}
\end{table}

As the line list we adopted the list from Strassmeier \& Steffen (\cite{xiboo}), which is based on Mel\'endez et al. (\cite{mel12}), except for the Li doublet with isotopic hyperfine components and the broadening constants, and the vanadium blend at 6708.1096\,\AA\ (Lawler et al. \cite{law}). Also added are lines of the TiO molecule of five Ti isotopes from the updated list based on Plez (\cite{tio}). The free parameters are A(Li), $^6$Li/$^7$Li, [Fe/H], a global wavelength adjustment, and a global Gaussian line broadening (FWHM), which are applied in velocity space to the synthetic interpolated line profiles to match the observational data as closely as possible. FWHM represents the full width half maximum of the applied Gaussian kernel and represents the combined instrumental plus (Gaussian) macro turbulence broadening. We note that we had to allow TurboMPfit to alter the continuum level as well. This has practically no impact for the minimization process for stars warmer than $T_{\rm eff}\approx 5000$~K but is needed for the coolest targets due to the dramatically increased molecular suppression of the continuum once $T_{\rm eff}<4500$~K. The micro turbulence remained fixed to the values in Table~\ref{T1}.

TurboMPfit fitting errors are interpreted as internal errors and are also listed in Table~\ref{T2}. The related $\chi^2$ from the fit range 6707.3--6708.5\,\AA\ (covered by 44 CCD pixels) is a good indicator for the reliability of the Li abundance but strongly depends on the completeness of the line list. The present high data quality excludes the limited S/N as the (usual) dominant error source. We note that the reason for the larger than expected $\chi^2$ with respect to the S/N values is certainly not some unknown systematics in the stellar spectra but is rather related to the imperfect model (synthetic spectra) we are using for fitting. We conclude that the external errors of our Li abundances are almost exclusively driven by the uncertainty of the stellar effective temperature, and only for the coolest stars also by the blending line list. Due to the increasing molecular contribution with lower temperature, the line list becomes more critical the cooler the target. Our line list allows excellent line-profile fits to within the present data quality for $T_{\rm eff}>4500$\,K but basically fails for $T_{\rm eff}<4200$\,K. Cummings et al. (\cite{cumm}) had also noted that their errors in A(Li) that result from errors in $T_{\rm eff}$ go fairly parallel to the A(Li) vs. $T_{\rm eff}$  trend in the sense the cooler the target the higher the uncertainty. We refer to their discussion and to Dumont et al. (\cite{dum:cha}) for a general summary of the extensive open cluster Li data including the Hyades.

We estimate external abundance errors by redoing the fits for all targets but with $T_{\rm eff}\pm100$\,K with respect to the nominal value. The maximum/minimum range for A(Li) and [Fe/H] is then adopted as a more realistic estimate of their true error. For A(Li), we obtain $\pm$0.08 to $\pm$0.15 for $T_{\rm eff}\approx6000$\,K to $\approx4500$\,K, respectively. For [Fe/H], we obtain $\pm$0.05 to $\pm$0.01 for the same $T_{\rm eff}$ range (smaller errors for the cooler stars in this case).

The only stars for which the formally best fit required $^6$Li were RSP~134, 439, 95, 429, and 216. Their isotopic ratios are between 10--15\% (Table~\ref{T2}). However,   we deem none of them to be a real detection and attribute all Li to $^7$Li. This is because the formal fitting errors provided by the Levenberg-Marquard algorithm are not relevant in this case because uncertainties are driven by the uncertainties of the line list rather than the fit to the data. The $^7$Li abundances range from a high of +3.01 (RSP\,340, $T_{\rm eff}$=6158\,K) to a low of negative (logarithmic) values for the most cool targets. The latter are then just upper limits. 3D~NLTE corrections are not available for all temperatures of our Hyades dwarfs and could thus not be applied uniformly. All Li abundances in this paper are thus LTE abundances. The real uncertainties for the targets with very low A(Li) values, basically all targets cooler than 4200\,K, are again driven by the molecular contamination (due to imperfect continuum setting, line profile blending, and partly wrong line parameters) and are merely estimates.

\begin{table}[!tbh]
\caption{Logarithmic Ca\,{\sc ii} IRT emission-line fluxes and radiative losses.}\label{T3}
\begin{flushleft}
\begin{tabular}{lllllll}
\hline\hline
\noalign{\smallskip}
RSP & $B-V$ & ${\cal F}_{\rm IRT1}$ & ${\cal F}_{\rm IRT2}$ & ${\cal F}_{\rm IRT3}$ & $R_{\rm IRT}$ & $R^\prime_{\rm IRT}$\\
    & (mag) & \multicolumn{3}{c}{(erg\,cm$^{-2}$s$^{-1}$)} & \multicolumn{2}{c}{($\sigma T^4_{\rm eff}$)} \\
\noalign{\smallskip}\hline\noalign{\smallskip}
 344 & 0.584 & 6.434 & 6.317 & 6.321 & --4.050 & --4.796 \\
 233 & 0.572 & 6.446 & 6.339 & 6.339 & --4.037 & --4.714 \\
 340 & 0.545 & 6.445 & 6.325 & 6.333 & --4.063 & --4.933 \\
 137 & 0.605 & 6.422 & 6.310 & 6.312 & --4.025 & --4.838 \\
 177 & 0.664 & 6.405 & 6.320 & 6.312 & --3.973 & --4.598 \\
 440 & 0.632 & 6.415 & 6.323 & 6.318 & --3.992 & --4.652 \\
 227 & 0.730 & 6.341 & 6.241 & 6.232 & --3.988 & --4.722 \\
 134 & 0.899 & 6.251 & 6.175 & 6.152 & --3.921 & --4.418 \\
 439 & 0.921 & 6.245 & 6.175 & 6.150 & --3.892 & --4.340 \\
 225 & 0.748 & 6.324 & 6.219 & 6.211 & --3.989 & --4.724 \\
  95 & 0.906 & 6.233 & 6.153 & 6.130 & --3.929 & --4.473 \\
 429 & 0.897 & 6.264 & 6.195 & 6.171 & --3.896 & --4.358 \\
 198 & 0.819 & 6.292 & 6.202 & 6.186 & --3.951 & --4.548 \\
  68 & 0.920 & 6.236 & 6.157 & 6.135 & --3.865 & --4.338 \\
 587 & 1.071 & 6.117 & 6.043 & 6.018 & --3.881 & --4.257 \\
 571 & 0.890 & 6.230 & 6.155 & 6.131 & --3.694 & --4.238 \\
 216 & 0.984 & 6.169 & 6.082 & 6.063 & --3.927 & --4.298 \\
 409 & 1.300 & 5.959 & 5.879 & 5.857 & --3.865 & --4.407 \\
 133 & 1.350 & 5.949 & 5.867 & 5.843 & --3.834 & --4.078 \\
 542 & 1.260 & 6.006 & 5.934 & 5.912 & --3.730 & --4.164 \\
 556 & 1.463 & 5.901 & 5.815 & 5.790 & --3.747 & --4.034 \\
\noalign{\smallskip}
\hline
\end{tabular}
\tablefoot{${\cal F}$ is the absolute emission-line flux, $R$ and $R'$ are the (logarithmic) total atmospheric and chromospheric radiative loss in units of the bolometric stellar flux, respectively. }
\end{flushleft}
\end{table}

\subsection{Chromospheric \ion{Ca}{ii} IRT flux}\label{S_IRT}

We measured the absolute line-core flux, ${\cal F}$, in the Ca\,{\sc ii} infrared triplet from our CD\,VI spectra at 8498, 8542 and 8662\,\AA\ (dubbed IRT-1, IRT-2, and IRT-3, respectively). The central portions of these spectra are visualized in Fig.~\ref{F_App2} in the Appendix. Fluxes in erg\,cm$^{-2}$s$^{-1}$ were computed from the measured relative flux in a 1-\AA\ band centered on the line core and scaled with the absolute continuum flux at these wavelengths. The latter is obtained for the $B-V$ range of our targets from the relations provided by Hall (\cite{hall96}). We followed the same procedure as in our recent spectroscopic survey of the ecliptic poles (Strassmeier et al.~\cite{vpnep}) or our search for Maunder-minimum candidates (J\"arvinen \& Strassmeier \cite{jar:str}) and refer to these papers for details.

The total atmospheric radiative loss in the Ca\,{\sc ii} IRT, $R_{\rm IRT}$, is determined from the sum of the absolute fluxes from the three lines in units of the stellar bolometric luminosity $\sigma T_{\rm eff}^4$. 
\begin{equation}\label{eqRIRT}
R_{\rm IRT} =  \log_{10} { \left( 
\frac{{\cal F}_{\rm IRT-1} + {\cal F}_{\rm IRT-2} + {\cal F}_{\rm IRT-3}}{\sigma \ T_{\rm eff}^4} 
\right) }, 
\end{equation}
It is an indicator of the star's atmospheric magnetic activity, photosphere and chromosphere combined. We also provide the chromospheric radiative loss, $R^\prime_{\rm IRT}$, by removing the expected photospheric contribution from the individual fluxes via bona-fide inactive stars. Such a correction is not unproblematic because it always makes the idealized assumption that inactive stars come without a chromosphere and that these star's photospheric contribution is unrelated to its activity, thus will introduce unspecified additional uncertainty. Nevertheless, Martin et al. (\cite{martin}) provided a convenient collection of such corrections based on 26 bona-fide inactive stars as a function of $B-V$ and $v\sin i$. These spectra were taken at a spectral resolution of $\sim$20,000 which is, however, over six times lower than the resolution of our present Hyades spectra. J\"arvinen et al. (\cite{jar:kor}) had provided IRT corrections based on 52 inactive stars observed with STELLA at a spectral resolution of 55,000. For the $B-V$ range of our Hyades sample, these corrections are on average 4--5\% lower than the ones by Martin et al. (\cite{martin}). Although this is only a small amount well within the expected flux uncertainties, the difference is systematic and likely due to the different spectral resolution (see J\"arvinen \& Strassmeier \cite{jar:str} for more details). Numerical photospheric corrections are between 1$\times$10$^6$ and 6$\times$10$^6$ erg\,cm$^{-2}$s$^{-1}$ for the sum of all three IRT lines for $B-V$ of 1.4 to 0.6\,mag, respectively. Table~\ref{T3} lists the corrected radiative losses based on the J\"arvinen et al. (\cite{jar:kor}) correction along with the uncorrected fluxes. We refrain from further corrections, for example of the basal flux due to acoustic heating (c/o Schrijver \cite{schrijver}, see discussion in Martin et al. \cite{martin}). In any case the photospherically corrected radiative losses are an indirect measure of the magnetic flux from the chromosphere.

Because only a single spectrum per target with CD\,VI is available, our fluxes are just snapshots from the range of expected variability due to rotational modulation. Measuring errors for $R_{\rm IRT}$ are driven by the error for $T_{\rm eff}$ and $B-V$ and are not explicitly listed in the table but are estimated to be of the order of 10--20\% following the $T_{\rm eff}$ evaluation in Cummings et al. (\cite{cumm}) and Douglas et al. (\cite{doug19}). We expect rotational IRT flux modulation that is significantly smaller than $\approx$15\% (as measured for the RS\,CVn binary $\lambda$\,And; Adebali et al. \cite{lamand}).

\begin{table}[!t]
\caption{Magnetic field strengths.}\label{T4}
\begin{flushleft}
\begin{tabular}{llllll}
\hline\hline
\noalign{\smallskip}
 RSP & Season & $\langle B_{\rm long}\rangle$ & $\Delta B_{\rm long}$ & $\mathrm{rms} B_{\rm long}$ & $\langle |B|\rangle$ \\
       &             & \multicolumn{4}{c}{(G)} \\
\noalign{\smallskip}\hline\noalign{\smallskip}
 344 & S20 & --6.5  & 73 & 10 & 13.3$\pm$1.7 \\
 233 & S20 & --1.3 & 41 & 9 & 11.6$\pm$1.5 \\
 340 & S20 & --3.0 & 30 & 12 & 11.4$\pm$1.6 \\
 137 & S20 & --1.6 & 47 & 12 & 13.5$\pm$2.6 \\
 177 & S20,S22 & --2.0 & 86 & 13 & 14.1$\pm$2.5 \\
 440 & S20 & --2.1 & 49 & 11 & 14.4$\pm$2.6 \\
 227 & S21 & --1.3 & 53 & 9 & 10.9$\pm$2.7  \\
 134 & S21 & +1.8 & 45 & 11 & 16.0$\pm$1.5 \\
 439 & S21 & --12 & 61 & 15 & 25.2$\pm$2.4  \\
 225 & S20,S22 & +2.0 & 90 & 15 & 15.8$\pm$2.5 \\
  95 & S20,S22 & --0.1 & 51 & 12 & 16.3$\pm$1.3  \\
 429 & S20,S22 & --1.5 & 57 & 18 & 20.5$\pm$2.5  \\
 198 & S20,S22 & +7.5 & 40 & 9 & 17.6$\pm$2.0 \\
  68 & S22 & --3.1 & 71 & 11 & 13.0$\pm$2.8  \\
 587 & S20,S22 & +8.0 & 85 & 23 & 37.2$\pm$4.8  \\
 571 & S20,S22 & +10 & 174 & 23 & 31.3$\pm$3.7  \\
 216 & S20,S22 & +0.2 & 84 & 15 & 17.6$\pm$1.4  \\
 409 & S20,S22 & +18 & 180 & 38 & 74.9$\pm$9.8  \\
 133 & S22 & +11 & 125 & 48 & 90.5$\pm$26  \\
 542 & S20,S22 & +10 & 233 & 60 & 95$\pm$9  \\
 556 & S22 & +67 & 103 & 24 & 217$\pm$27 \\
\noalign{\smallskip}
\hline
\end{tabular}
\tablefoot{Column Season identifies the observing season for a particular dataset (S20 = season 2020/21, S21 = season 2021/22, S22 = season 2022/23). Magnetic values are from all observing seasons and phases combined. $\langle B_{\rm long}\rangle$ is the phase-average longitudinal (signed) field, $\Delta B_{\rm long}$ its min-max variability amplitude, $\mathrm{rms} B_{\rm long}$ its rms from the mean, and $\langle |B|\rangle$ the unsigned average longitudinal field. The latter errors are scaled rms errors. }
\end{flushleft}
\end{table}

\begin{figure*}
{\bf a.}\hspace{60mm}{\bf b.}\hspace{60mm}{\bf c.}\\
\includegraphics[angle=0,width=6cm,clip]{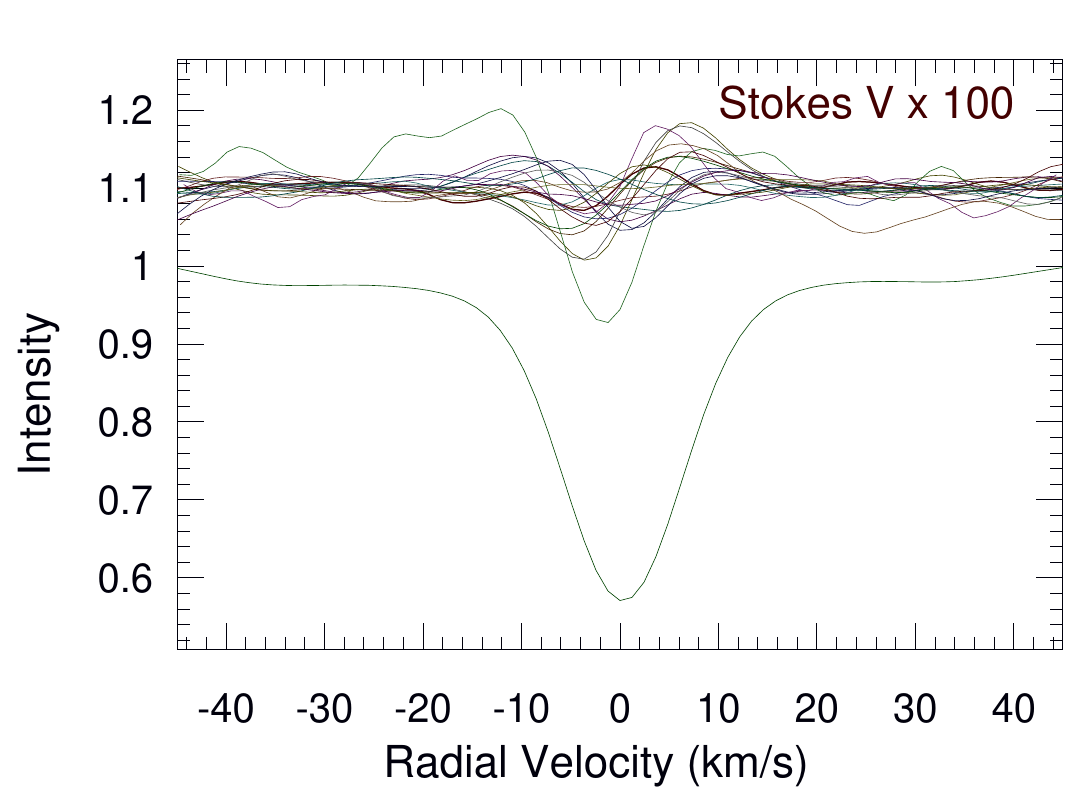}
\includegraphics[angle=0,width=6cm,clip]{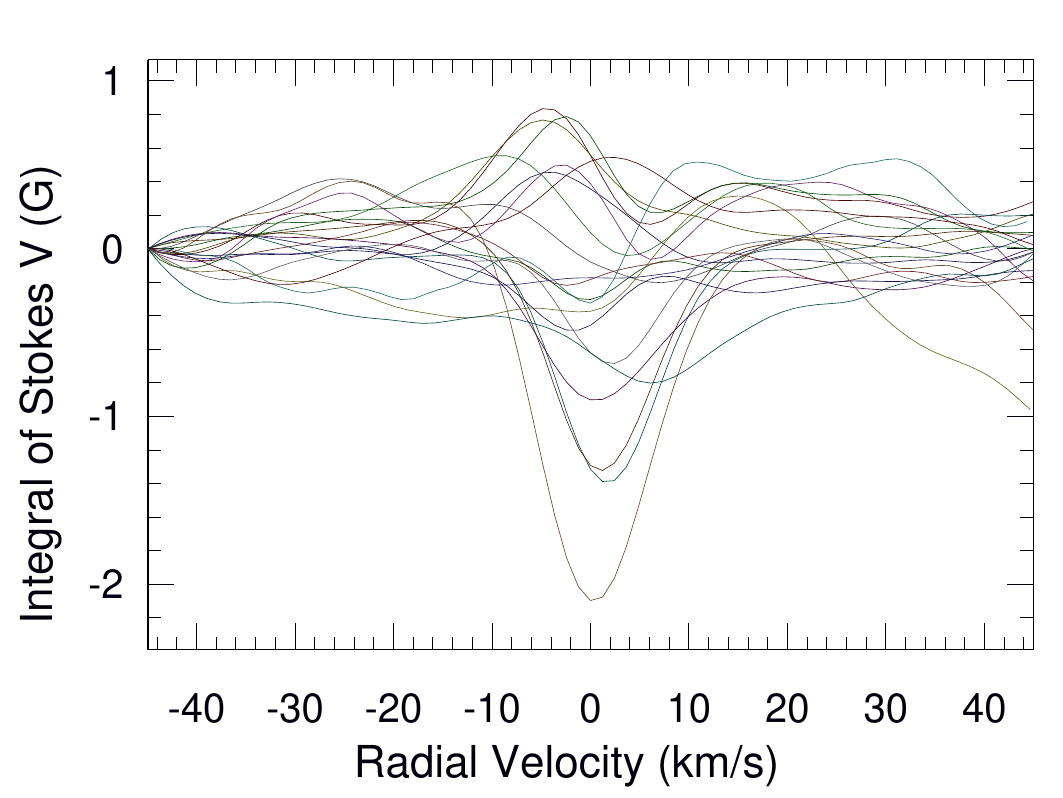}
\includegraphics[angle=0,width=6cm,clip]{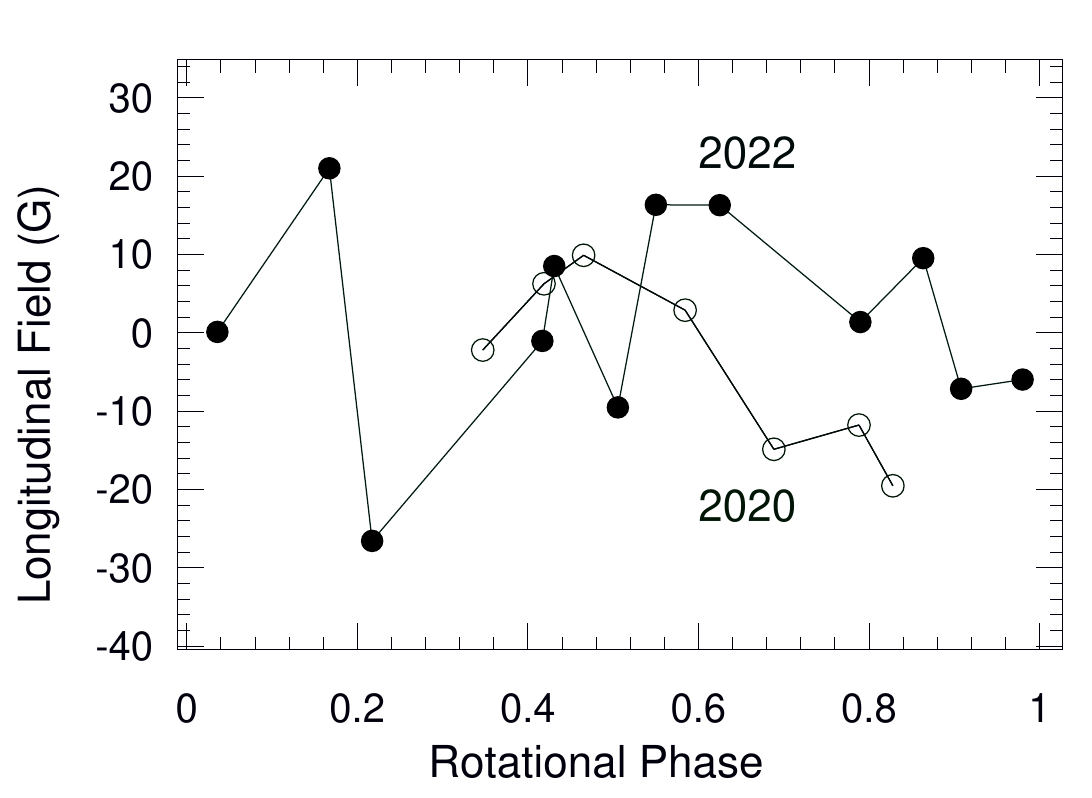}
\caption{Representative Stokes-V data and results for RSP\,177. $a.$ Top spectra: Overplot of all available 19 LSD Stokes-V line profiles in units of the continuum expanded by a factor 100 and shifted in intensity by 0.1 for better visibility. Bottom spectrum: Example LSD Stokes-I profile. $b.$ Phase-resolved disk-integrated Stokes-V LSD line profiles in units of Gauss. $c.$ Longitudinal magnetic field versus rotational phase. Indicated are the two observing seasons for this target (S22: filled dots, S20: rings) .The Appendix provides similar figures for the other targets (Fig.~\ref{F_App3}).}
\label{F4}
\end{figure*}

\subsection{Magnetic field measurements}\label{SMF}

We first reconstruct Stokes-V LSD spectral lines from the combined spectra from CD\,III and CD\,V. Figure~\ref{F4}a shows a comparison of the LSD Stokes-V (and Stokes-I) profiles for one representative target (RSP\,177, $T_{\rm eff}$=5770\,K). These LSD profiles are then converted to a disk-integrated Stokes-V profile in units of Gauss (Fig.~\ref{F4}b) following the procedure laid out in Kochukhov et al. (\cite{koclsd}) based on the earlier notions from Solanki \& Stenflo (\cite{sol:ste}). Spectral lines for the deconvolution were selected from the most-recent version of the Vienna Atomic Line Database (VALD-3; Ryabchikova et al. \cite{vald}) based on line strength, Land\'e factor, and blending. The number of available lines varies with stellar effective temperature and ranges from a minimum of 3326 for RSP\,340 to 27,158 for RSP\,542. 

The LSD wavelength integration gives the longitudinal field component $B_{\rm long}$, with either positive or negative polarity (Fig.~\ref{F4}c) by assuming the standard weak-field approximation (e.g., Stenflo \cite{stenflo89}). Its absolute value, $|B|$, is obtained by integrating the positive and negative polarities and is commonly called the ``unsigned'' magnetic field. Longitudinal fields are extracted for every available LSD Stokes-V line profile. Individual errors typically range from $\pm$0.7\,G to $\pm$1.0\,G for the majority of targets depending on the number of available lines and the S/N per pixel. The five targets with $T_{\rm eff}<4500$~K have errors of up to $\pm$10~G.  Rotational phases, $\varphi$, were calculated with the respective rotation periods from Table~\ref{T1} and an arbitrary zero point assumed to be the time of the first integration of our first Hyades target; BJD\,2,459,185.7700 (= Dec.\,2, 2020). Whenever there is more than one period determination, we preferred the period from K2 data or, if no K2 period was available, the one with the smaller error. For the six stars where this is the case the period used is the one listed in Table~\ref{T1} and referenced in Table~\ref{T1-App} in the Appendix. Phase gaps due to bad weather are unavoidable on Mt.~Graham and are identified in the last three columns of Table~\ref{T1-App}. We consider the phase coverage ``partial'' (indicated there by a letter P) if a gap of $\geq$0.30 phases exists.

Figure~\ref{F4}c shows the phase variations of the measured longitudinal fields for one example target; RSP\,177. Plots for the other targets are collected in Fig.~\ref{F_App3} in the Appendix. We apply a simple harmonic (multi-periodic) fit to all rotationally-phased measurements of all targets in order to estimate the distribution of isolated polarities, dubbed magnetic spots. From Fig.~\ref{F_App3} in the Appendix, we see that most of our targets exhibit a relatively high frequency of variability during a rotation cycle. Among the most extreme are RSP\,134 with four maxima (i.e., positive polarity dominating) and four minima (negative polarity dominating) and RSP\,177 in 2020 with only one maximum and one minimum. This is not unexpected when compared to the Sun. In a recent technical paper on Sun-as-a-star Stokes-V polarimetry with PEPSI (Strassmeier  et al. \cite{sdi}), we observed changes of the shape of its Stokes-V line profile within four days ($\approx$0.15 rotational phases; also with the same sampling of one spectrum per day) due to the change in dominance from positive to negative polarities and back again. Our more active Hyads apparently show a similar and likely even quikier variability than today's Sun.   

Table~\ref{T4} lists the mean longitudinal fields observed from the available data sets, $\langle B_{\rm long}\rangle$, its variability amplitude $\Delta B_{\rm long}$ and rms, and the time/phase-averaged unsigned magnetic fields $\langle|B|\rangle$ with errors. We note that these errors are simply rms values scaled by the square root of the number of available measurements. 

\begin{sidewaysfigure*}
\hspace*{5mm}{\bf a.}\hspace{60mm}{\bf b.}\hspace{60mm}{\bf c.}\hspace{60mm}{\bf d.}\\
\begin{center}
\includegraphics[angle=0,width=6cm,clip]{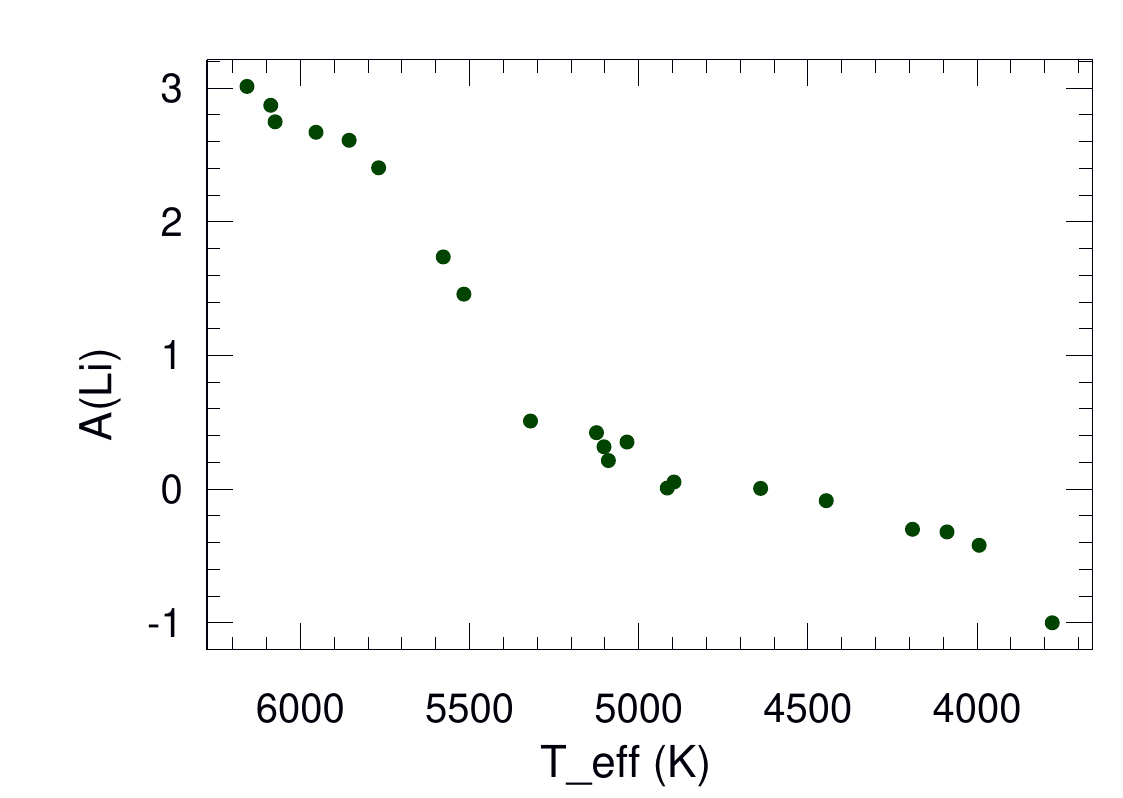}
\includegraphics[angle=0,width=6cm,clip]{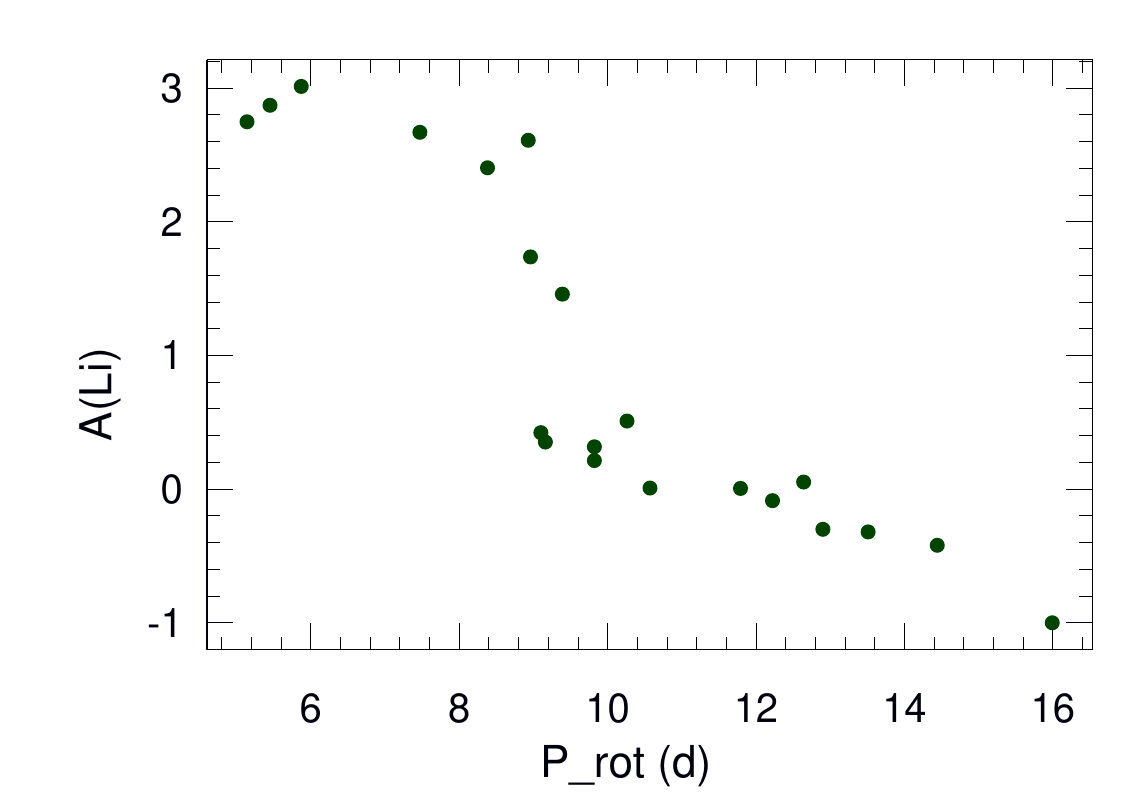}
\includegraphics[angle=0,width=6cm,clip]{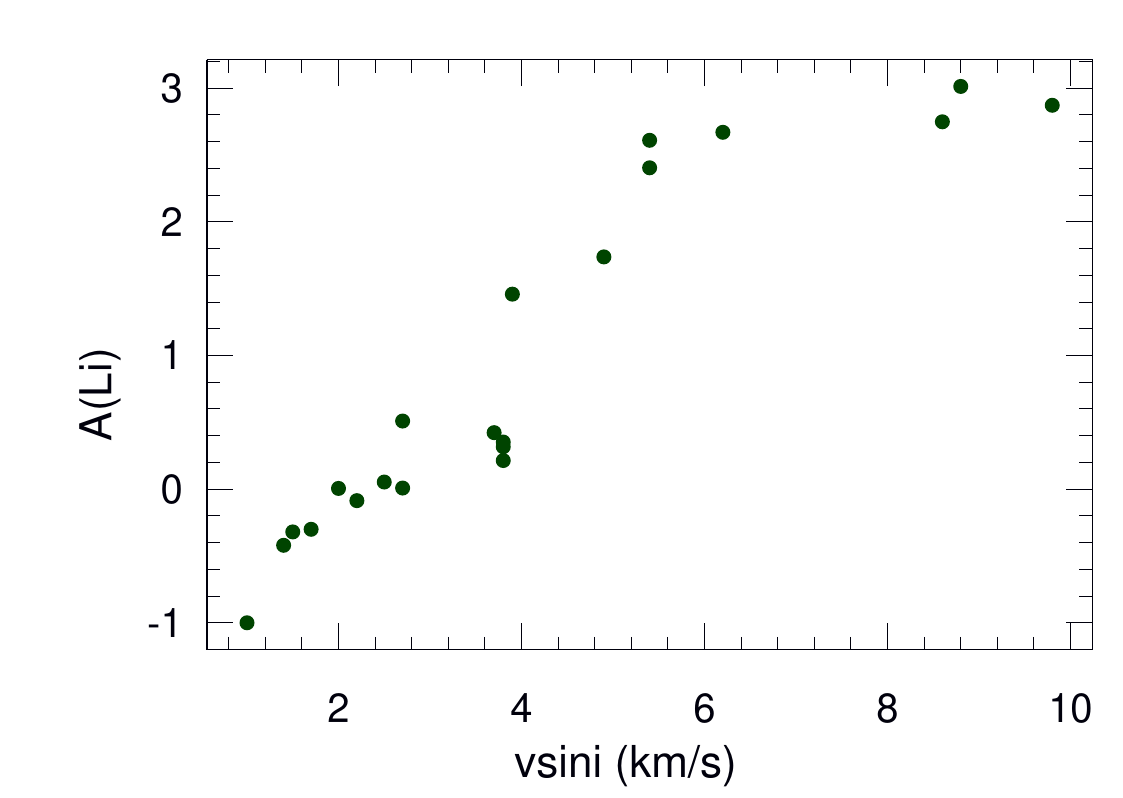}
\includegraphics[angle=0,width=6cm,clip]{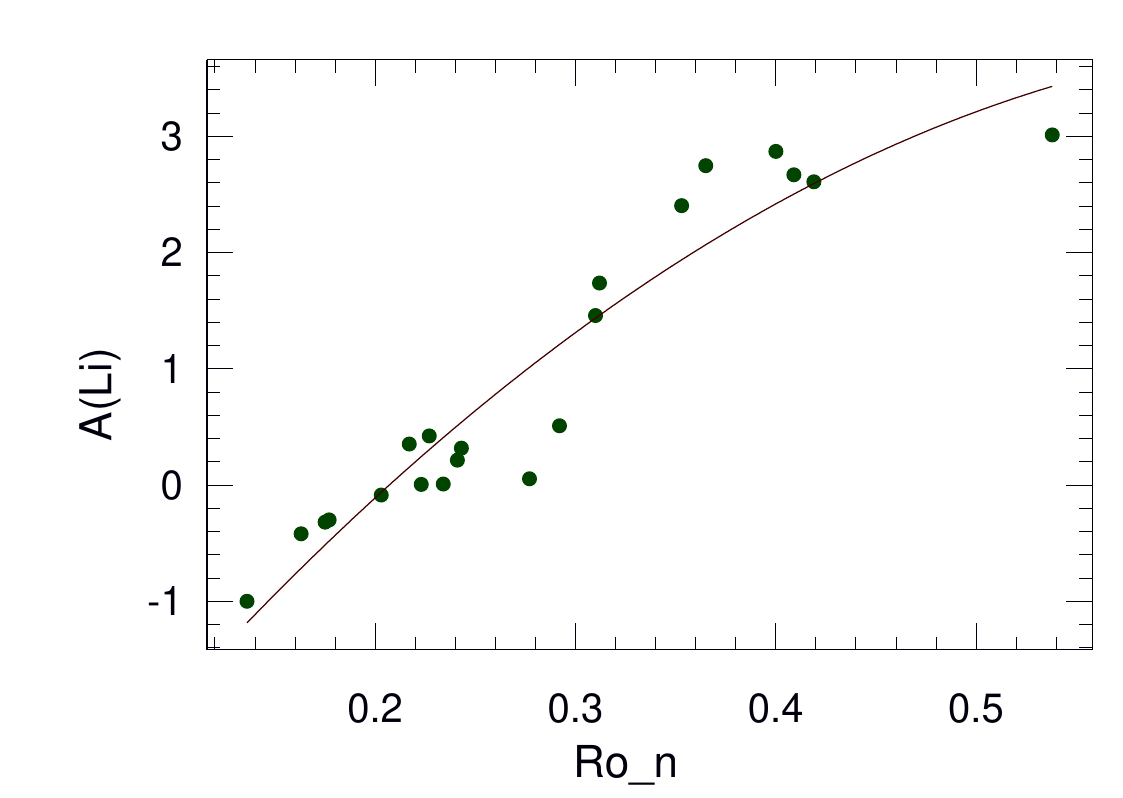}
\end{center}
\hspace*{5mm}{\bf e.}\hspace{60mm}{\bf f.}\hspace{60mm}{\bf g.}\hspace{60mm}{\bf h.}\\
\begin{center}
\includegraphics[angle=0,width=6cm,clip]{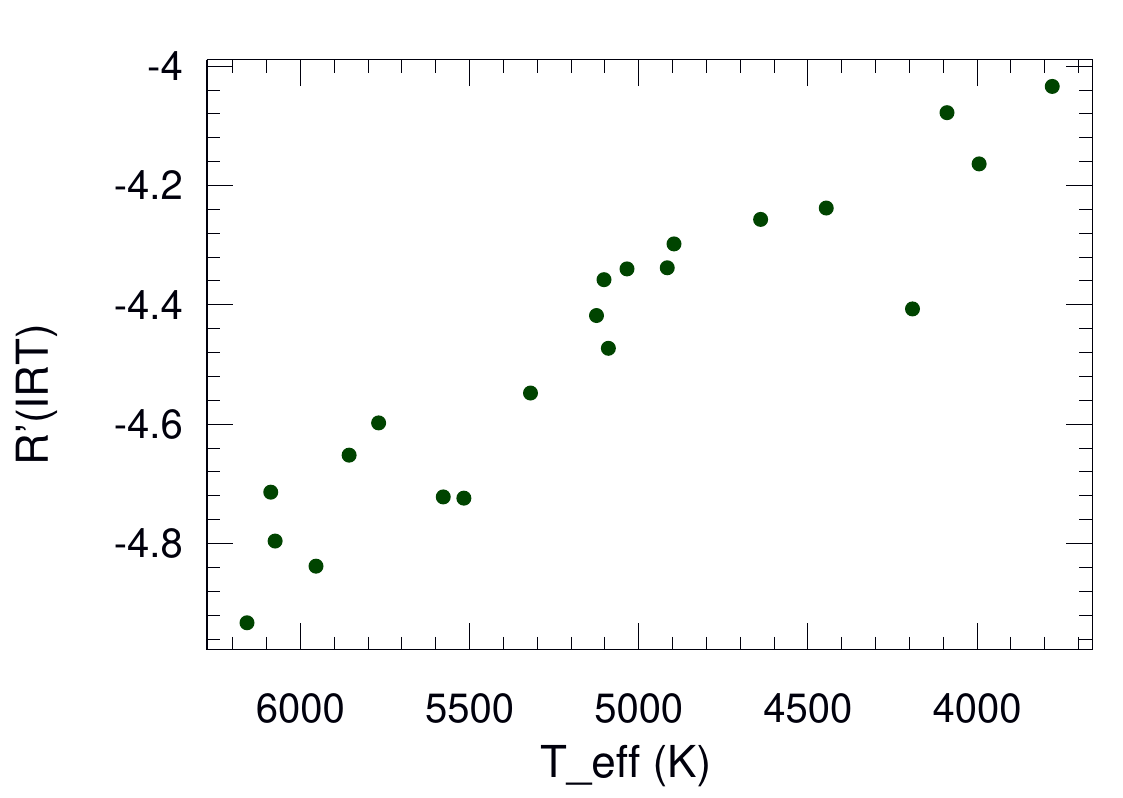}
\includegraphics[angle=0,width=6cm,clip]{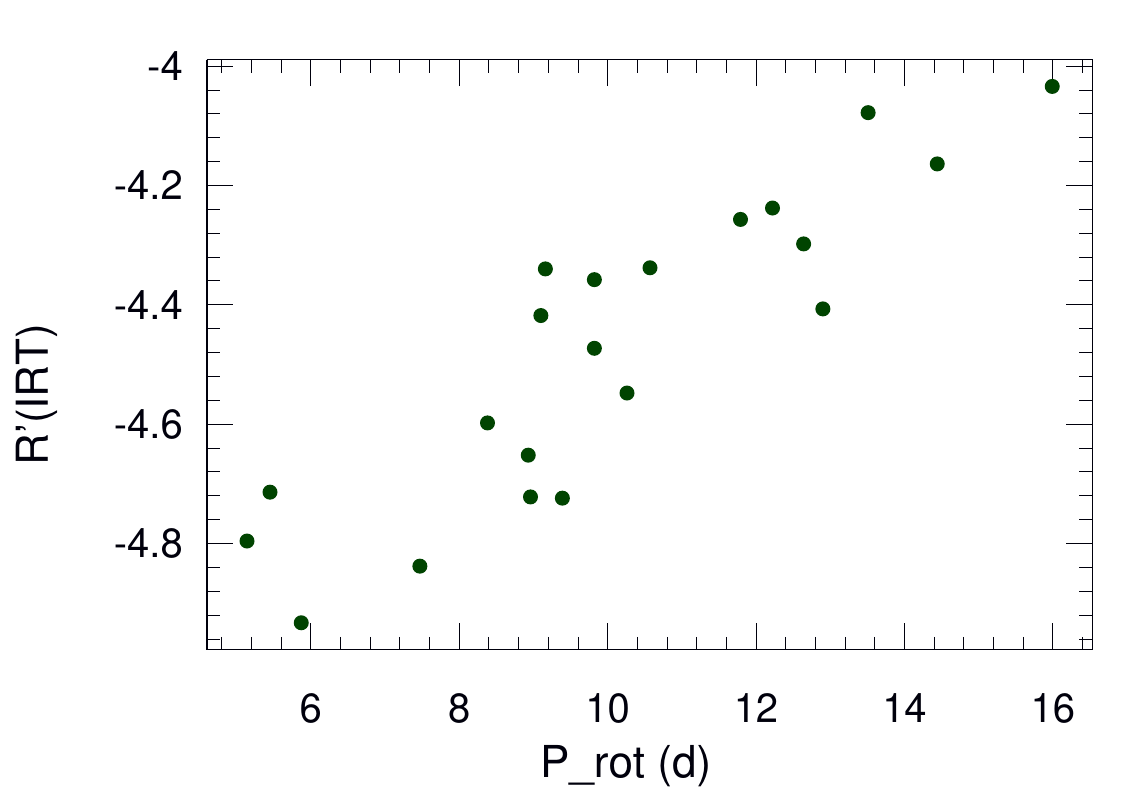}
\includegraphics[angle=0,width=6cm,clip]{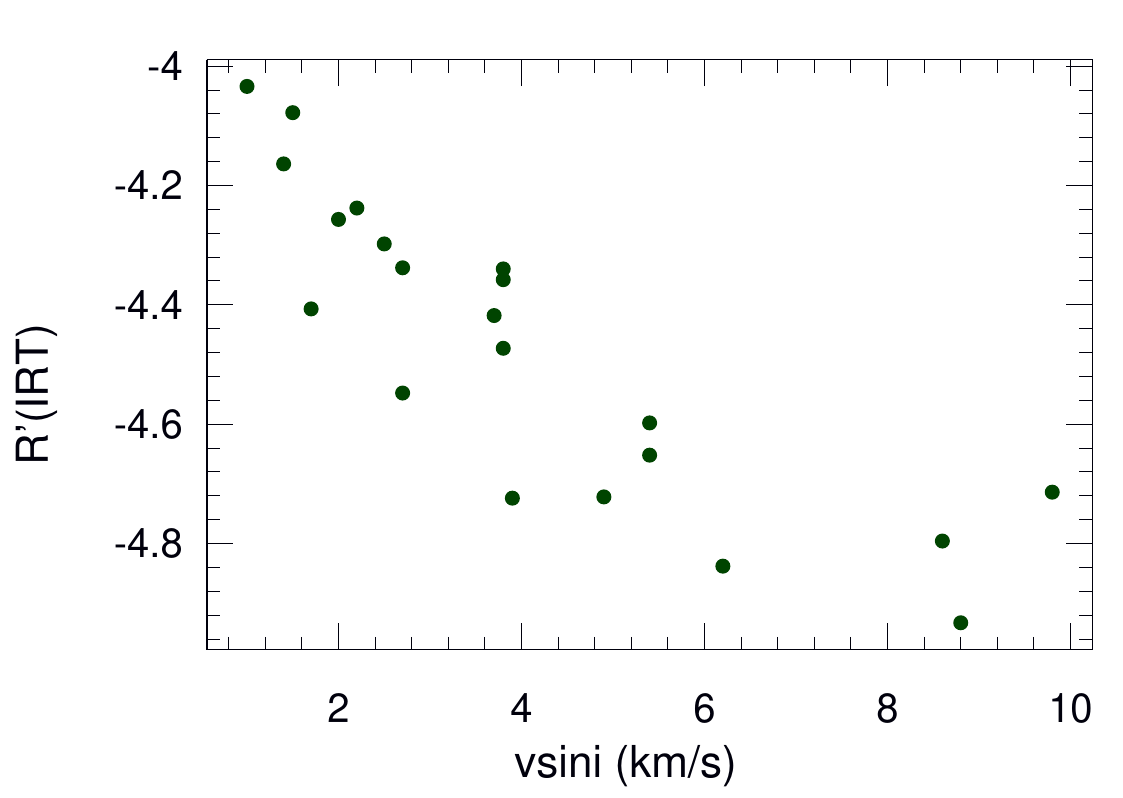}
\includegraphics[angle=0,width=6cm,clip]{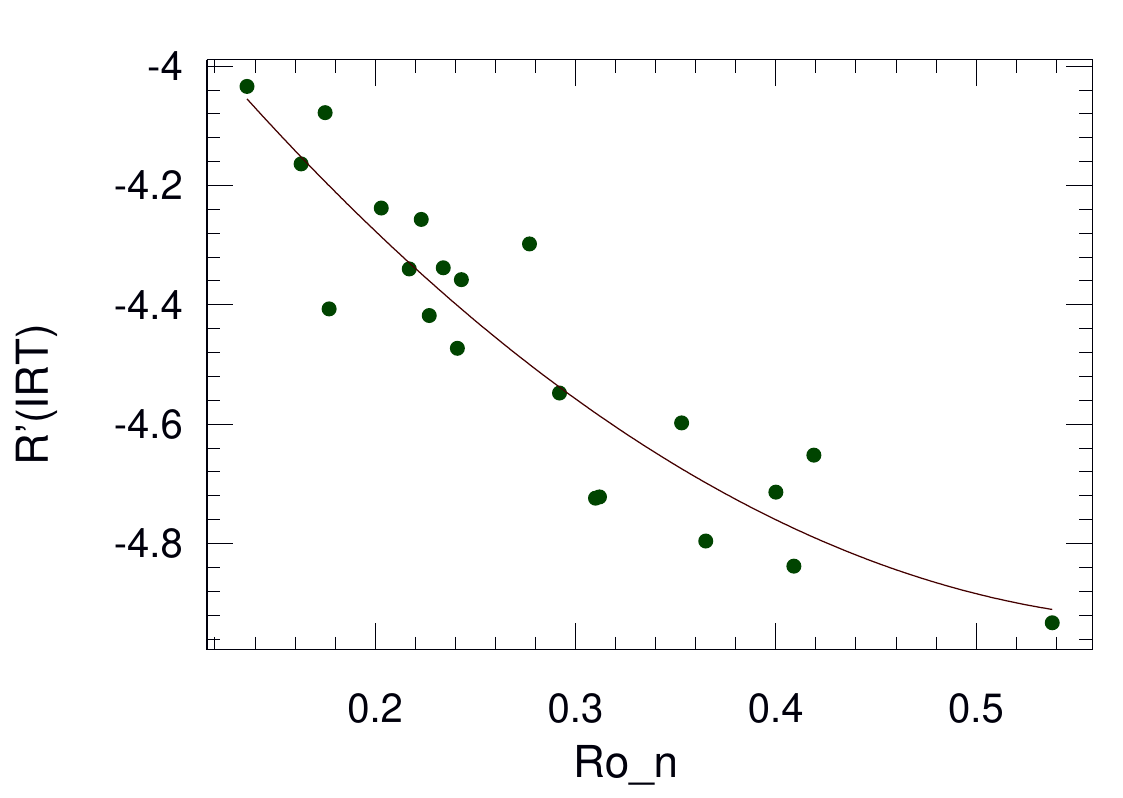}
\end{center}
\hspace*{5mm}{\bf i.}\hspace{60mm}{\bf j.}\hspace{60mm}{\bf k.}\hspace{60mm}{\bf l.}\\
\begin{center}
\includegraphics[angle=0,width=6cm,clip]{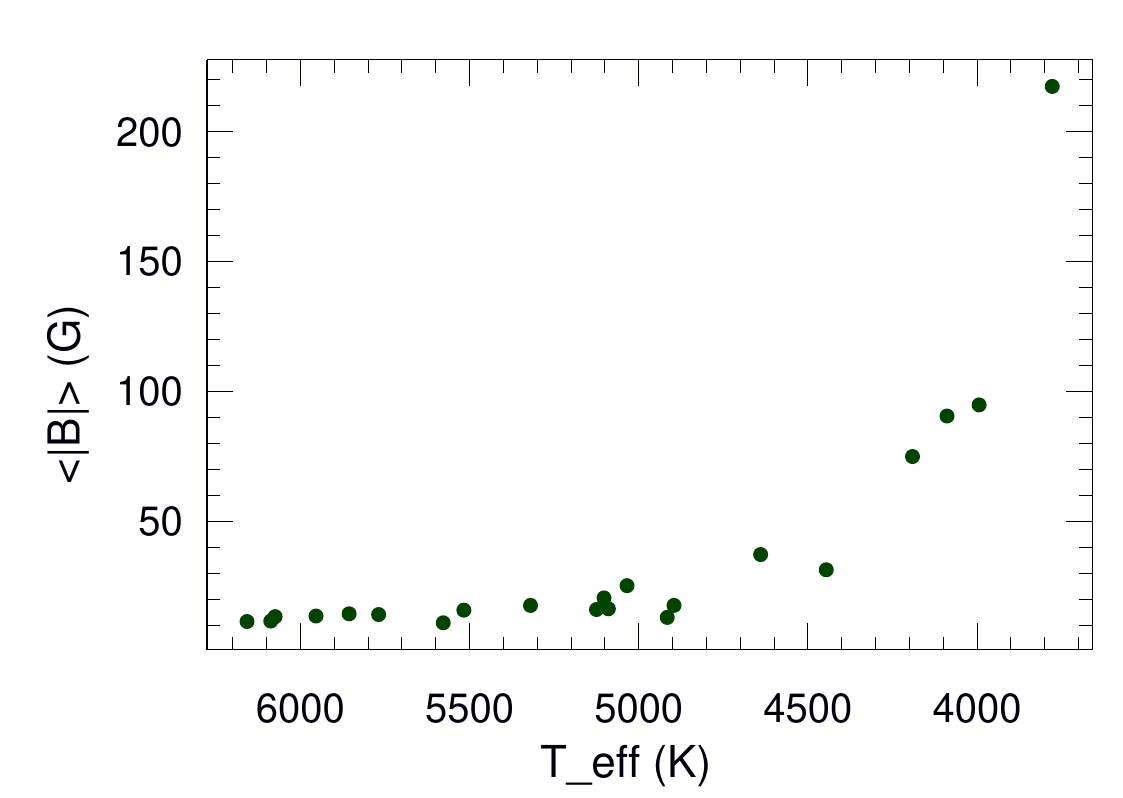}
\includegraphics[angle=0,width=6cm,clip]{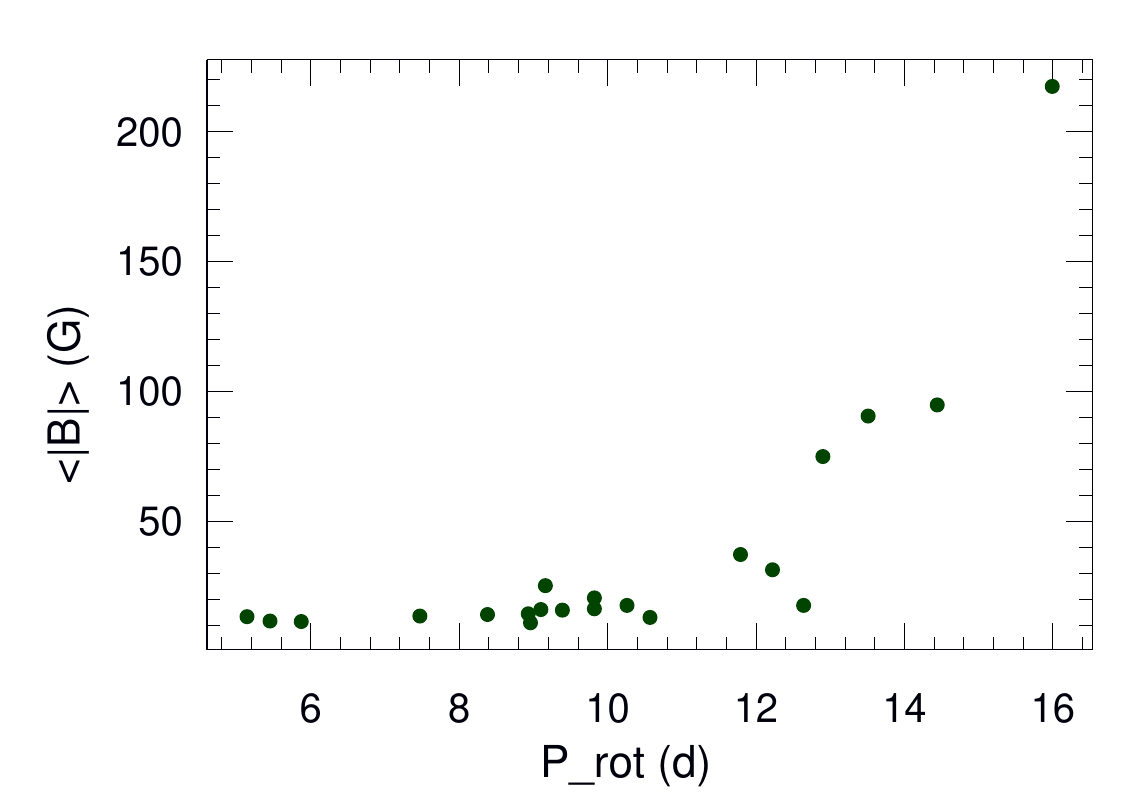}
\includegraphics[angle=0,width=6cm,clip]{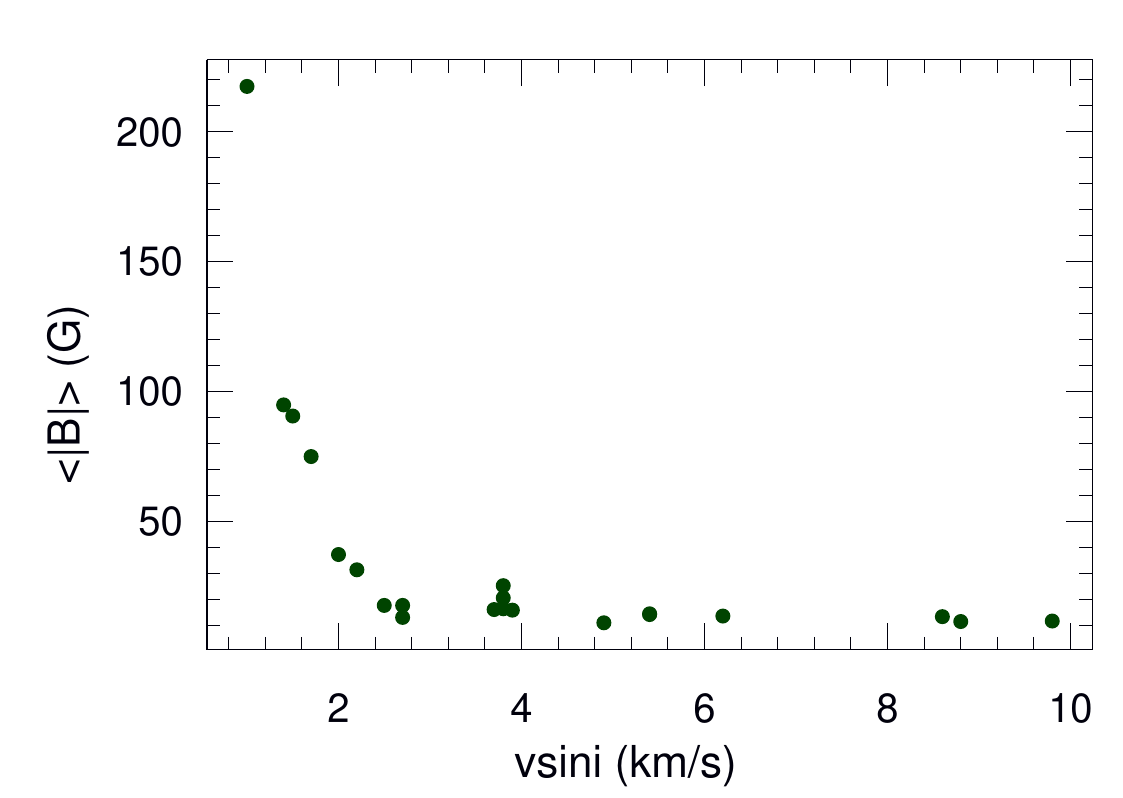}
\includegraphics[angle=0,width=6cm,clip]{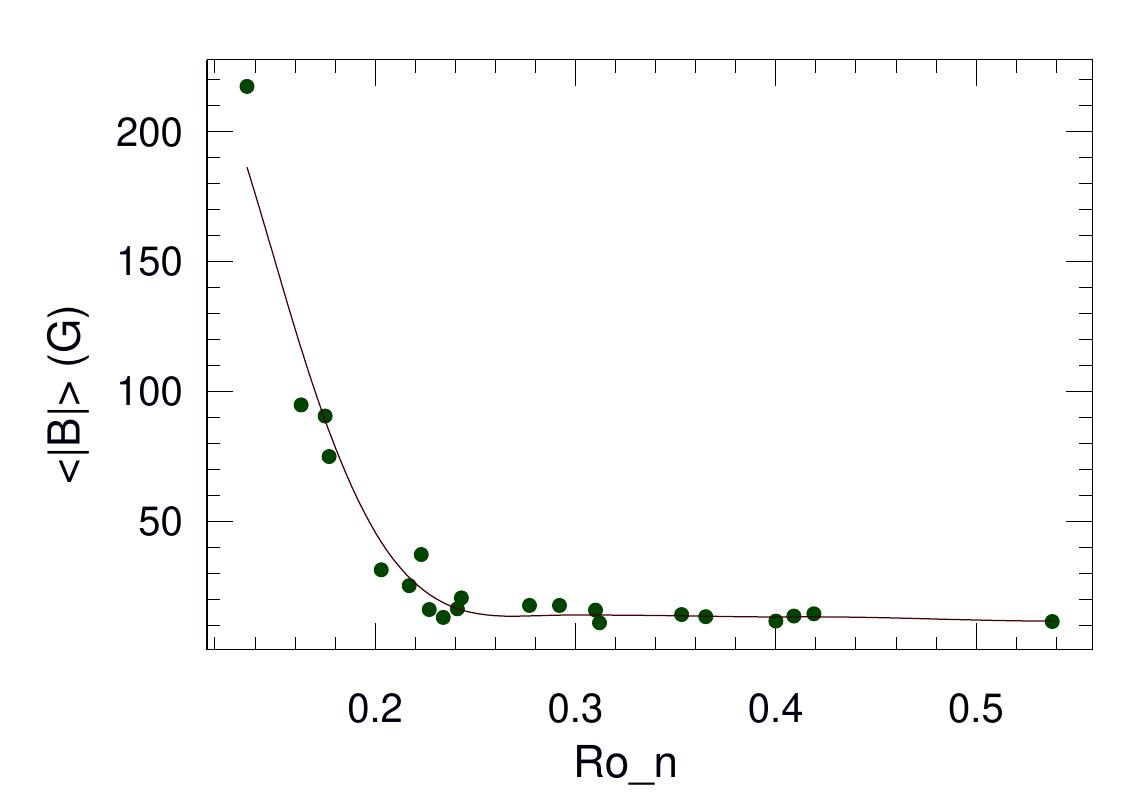}
\end{center}
\caption{Lithium-Activity-Rotation relations. The figure compares various relations in a matrix form (rows vs. columns) of A(Li), $R'_{\rm IRT}$, and $\langle |B|\rangle$ (from top to bottom) vs. $T_{\rm eff}$, $P_{\rm rot}$, $v\sin i$, and $\mathrm{Ro_n}$ (from left to right). Fits are shown for the panels for Rossby-number dependency. The line in panel $l$ is a combined and interpolated fit from the two-piece linear fit in Eq.~\ref{fitB}.}
 \label{F5}
\end{sidewaysfigure*}

\section{Hyades activity-rotation relations}\label{S6}

\subsection{Lithium dependency}\label{S61}

The high resolution and S/N of our data allow a new look at the Li dependence on stellar $T_{\rm eff}$ and $P_{\rm rot}$ for the age of the Hyades. Cummings et al. (\cite{cumm}) had already found a very tight A(Li) vs. $T_{\rm eff}$ relation for the Hyades G-dwarfs (5400--6200\,K) from their WIYN/Hydra sample of 34 targets. These data had a comparably low resolution of 13,500 but high S/N of typically several hundreds. For a direct comparison with our data, we have 8 targets in common (RSP\,137 = vB10, RSP\,177 = vB15, RSP\,225 = vB26, RSP\,227 = vB27, RSP\,233 = vB31, RSP\,340 = vB65, RSP\,344 = vB66, RSP\,440 = vB97). Their individual A(Li) differences scatter with a rms of 0.06\,dex with an average difference of just 0.016\,dex (ours on average lower with respect to Cummings et al.). While Cummings et al. obtained A(Li) from individual spectra, our A(Li) are from phase-averaged spectra. As we have shown in Fig.~\ref{F3} for one target (RSP\,177), the line-core changes due to rotational modulation are very small (typically $<0.8$\%) and impact on A(Li) only on the one-percent level. From this, we conclude that no zero-point shift between our two data sets must be applied despite of the time difference of the observations of more than a decade.

The first row  in the matrix-like Fig.~\ref{F5} shows plots of A(Li) versus $T_{\rm eff}$ (Fig.~\ref{F5}a), $P_{\rm rot}$ (Fig.~\ref{F5}b), $v\sin i$ (Fig.~\ref{F5}c), and $\mathrm{Ro_n}$ (Fig.~\ref{F5}d). Quite intriguing is the tight relation of A(Li) with $T_{\rm eff}$ for our full temperature range from 6160\,K to 3780\,K, and not just for the G-dwarfs as demonstrated by Cummings et al. (\cite{cumm}). We do not observe a continuous steep depletion trend for the stars cooler than approximately 5000\,K but more a weakened decrease towards a lower limit at around 1/25  ($\approx -0.4$\,dex) of the solar A(Li). Such a depletion change and minimum abundance, if real, remains unexplained in the standard Li isochrones (referring to the reviews by Pinsonneault \cite{P97} and Bouvier \cite{bou20}, that is, models without rotation, diffusion, mass loss, etc.) but had already been noted in the early Li study by Thorburn et al. (\cite{thor}). Whether the abundance raeches truly zero or some sort of a minimum steady-state abundance remains undetermined from our spectra because of the uncertain molecular contribution to the synthesis line list. The A(Li) values for the four coolest stars in our sample, and in particular the most cool target (RSP\,556; $T_{\rm eff}$=3780\,K, A(Li)$\approx$--1) are thus likely upper limits. For these, we estimate a more realistic $\pm$0.3\,dex internal uncertainty, which then would transform to a possible steady-state limit of the above mentioned $\approx$1/25 of the solar Li abundance. The Li-$T_{\rm eff}$ relation from Cummings et al. (\cite{cumm}) already demonstrated the significant surface depletion down to solar levels for the late G dwarfs around 5400\,K. Our G-M star abundances verify and extend this trend and the current observations may indicate a minimum Li level for stars cooler than 4100\,K (M1 or $\approx$0.5\,M$_\odot$). Its interpretation is pending but could be related to a minimal efficiency of the transportation capacity of the internal mixing process. 

Is there also a dependency of A(Li) on stellar rotation? The common understanding is that faster-rotating stars lose more angular momentum and accordingly induce greater internal mixing and thus surface Li depletion (e.g., Butler et al. \cite{but:coh}, Balachandran et al. \cite{bal:lam}, Soderblom et al. \cite{soder}, Bouvier et al. \cite{bou:bar}, Llorente de Andres et al. \cite{lorente}). This is what is seen for Hyades stars cooler than the Li dip (G dwarfs and cooler) where higher A(Li) is measured for the faster rotators. We note that Boesgaard's (\cite{boes87}) study of Hyades F dwarfs showed that the decline in $v\sin i$ from $T_{\rm eff}$ of 6600\,K to about 6300\,K coincided with an increase in A(Li), that is, an inverse rotation-lithium relation leading to a second A(Li) maximum near 6200\,K in the A(Li) vs. $T_{\rm eff}$ plane (stars cooler than $\approx$6000\,K were not covered then). More recently, Takeda et al. (\cite{tak:hon}) investigated 68 Hyads warmer than 5100\,K at high spectral resolution ($R\sim67,000$) which they used to also search for a $v\sin i$ dependence. They concluded that A(Li) in their sample is essentially controlled by $T_{\rm eff}$ and shows no positive correlation with stellar rotation. However, the rotational velocity range of their studied subsample with $T_{\rm eff}<6000$\,K was just 3.5--6.0\,\kms. This finding would be somewhat in contradiction with the finding of Llorente de Andres et al. (\cite{lorente}) of a minimum (critical) rotational velocity of $\approx$5\,\kms\ that separates Li-rich from Li-poor stars at A(Li)$\approx$2.5. Their sample included field stars as well as open-cluster stars from many different sources. Our Hyades sample shows the same strong A(Li)-$T_{\rm eff}$ dependence as found previously but also a clear relation to rotation described by the three functionals $v\sin i$, $P_{\rm rot}$, and $\mathrm{Ro_n}$, themselves implicitly related to $T_{\rm eff}$ and radius though. A second order polynomial fit to A(Li) vs. $\mathrm{Ro_n}$ in Eq.~\ref{fitALi} gives
\begin{equation}\label{fitALi}
\mathrm{A(Li)} = -3.90\pm0.25 \ + \ 22.08\pm0.19 \ \mathrm{Ro_n} \ - \ 15.73\pm0.15 \ \mathrm{Ro_n}^2  .
\end{equation}
This fit is shown in Fig.~\ref{F5}d as a solid line quantifying higher lithium abundance for targets with larger Rossby numbers. Figure~\ref{F5}b shows the commonly recognized behavior once rotation period is plotted alone, that is, higher lithium abundance for the faster rotating targets, basically the same behavior than for $v\sin i$ in  Fig.~\ref{F5}c. The inverse trend with Rossby number in Fig.~\ref{F5}d thus just reflects that the respective convective turn-over times dominate over the rotational periods. While the latter span in our target sequence from 5 to 16d, that is a factor three, the convective turn-over times ($\tau$ in Table~\ref{T1}) span over a factor 14. Therefore, the plot versus rotation period (Fig.~\ref{F5}b) mirrors just the temperature dependency while the $v\sin i$ plot mirrors the true rotational dependency.  
We also note that Fig.~\ref{F5}c indicates a $v\sin i$ distribution with a critical velocity at $\approx$6.0~\kms\ ($P_{\rm rot} \approx 8$\,d, $\mathrm{Ro_n}\approx$0.4) and A(Li) of 2.6, where ``critical'' is defined as the onset of the A(Li) plateau towards higher rotation. This is in good agreement with the observations of Llorente de Andres et al. (\cite{lorente}). However, we emphasize that our sample is still restricted to $v\sin i$'s between 1--10~\kms\ (periods between 5--16\,d) and thus does not cover the full Li plateau.

\subsection{Chromospheric activity dependency}\label{S62}

Compared to the range of three orders of magnitude for A(Li), the Ca\,{\sc ii} IRT radiative losses span only one order of magnitude. Independent of a photospheric correction, it indicates the relative increase of the chromospheric contribution as one goes from late-F/early-G stars to early M-stars.

The second row in Fig.~\ref{F5} plots the photospherically corrected radiative loss $R^\prime_{\rm IRT}$ versus $T_{\rm eff}$ (Fig.~\ref{F5}e), $P_{\rm rot}$ (Fig.~\ref{F5}f), $v\sin i$ (Fig.~\ref{F5}g), and $\mathrm{Ro_n}$ (Fig.~\ref{F5}h), respectively. We recall that $R^\prime_{\rm IRT}\propto \sum {\cal F^\prime_{\rm IRT}}/\sigma T_{\rm eff}^4$ was computed from all three Ca\,{\sc ii} IRT lines subtracted by the respective photospheric flux and normalized to the bolometric luminosity in logarithmic form. It indicates the reaction of the chromosphere on the presence of a permeating magnetic field.  

Fritzewski et al. (\cite{fritz}) had demonstrated a tight $R^\prime_{\rm IRT}$ versus Ro relationship for the 300\,Myr open cluster NGC\,3532. This cluster is more than two times younger than the Hyades and its chromospheric IRT radiative losses on average two-and-a-half times larger ($\approx$0.4\,dex) than the ones we measured for the Hyades. NGC\,3532 stars with Ro$>$0.06 (and less than $\approx$0.2) were defined as unsaturated and fitted with a linear function (Eq.~4 in Fritzewski et al. \cite{fritz}). This fit is partly outside of our Hyades Ro range of 0.1--0.4 and is thus difficult to compare. Nevertheless, a similar linear fit to the Hyades versus unnormalized Rossby numbers like in  Fritzewski et al. (\cite{fritz}) is given in the following Eq.~\ref{fitIRTa}: 
\begin{equation}\label{fitIRTa}
R^\prime_{\rm IRT} = -3.833 \pm 0.057 \ - \ 3.035 \pm 0.050 \ \mathrm{Ro} \ . 
\end{equation}
It shows basically good agreement for the overlapping Ro regime but with the above mentioned offset in  $R^\prime_{\rm IRT}$  of $\approx$--0.4\,dex for the Hyades. A full-range second-order polynomial fit, like for A(Li) in Eq.~\ref{fitALi}, is applied to the normalized Rossby numbers ($\mathrm{Ro_n}$) and is given in Eq.~\ref{fitIRTb}: 
\begin{equation}\label{fitIRTb}
R^\prime_{\rm IRT} = -3.478\pm0.064 \ - \ 4.781\pm0.048 \ \mathrm{Ro_n} \ + \ 3.935\pm0.039 \ \mathrm{Ro_n}^2  .
\end{equation}

Figure~\ref{F5}e shows a remaining temperature dependency even after photospheric correction verifying that the cooler targets tend to have more active chromospheres. It is basically mirrored into the rotation-period dependency (Fig.~\ref{F5}f): Hyads with longer rotation periods exhibit higher chromospheric IRT losses, which is a sort-of inverse rotation-activity relation. However, this is driven by the natural dependency of $\tau$ on $T_{\rm eff}$ because the longer-period targets are also the coolest. 

The dependency on projected rotational velocity (Fig.~\ref{F5}g) is comparable and in the sense higher radiative losses are seen for targets with lower rotational velocities. The latter is also opposite to what is seen for A(Li) where higher abundances are seen for the targets with higher rotational velocities. The $v\sin i$ dependency of $R^\prime_{\rm IRT}$ is again mirrored in the one versus Rossby number in Fig.~\ref{F5}h, that is, higher radiative losses for targets with smaller Rossby numbers, which follows the expected rotation-activity relation for unsaturated stars. It is worth noting here that the three relations versus $v\sin i$ in Figs.~\ref{F5}cgk are qualitatively the same if plotted versus angular momentum instead. We recall that the latter ($M R v$) explicitly includes the masses, radii, and the most probable inclinations of the rotational axis, $i$,  from Table~\ref{T1}. 

\subsection{Magnetic field dependency}\label{S63}

The longitudinal component of the large-scale surface magnetic field of our Hyads is reconstructed as a function of rotational phase from Stokes-V LSD line profiles. Because an LSD profile could be considered a mean photospheric line, our magnetic-field measurements are thus for a mean photosphere. At this point, we (re)emphasize that CP-based data measure the topological field strengths rather than the expectedly much higher local field strengths and thus much better represent the global scale of the field. However, the disk averaging makes it prone to flux cancellation in regions of opposite magnetic polarity within a surface resolution element. Among a subsample of slowly rotating M stars, Reiners et al. (\cite{reiners}) showed that the surface averaged small-scale magnetic field strength is proportional to Rossby number, and thus in qualitative agreement with the global-scale field from polarimetric spectra, albeit with higher field strengths by a factor of ten compared to polarimetric data. The underlying cause for a correlation with rotation and age is the magnetic-field decay as stars age and rotation decreases. As emphasized in the review by Linsky (\cite{lin17}) it is difficult to quantify these relations as they likely depend also on local magnetic field strength and corresponding magnetic energy which usually can only be directly measured on the Sun. An actual relation between local and global field remains an open dynamo flux-transport issue (see, e.g., Brun et al. \cite{brun}). 

We obtained longitudinal field strengths in the range $-100$\,G (e.g., RSP\,542) and $+150$\,G (e.g., RSP\,409 and 571), converted to unsigned average fields ranging between 11.4 and 217\,G, thus covering rougly a range of a factor twenty. One target, RSP\,556, was observed with only positive polarity but had two significant phase gaps which prevented complete surface visibility. For the two targets in common with the study of Folsom et al. (\cite{fol}), RSP\,68=Mel25-5 and RSP\,198=Mel25-21, we measured approximately twice as high $B_{\rm long}$ values: 20\,G with a range of 46\,G (compared to Folsom et al. \cite{fol} of 11\,G with a range of 15\,G) and 35\,G with a range of 43\,G (compared to 14\,G with range 18\,G), for the two targets, respectively. We note that there were seven years in between our observations and those of Folsom et al. (\cite{fol}). The surface-averaged unsigned magnetic field strengths for above two targets are nevertheless quite comparable: 13.0\,G vs. 13.0\,G, and 17.6 vs. 12.7\,G, for ours vs. Folsom et al. (\cite{fol}), respectively. For target-to-target comparisons prior to ZDI in this paper, we thus focus on the unsigned average fields listed in Table~\ref{T4} in column $\langle |B|\rangle$. 

The third row in the matrix figure Fig.~\ref{F5} plots the phase-averaged unsigned field strength $\langle |B|\rangle$ versus $T_{\rm eff}$ (Fig.~\ref{F5}i), $P_{\rm rot}$ (Fig.~\ref{F5}j), $v\sin i$ (Fig.~\ref{F5}k), and $\mathrm{Ro_n}$ (Fig.~\ref{F5}l), from left to right, respectively. It is obvious that it is the coolest and smallest stars that exhibit the strongest surface-averaged large-scale field densities (as well as higher chromospheric radiative losses). In our sample it is the slow rotators defined by small $v\sin i$ and large $P_{\rm rot}$ that appear with higher global field densities and thus present us an apparent inverse rotation-activity relation. This is also seen in the Ro-panel in Fig.~\ref{F5}l where the coolest stars with the longest rotation periods are the ones with the lowest (normalized) Rossby numbers ($\mathrm{Ro_n}\approx 0.2$) and highest field densities. Usually, smaller Rossby numbers indicate faster surface rotation. However, the $\langle |B|\rangle - \mathrm{Ro_n}$ relation in Fig.~\ref{F5}l appears bimodal with two slopes. A more or less constant (or only slightly decreasing) field density for $\mathrm{Ro_n}\geq 0.25$ with an average of 15.4$\pm$3.6(rms)\,G, and a steeply increasing field density for $\mathrm{Ro_n}< 0.25$ with an average of 91$\pm$61(rms)\,G. We quantify it with a two-piece linear fit in Eq.~\ref{fitB}:  
\begin{eqnarray}\label{fitB}
\langle |B|\rangle &=& 376 \pm 17 \ - \ 1561 \pm 12 \ \times \mathrm{Ro_n} \ \dots \ \mathrm{for} \ \mathrm{Ro_n}<0.25, \nonumber \\
\langle |B|\rangle &=& 23 \pm 0.9 \ - \ 24 \pm 0.7 \ \times \mathrm{Ro_n} \ \dots \ \mathrm{for} \ \mathrm{Ro_n}>0.25 .
\end{eqnarray}

At this point it may be worth mentioning that our sample contains six stars of roughly the same effective temperature around 5000\,K (4700-5200\,K) but with rotation periods between 9.10\,d (RSP\,134) and 12.64\,d (RSP\,216). While the mean longitudinal surface magnetic field, $\langle |B|\rangle$, is relatively small for all of them ($\approx18\pm4$\,G), its individual variability amplitudes increase from 45\,G (RSP\,134) to 84\,G (RSP\,216) towards the slower rotators, that is, nearly a factor two. It not only suggests that $\langle |B|\rangle$ may be a too simplified parameter for the quantification of a stellar magnetic field but, more importantly, that there is a remaining weak (inverse) rotational dependency also for the stars with $\mathrm{Ro_n}\approx 0.25$ (spectral type $\approx$K1).

\section{Summary and conclusions}\label{S7}

The basis of our analysis are high-resolution, high-S/N spectra that allowed the precise determination of three distinct surface-activity tracers for each of the 21 sample dwarfs. The three tracers were lithium abundance, A(Li), radiative loss in the chromospheric emission lines of the Ca\,{\sc ii} infrared triplet, $R^\prime_{\rm IRT}$, and the average, unsigned, surface magnetic-field strength, $\langle |B|\rangle$. All three parameters were found to vary with other observables, $T_{\rm eff}$, $P_{\rm rot}$, and $v\sin i$, but not in the same way. Our initial target selection of Hyades dwarfs implicitly sampled a stellar mass sequence (and thus also a radius and $T_{\rm eff}$ sequence) at an age of $\approx$700\,Myr and supersolar metallicity of +0.18. It is thus expected that we see parameter dependencies along this sequence. The addition of rotational parameters made this a true Rossby-type sequence. 

In particular, we find a clear Rossby-number dependency of A(Li), $R^\prime_{\rm IRT}$, and $\langle |B|\rangle$. But only A(Li)  increases with increasing Rossby number while the other two parameters basically decline with increasing Rossby number. At the same time A(Li) shows also an increase with projected rotational velocity, $v\sin i$, and a decrease with rotational period, basically reflecting the mass/radius dependency of the Rossby sequence, while the other two parameters, $R^\prime_{\rm IRT}$ and $\langle |B|\rangle$, indicate the opposite, that is, a decrease with increasing rotation.  

We interpret this in the sense that the observed trend of Li depletion is not primarily caused by some rotation-induced mixing process. It is rather the mass dependence of the depth of the outer convection zone that enforces a mass-dependent Li depletion rate. We see a (surface) rotational and angular-momentum dependency as well but argue that rotational mixing exists in parallel but only modifies the A(Li) vs. $T_{\rm eff}$ relation, in particular for the faster rotators. We speculate that the Li dip (at $T_{\rm eff}$ higher than our sample) is somehow a complex consequence of rotational mixing.

The dependency of averaged unsigned magnetic field strength with Rossby number appears having two very different slopes in the sense that only the fastest, of our generally slow, rotators show a clear dependency. Stars with (normalized) Rossby numbers greater than $\approx$0.25 exhibit on average a factor $\approx$6 weaker fields, and no or only weak dependency on rotation. The inverse rotation-activity relation is likely due to a change of surface magnetic-field morpholgy with effective temperature/mass taking place at around $\mathrm{Ro_n}\approx 0.25$.  We will revisit this with our future ZDI. It appears that the global surface activity on targets of Hyades age, in particular when cooler than $\approx$5000\,K, is dominated by its convective motions, in our case expressed by its turnover time which itself is inverse proportional to stellar mass and effective temperature, rather than stellar rotation.

\begin{acknowledgements}
We thank an anonymous referee for the many very helpful suggestions. It is a pleasure to thank the German Federal Ministry (BMBF) for the year-long support for the construction of PEPSI through their Verbundforschung grants 05AL2BA1/3 and 05A08BAC as well as the State of Brandenburg for the continuing support of AIP and PEPSI for the LBT (see https://pepsi.aip.de/). The LBT is an international collaboration among institutions in the United States, Italy and Germany. LBT Corporation partners are: The University of Arizona on behalf of the Arizona Board of Regents; Istituto Nazionale di Astrofisica, Italy; LBT Beteiligungsgesellschaft, Germany, representing the Max-Planck Society, The Leibniz Institute for Astrophysics Potsdam, and Heidelberg University; The Ohio State University, representing OSU, University of Notre Dame, University of Minnesota and University of Virginia. \\ This work has made use of NASA's Astrophysics Data System and of CDS's Simbad database which we all gratefully acknowledge.
\end{acknowledgements}

\appendix

\section{RSP 348}\label{App-A}

\begin{figure*}
\begin{minipage}{0.3\textwidth}
{\bf a.}\\
\includegraphics[angle=0,width=45mm,clip]{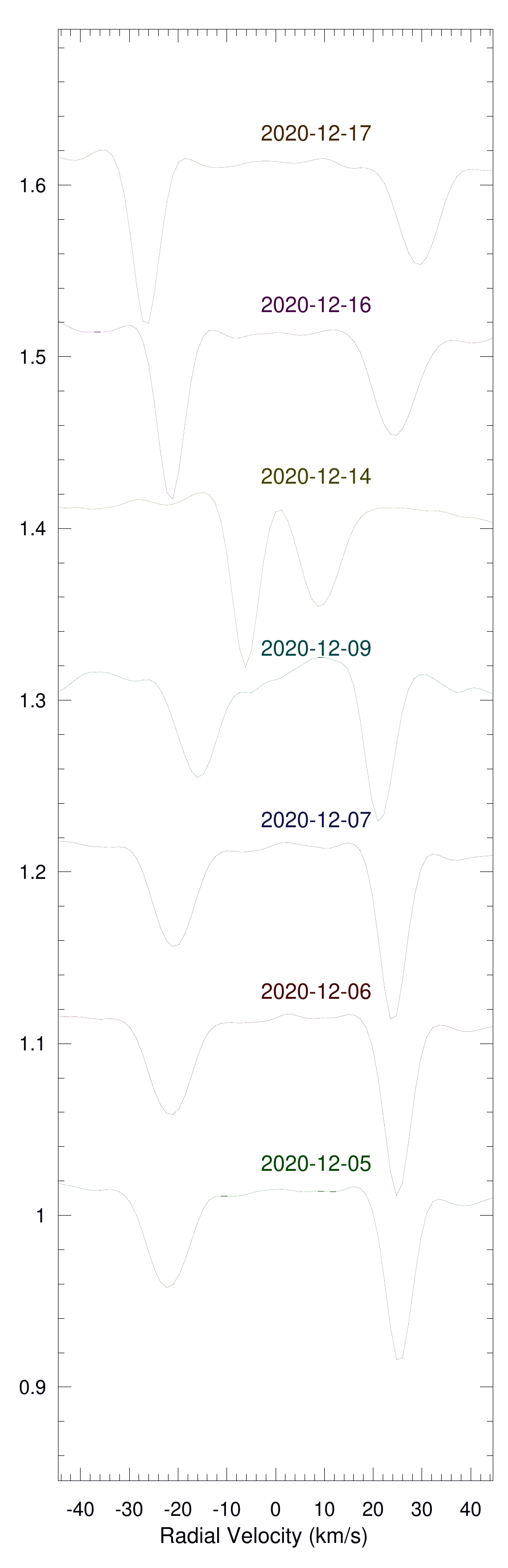}
\end{minipage}
\begin{minipage}{0.5\textwidth}
{\bf b.}\\
\includegraphics[angle=0,width=86mm,clip]{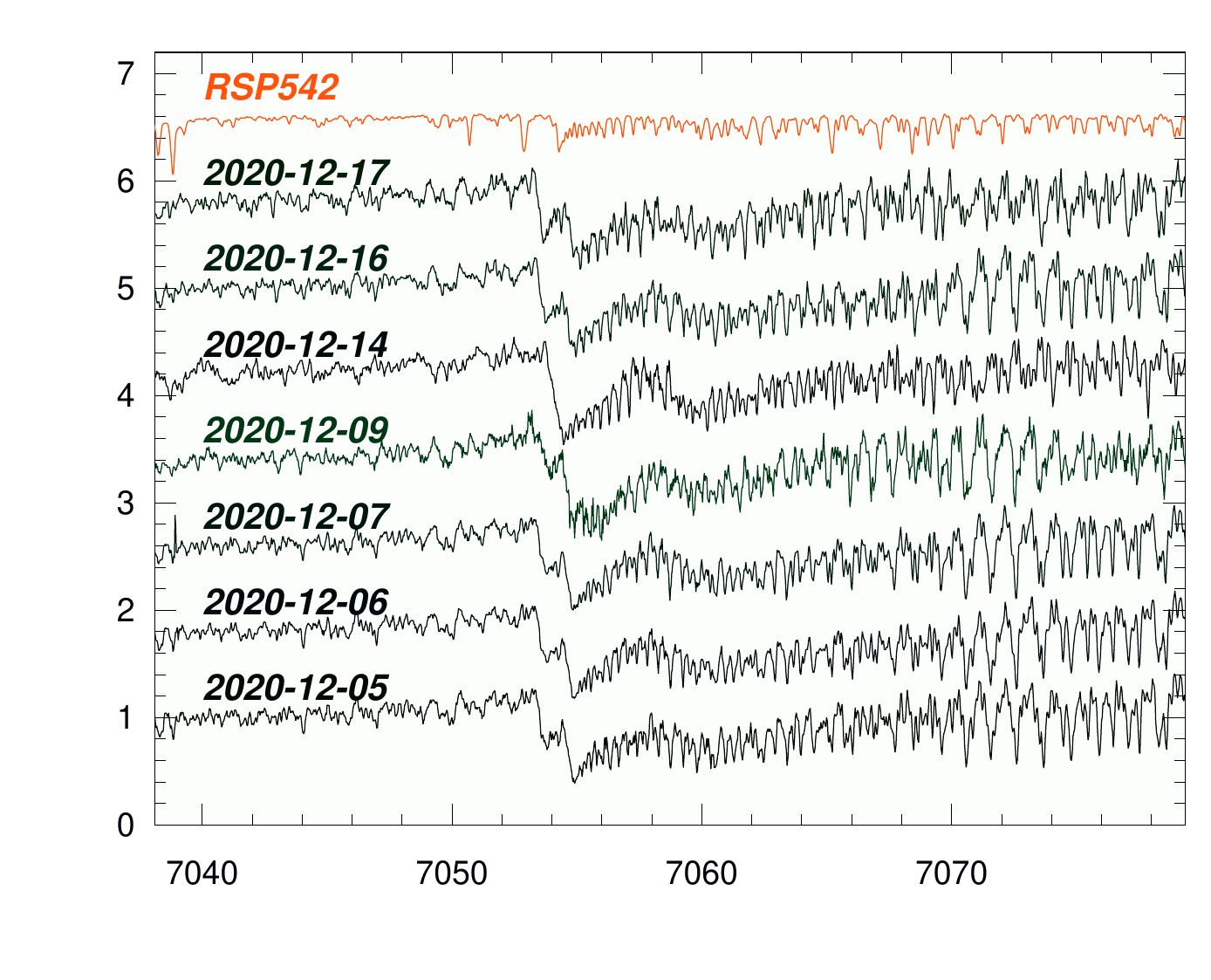}

{\bf c.}\\
\includegraphics[angle=0,width=86mm,clip]{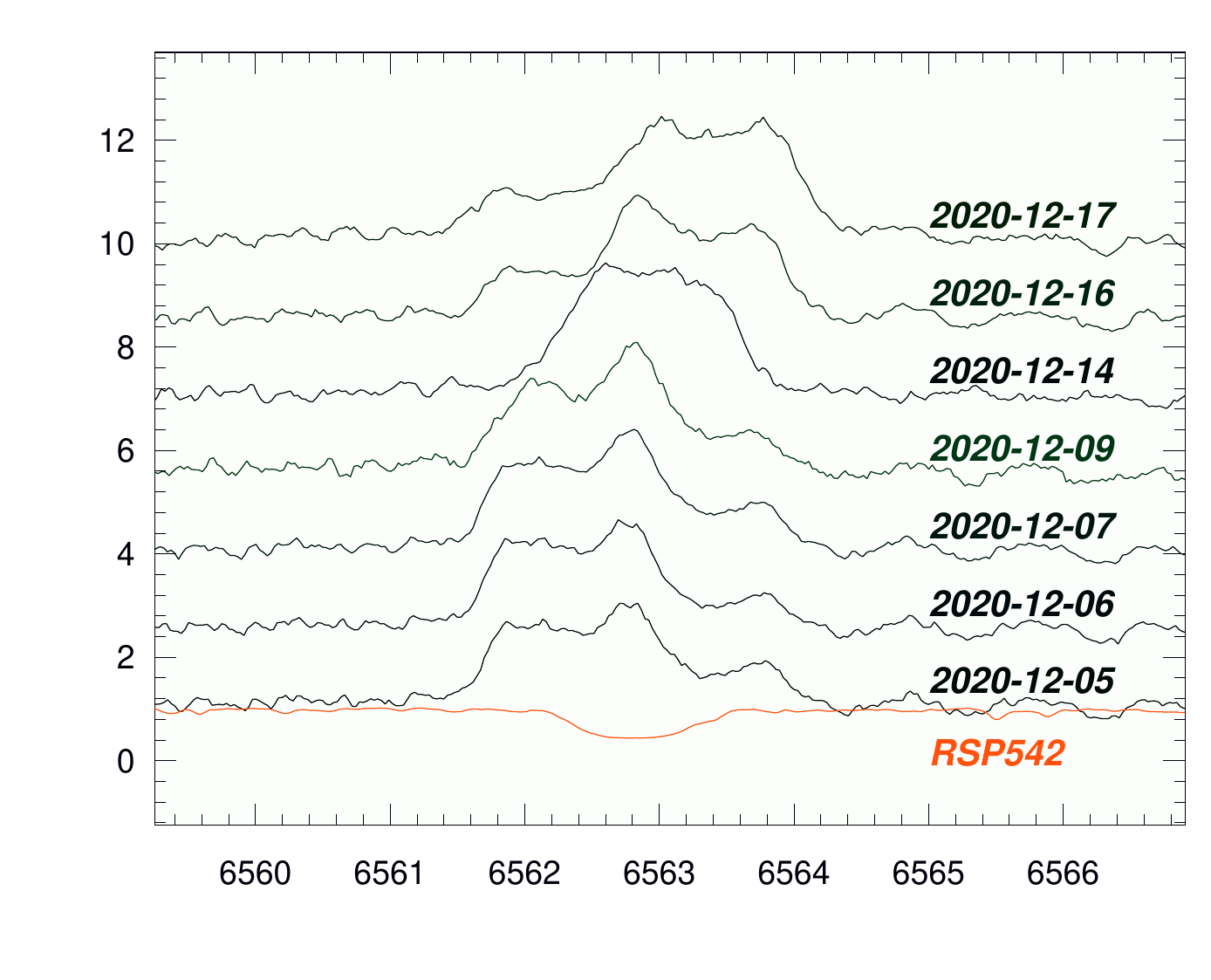}
\end{minipage}\hspace{1mm}
\caption{Time series Stokes-I spectra of RSP\,348 (photometric $P=3.30$\,d, black lines) showing it to be a hitherto unknown SB2. $a$. LSD photospheric lines, $b$. TiO $\gamma$(0,0) band head at 7055\,\AA, $c$. \Halpha . Each spectrum is identified with the UT date. The red spectrum in panels $b$ and $c$ is the single star RSP\,542 ($T_{\rm eff}\approx$4000\,K) for comparison. Spectra are shifted in intensity for better visibility. }\label{F-A1}
\end{figure*}

This subsection is dedicated to the SB2 binary RSP\,348 because we rejected this target from the main sample. Our seven spectra do not yet allow an orbit determination. However, the data cover one conjunction passage on UT2020-12-14 with doubled lines five days before and two days after, which is indicative of an orbital period clearly longer than the photometric period of 3.30\,d (Douglas et al. \cite{doug19}), maybe twice as long.

The two components are similar but have different line depths and broadening and are thus likely not of exactly equal temperature. We define the stronger-line component as the primary with suffix $a$, the other with suffix $b$. The LSD-based primary line system is $\sim$40\% deeper than the secondary lines (Fig.~\ref{F-A1}a). If we set the measured equivalent-width ratio (from Gaussian fits) equal to the expected continuum ratio, we obtain a secondary:primary continuum ratio of 0.86$\pm$0.01 at $\sim$6450\,\AA. The total line broadening from the LSD profiles is FWHM $a$ 6.7\,\kms\ and $b$ 10.2\,\kms. The primary's FWHM is comparable to the respective value of the comparison single star RSP\,542 (M0, $v\sin i$=1.4\,\kms). Rotational broadening is thus basically only detected for the $b$ component ($v\sin i \approx 5\pm1$\,\kms). Also, no Li lines from either component could be identified, mostly because the region is plastered with TiO absorption lines from both stellar components. From the depth of the TiO-band heads at 7055\,\AA\ (shown in Fig.~\ref{F-A1}b), 7088\,\AA, and 7126\,\AA\ of the $\gamma$(0,0) system, we estimate effective temperatures with the relation given in Strassmeier \& Steffen (\cite{xiboo}). We emphasize that the four spectra from UT2020-12-05 to 2020-12-09 show the TiO-7055\,\AA\ band head of the secondary component blueshifted, and thus unblended from the TiO-line system of the primary, while the two spectra from UT2020-12-016 and 2020-12-17 show the inverse situation with the primary's band head blueshifted and thus unblended. The other two $\gamma$(0,0) band heads likely remain blended during the orbital motion and were not used. Assuming the continuum ratio of 0.86 also for 7055\,\AA, we obtain a band head line depth of 0.71 for component $a$ and 0.78 for component $b$. Its straight-forward comparison with the synthetic TiO spectrum library computed with Turbospectrum (Appendix~C in Strassmeier \& Steffen \cite{xiboo}) then gives $T_{\rm eff}$ of $\approx$3700\,K and $\approx$3550\,K for $a$ and $b$, respectively. However, these values are still just estimates because of the inherent difficulty of continuum definition at these temperatures, combined with the low S/N of the spectra.

Figure~\ref{F-A1}c reveals both components with strong \Halpha\ emission. The primary's emission (peak intensity of 3.45 above continuum) is almost twice as strong as the secondary's (peak intensity of 2.07) and shows a pronounced central self reversal indicative of a very active dMe chromosphere (e.g., Houdebine \& Stempels \cite{hou:ste}). The self-reversal depth in component $a$ is 0.41\,\AA\ (the secondary emission profile remains blended and is not measurable) while the equivalent widths are $-2.35$\,\AA\ and $-1.00$\,\AA\ for $a$ and $b$, respectively. The line widths at continuum level are comparable ($\sim$1.5\,\AA\ for both components) and are significantly larger than the absorption profile width of the comparison dM0 RSP\,542 ($\sim$1.0\,\AA). The respective \Halpha\ FWHM are 1.3\,\AA\ and 1.1\,\AA\ for $a$ and $b$, respectively, and allow an estimate of the absolute magnitude based on the  Wilson-Bappu like relation as defined in Houdebine \& Stempels (\cite{hou:ste}). It gives $M_{\rm V}(a)\approx 10.1$\,mag and $M_{\rm V}(b)\approx 12.2$\,mag, or M3 and M5 according to ``standard'' tables, in good agreement with the TiO-based effective temperatures.

\section{Extra figures}

\begin{figure*}
 \centering
\includegraphics[angle=0,width=5.cm,clip]{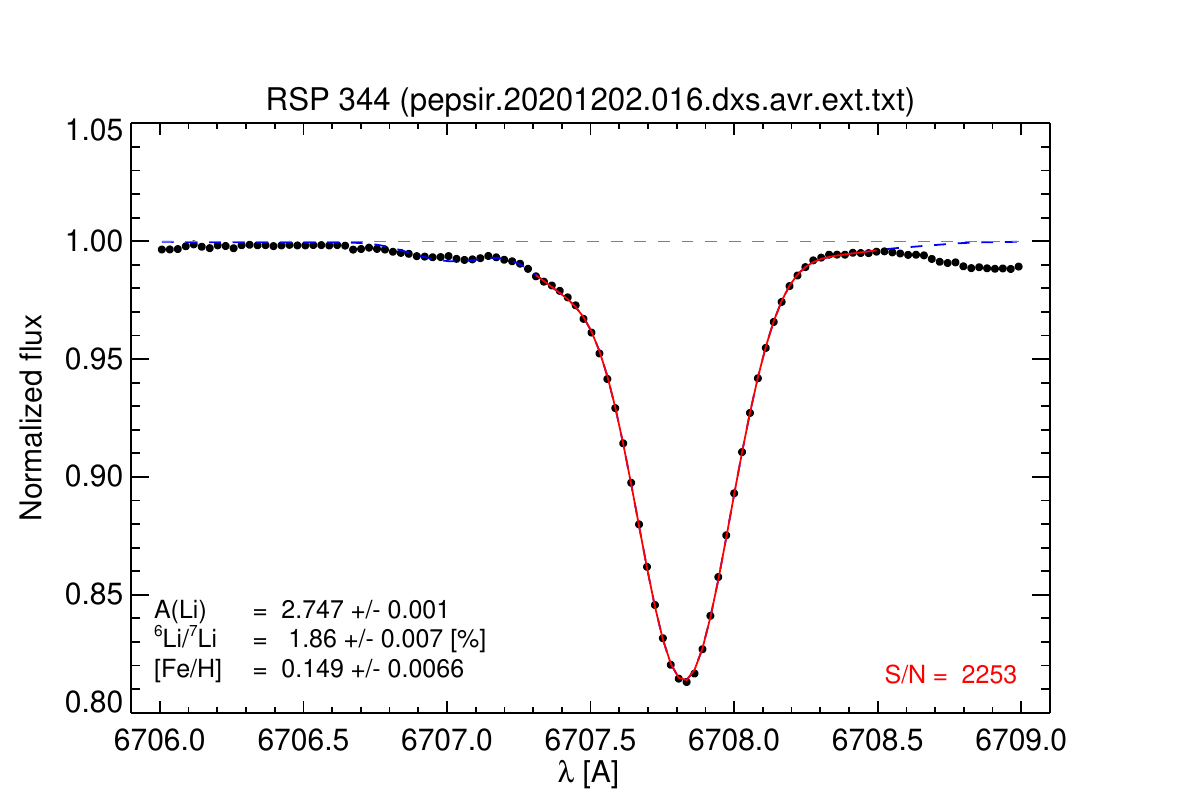}
\includegraphics[angle=0,width=5.cm,clip]{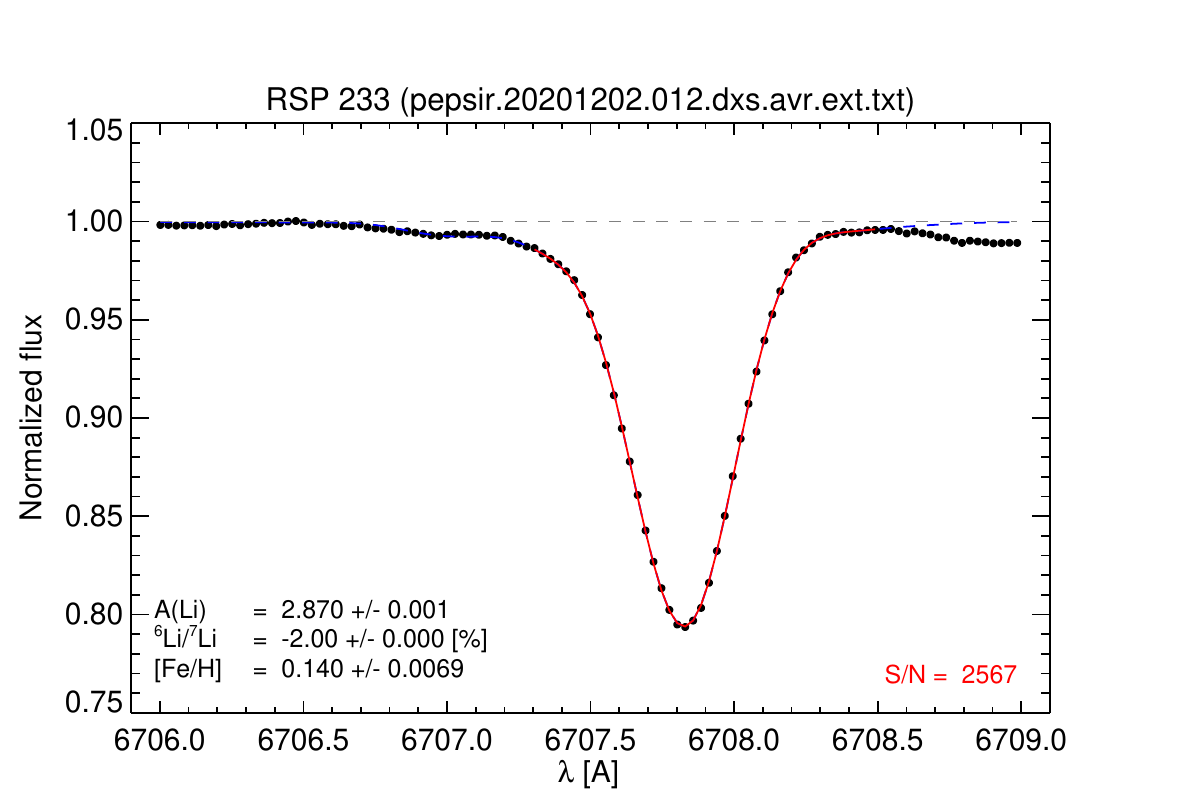}
\includegraphics[angle=0,width=5.cm,clip]{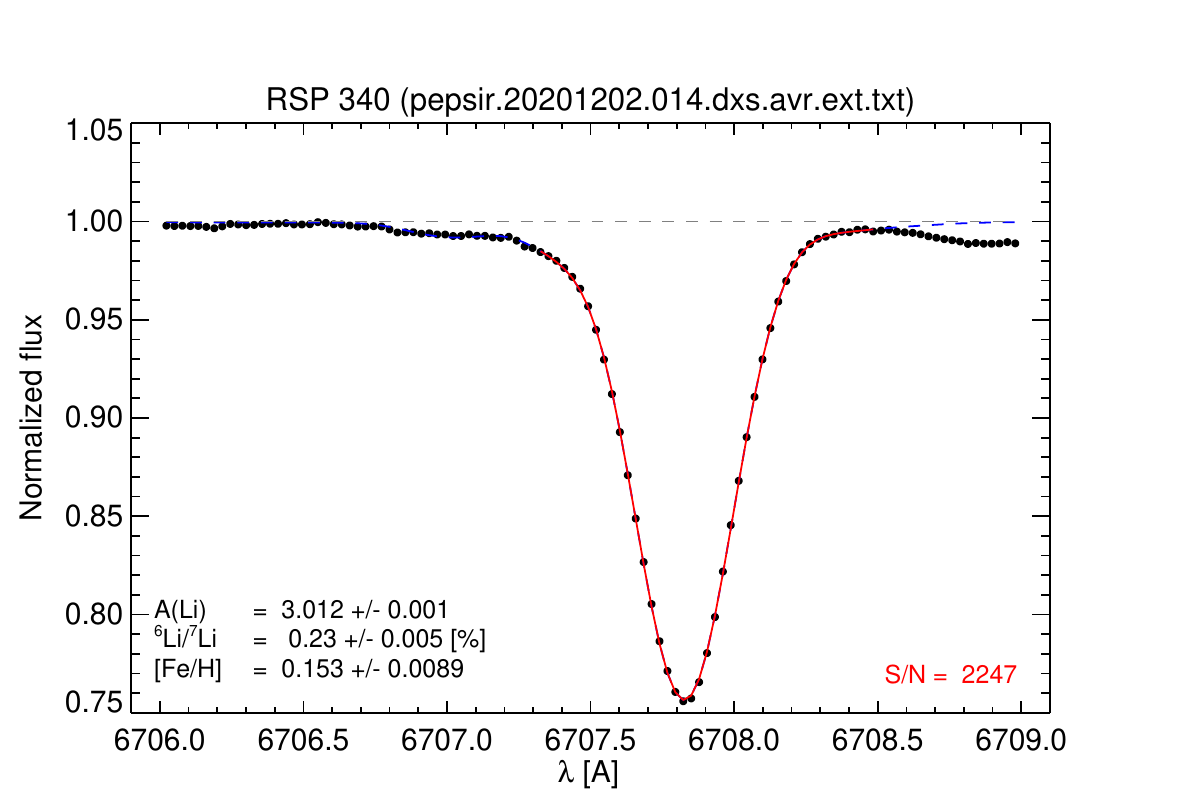}
\includegraphics[angle=0,width=5.cm,clip]{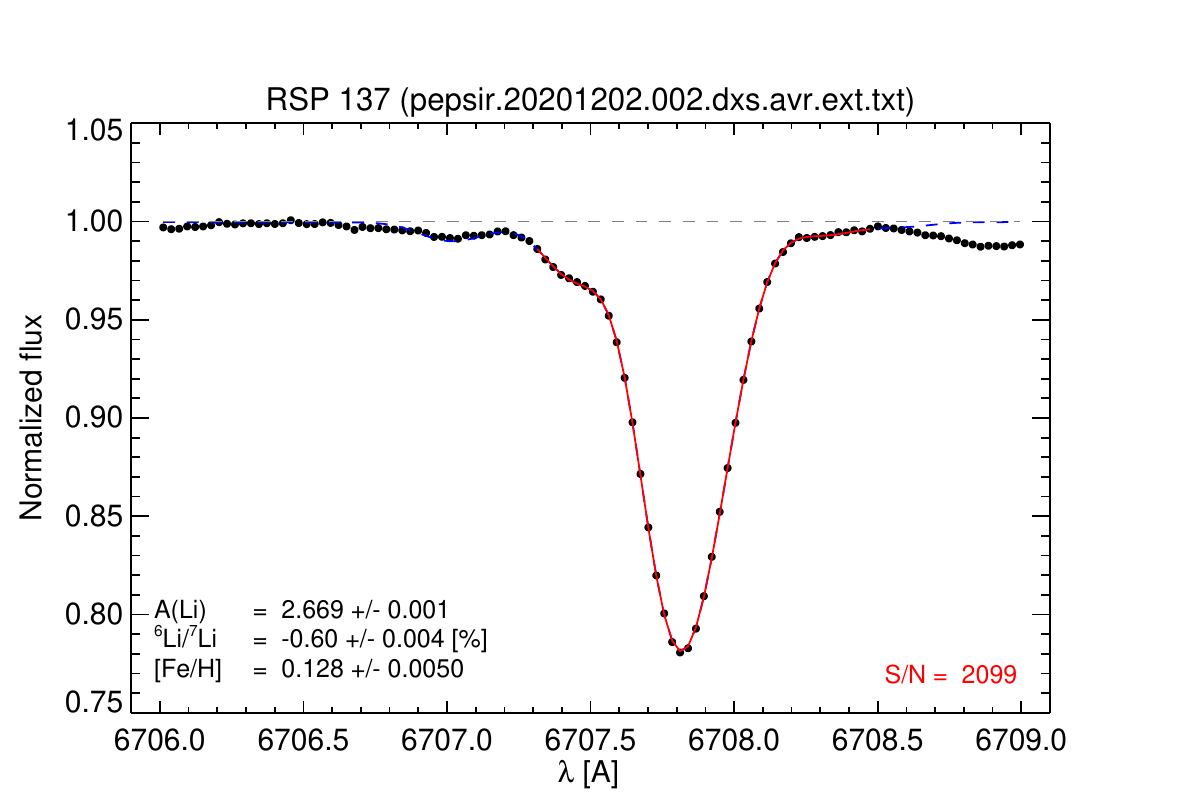}
\includegraphics[angle=0,width=5.cm,clip]{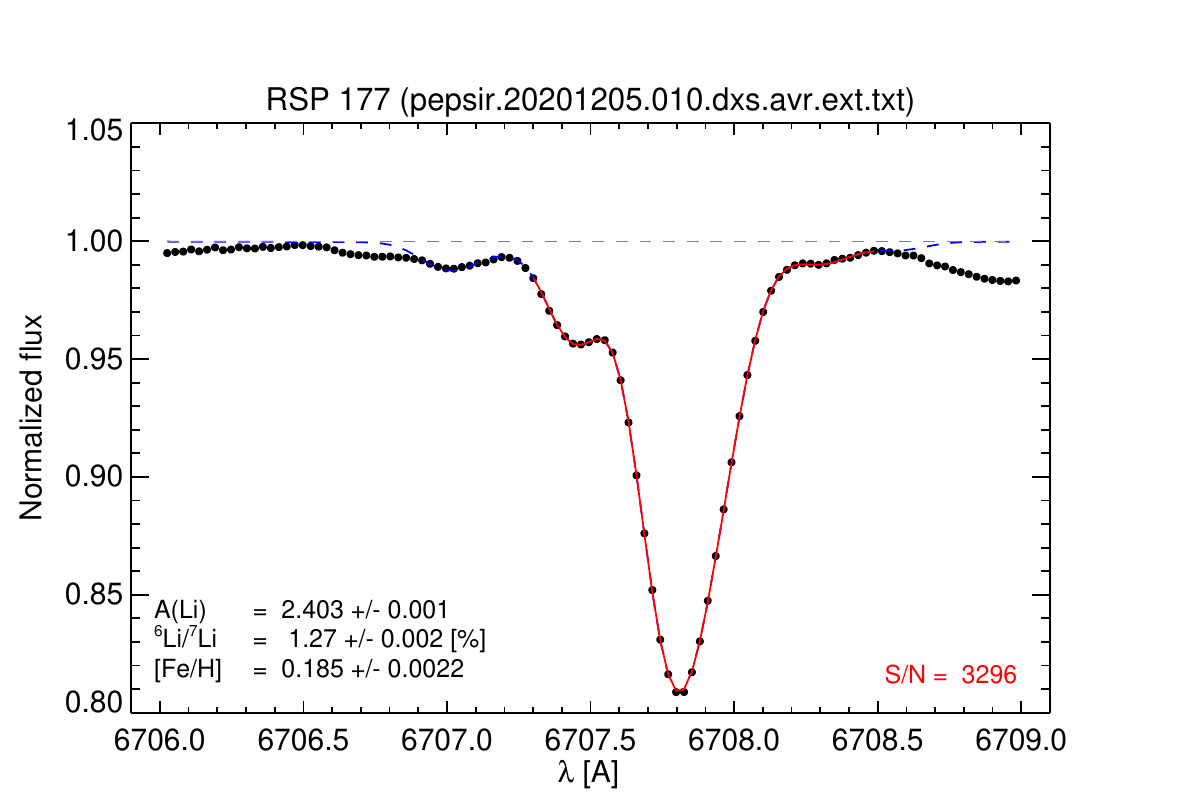}
\includegraphics[angle=0,width=5.cm,clip]{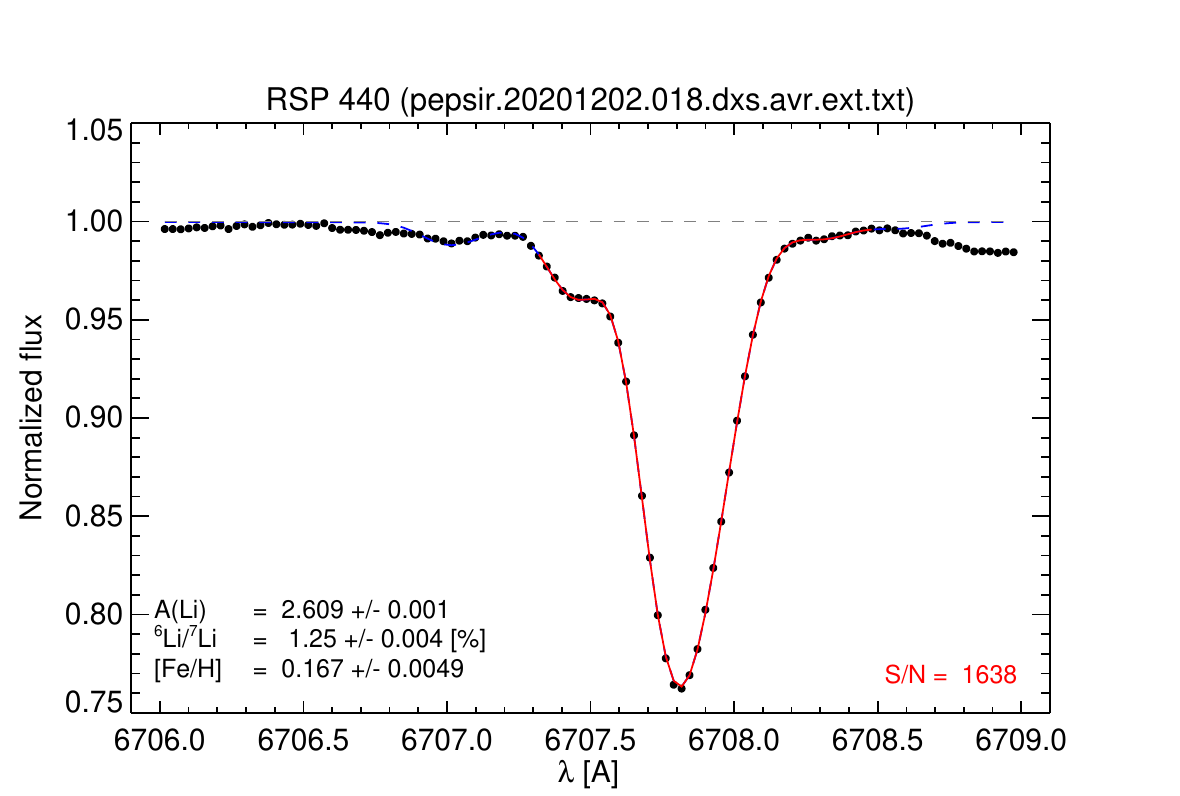}
\includegraphics[angle=0,width=5.cm,clip]{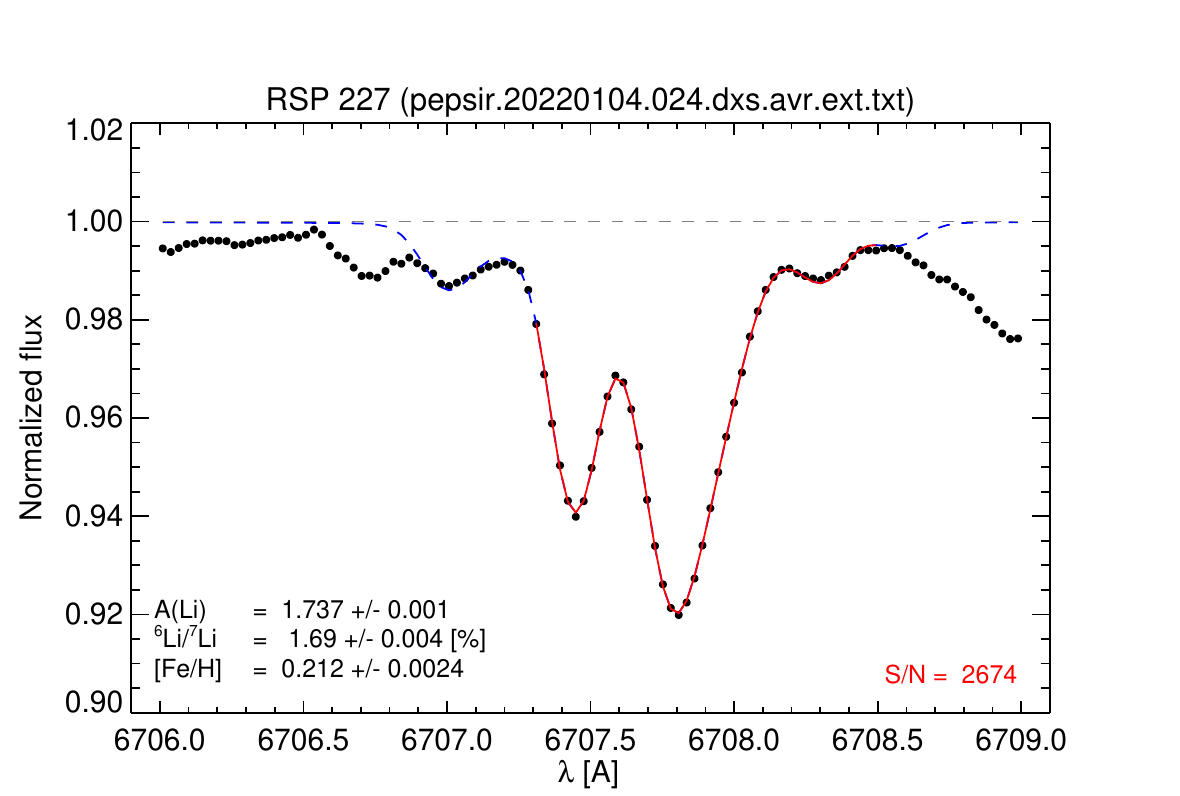}
\includegraphics[angle=0,width=5.cm,clip]{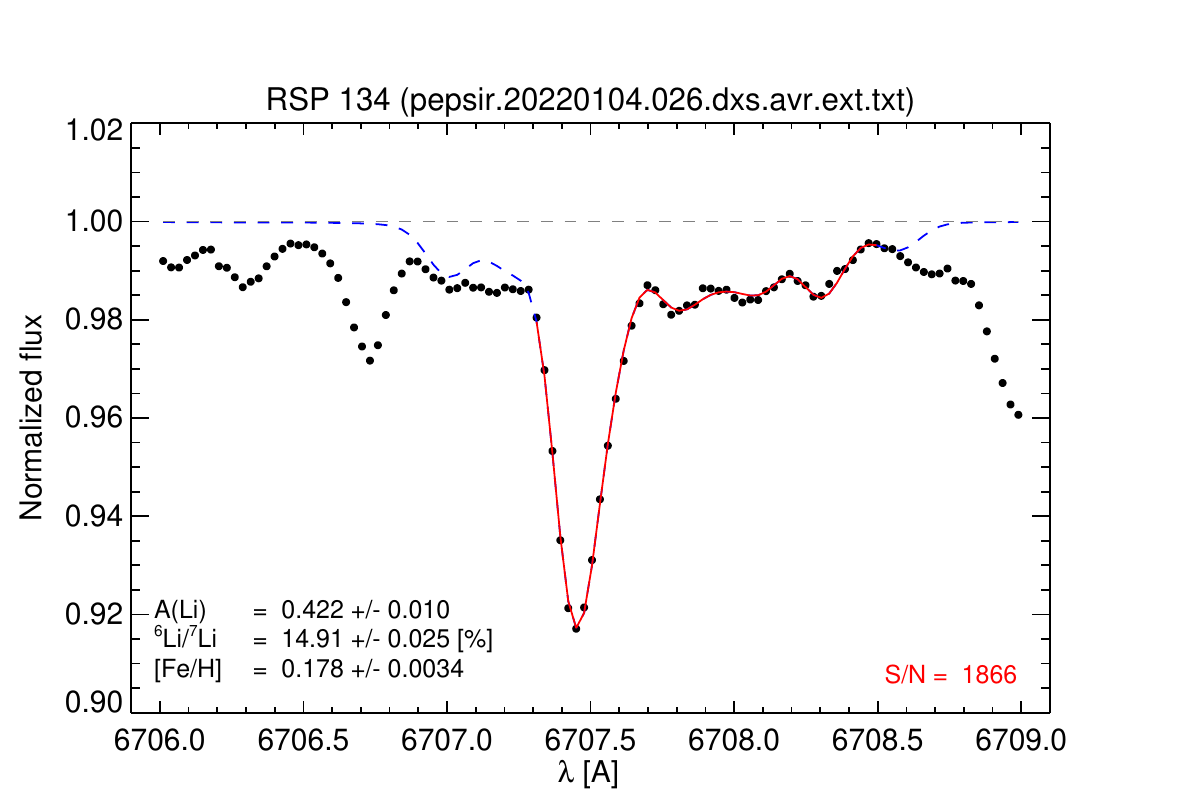}
\includegraphics[angle=0,width=5.cm,clip]{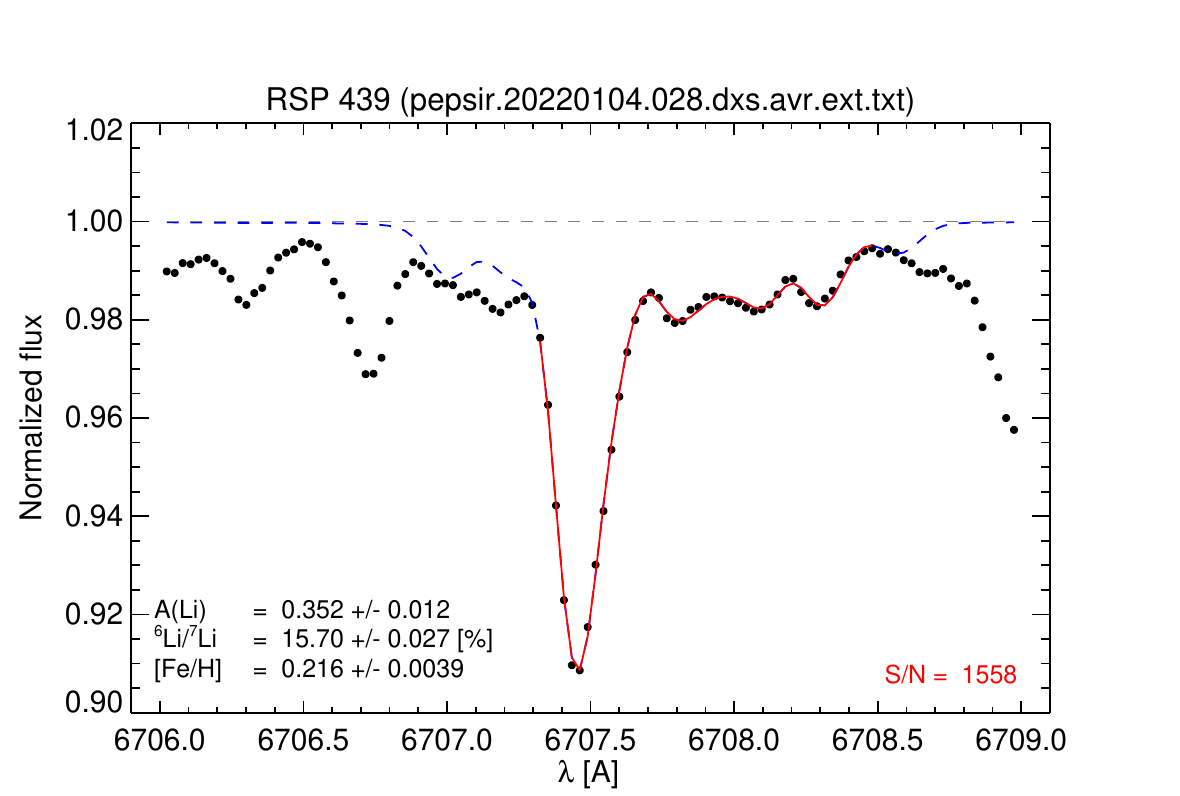}
\includegraphics[angle=0,width=5.cm,clip]{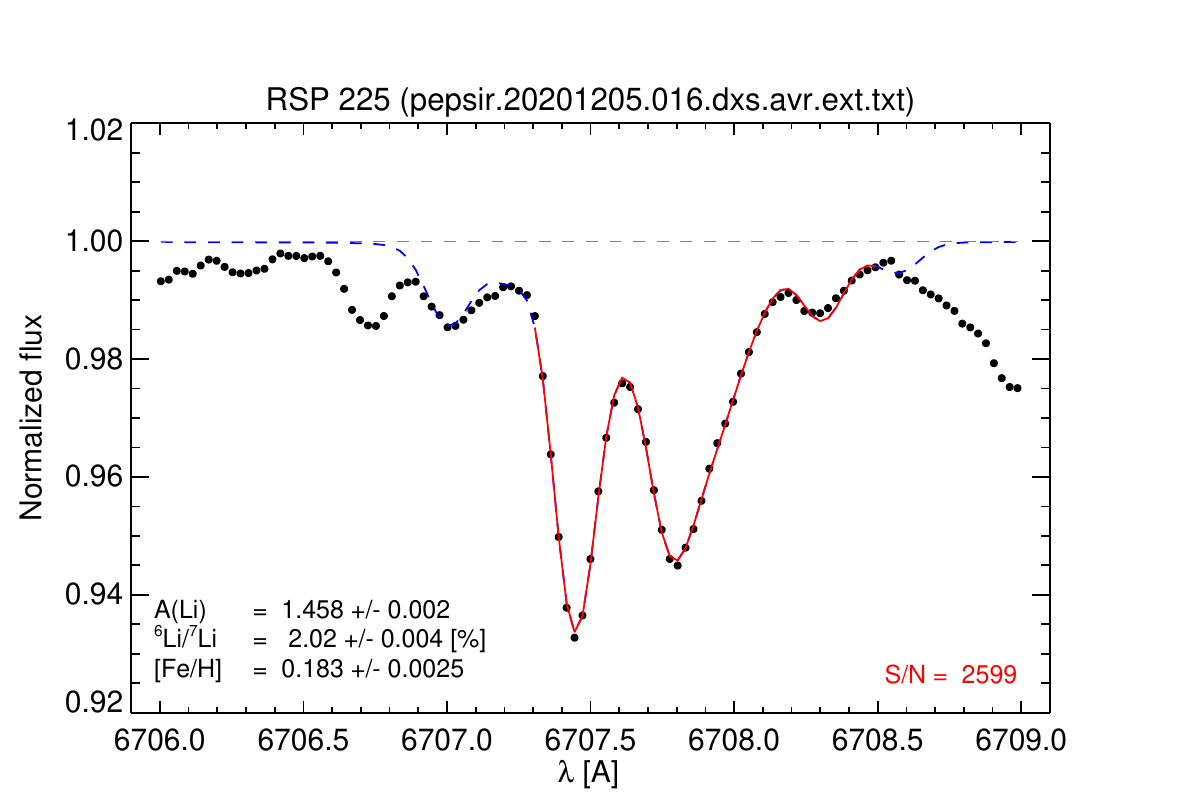}
\includegraphics[angle=0,width=5.cm,clip]{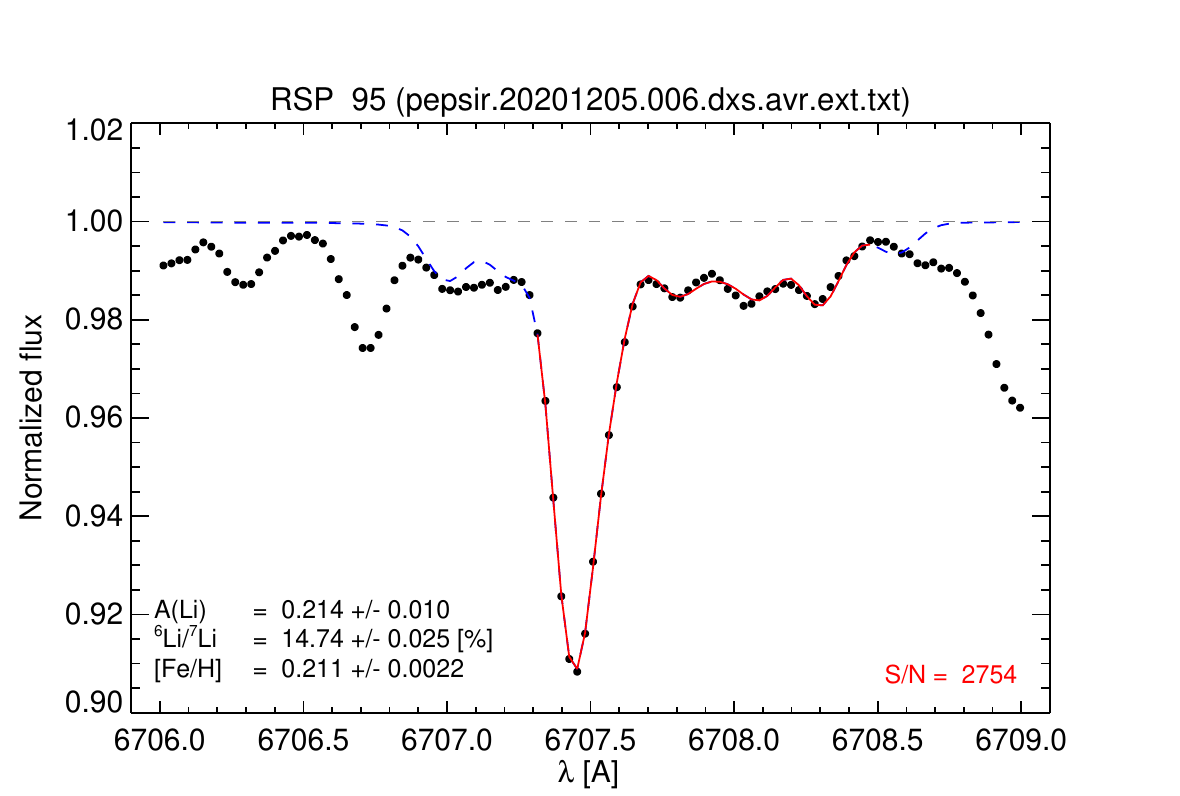}
\includegraphics[angle=0,width=5.cm,clip]{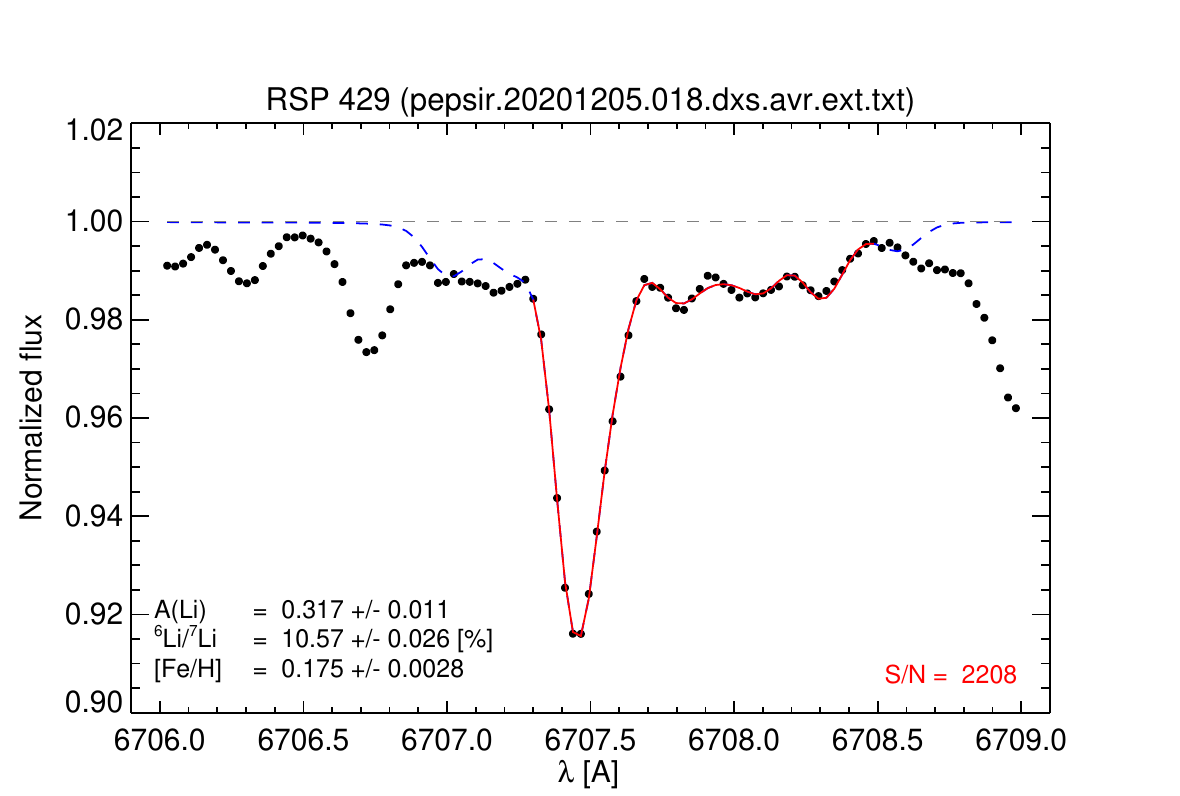}
\includegraphics[angle=0,width=5.cm,clip]{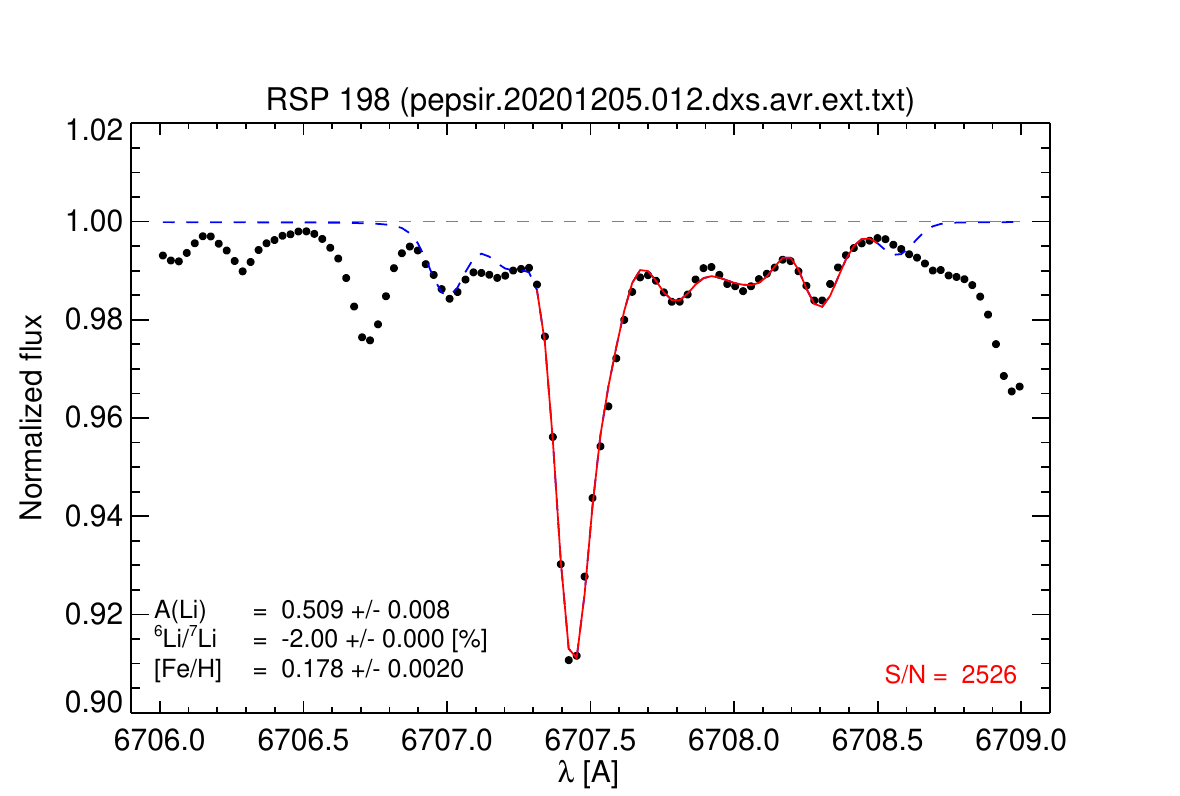}
\includegraphics[angle=0,width=5.cm,clip]{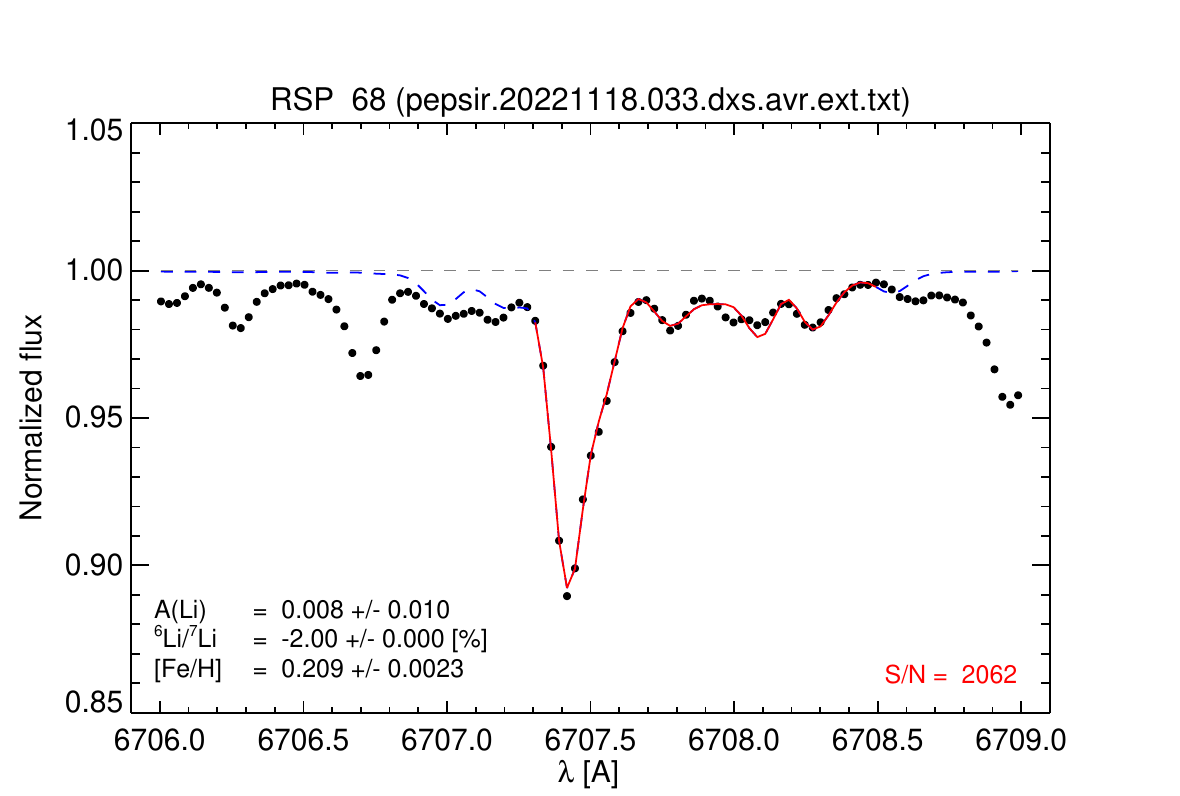}
\includegraphics[angle=0,width=5.cm,clip]{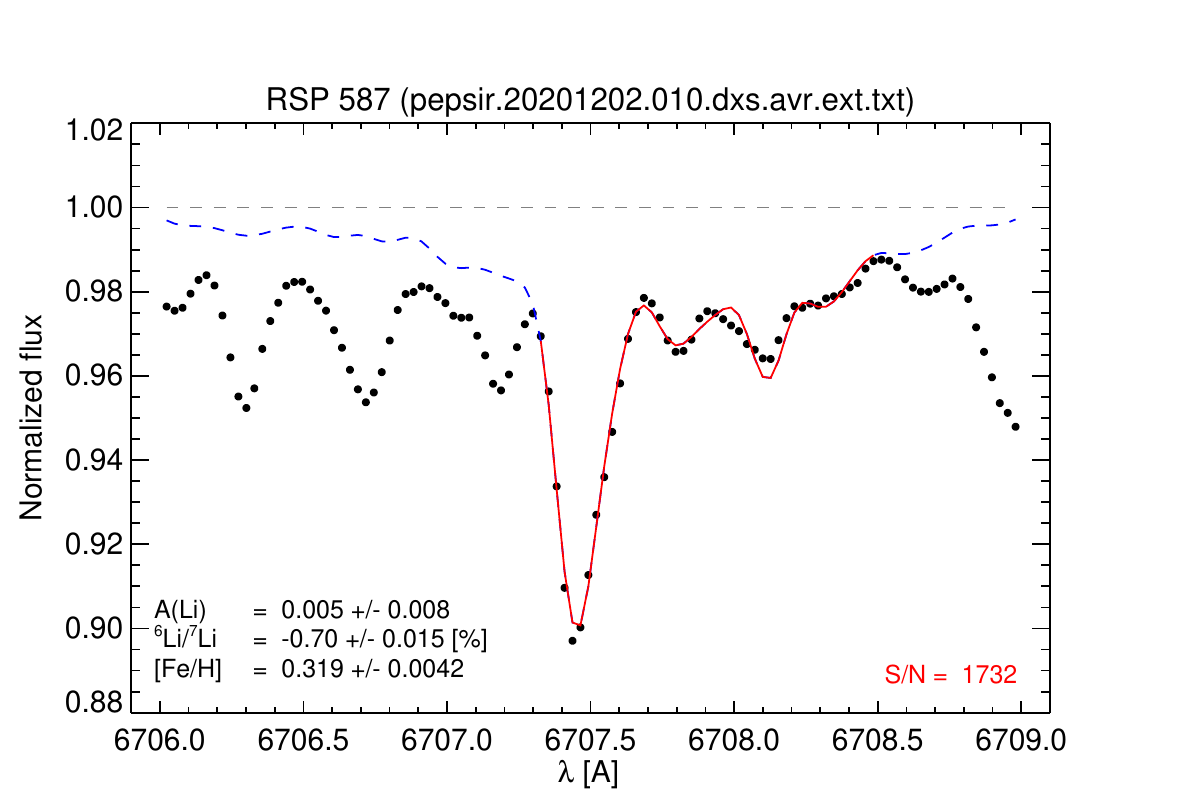}
\includegraphics[angle=0,width=5.cm,clip]{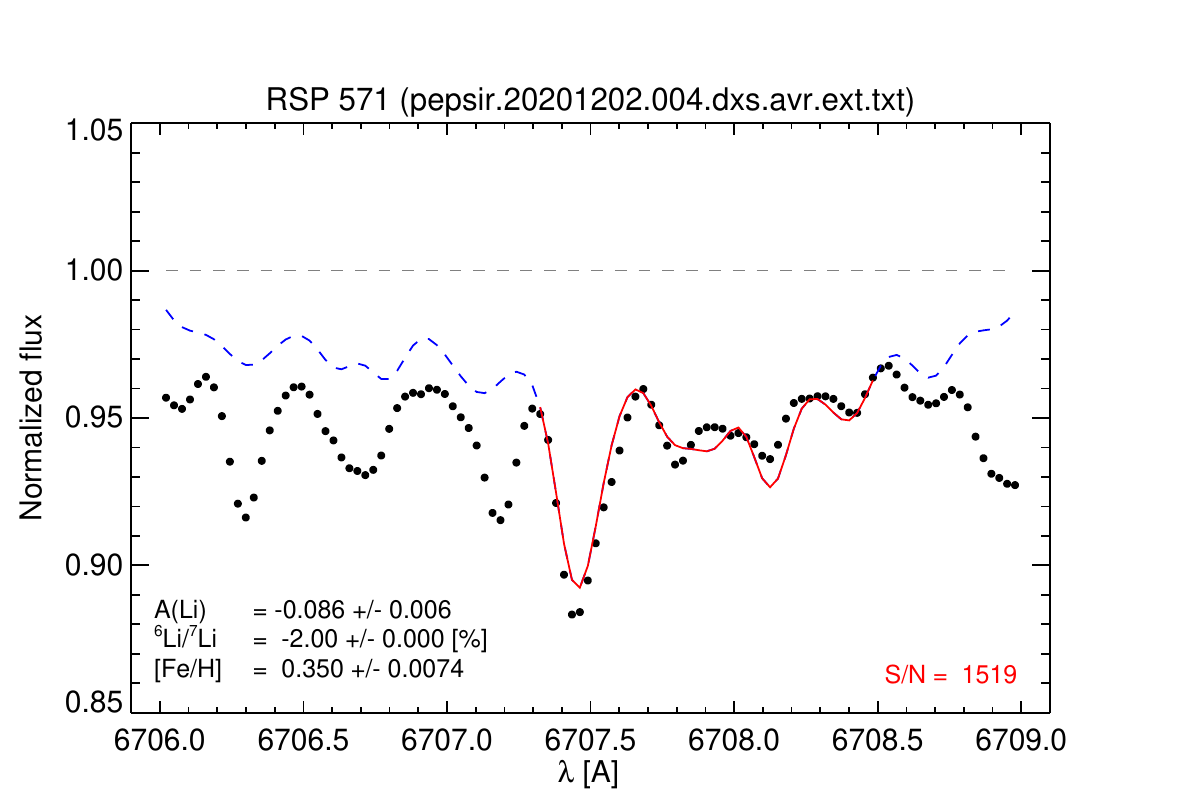}
\includegraphics[angle=0,width=5.cm,clip]{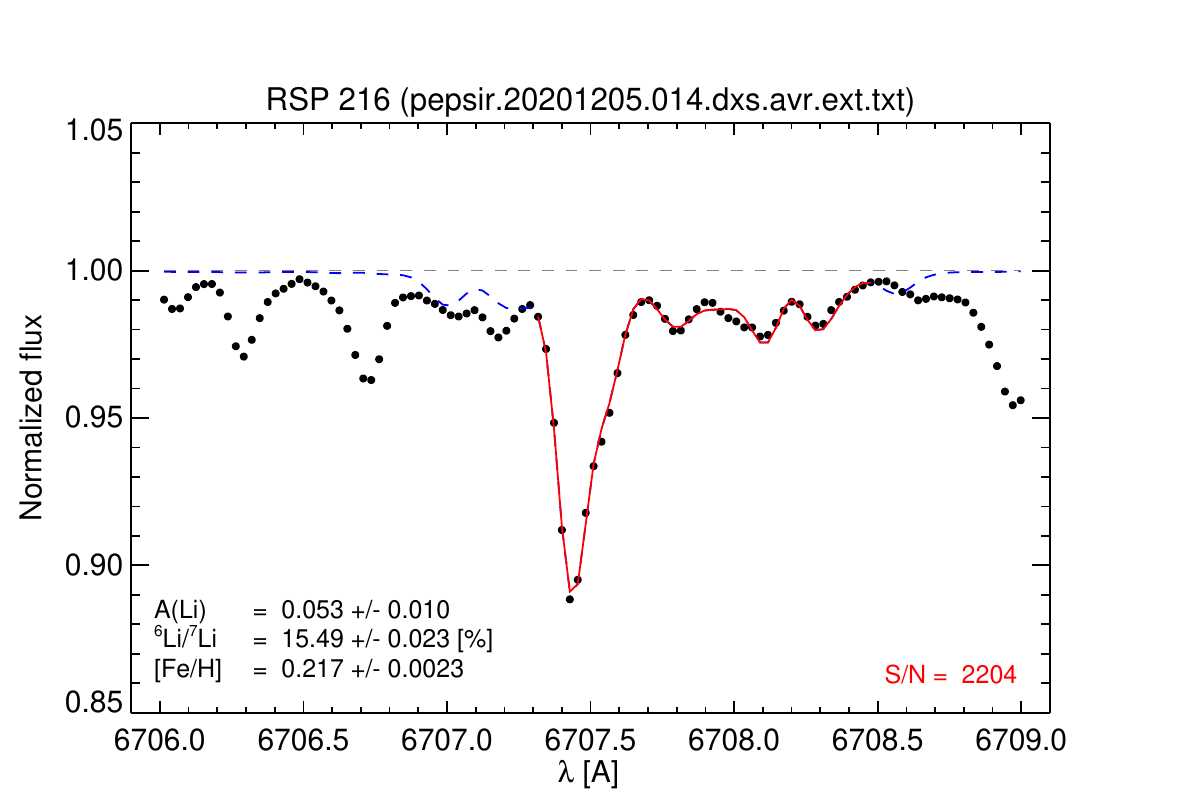}
\includegraphics[angle=0,width=5.cm,clip]{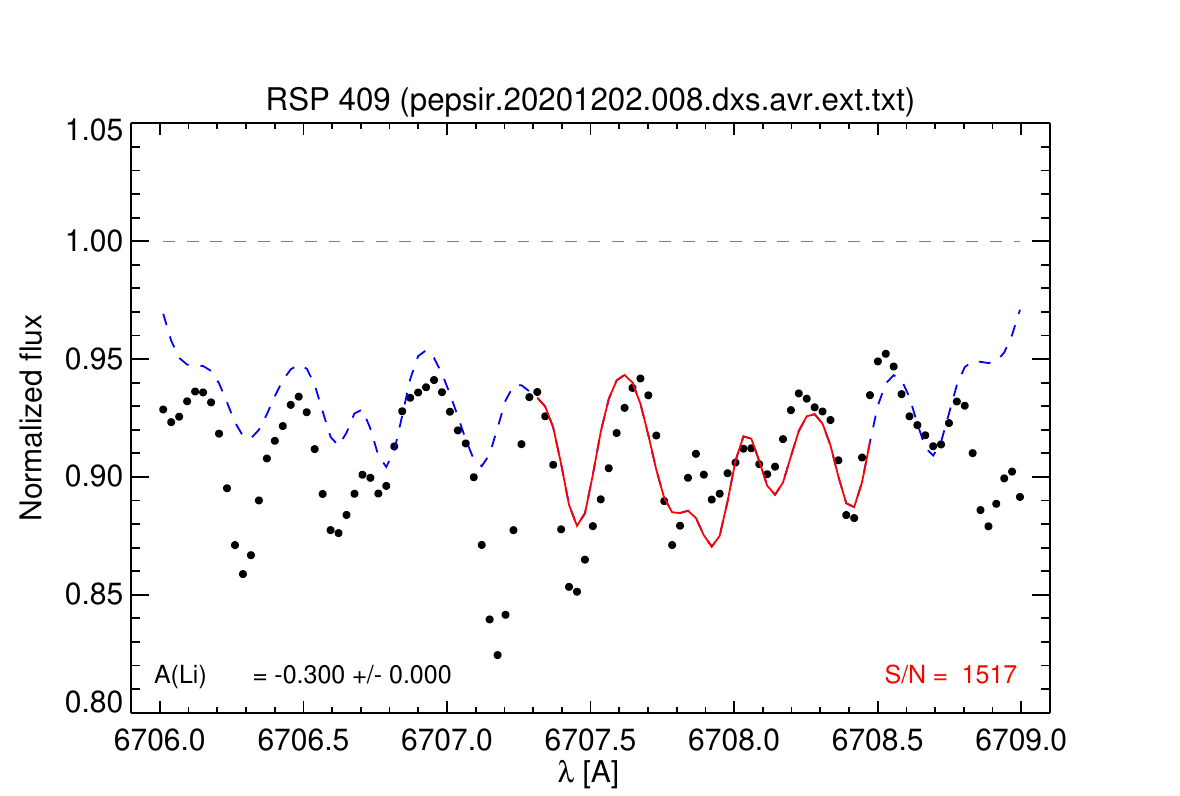}
\includegraphics[angle=0,width=5.cm,clip]{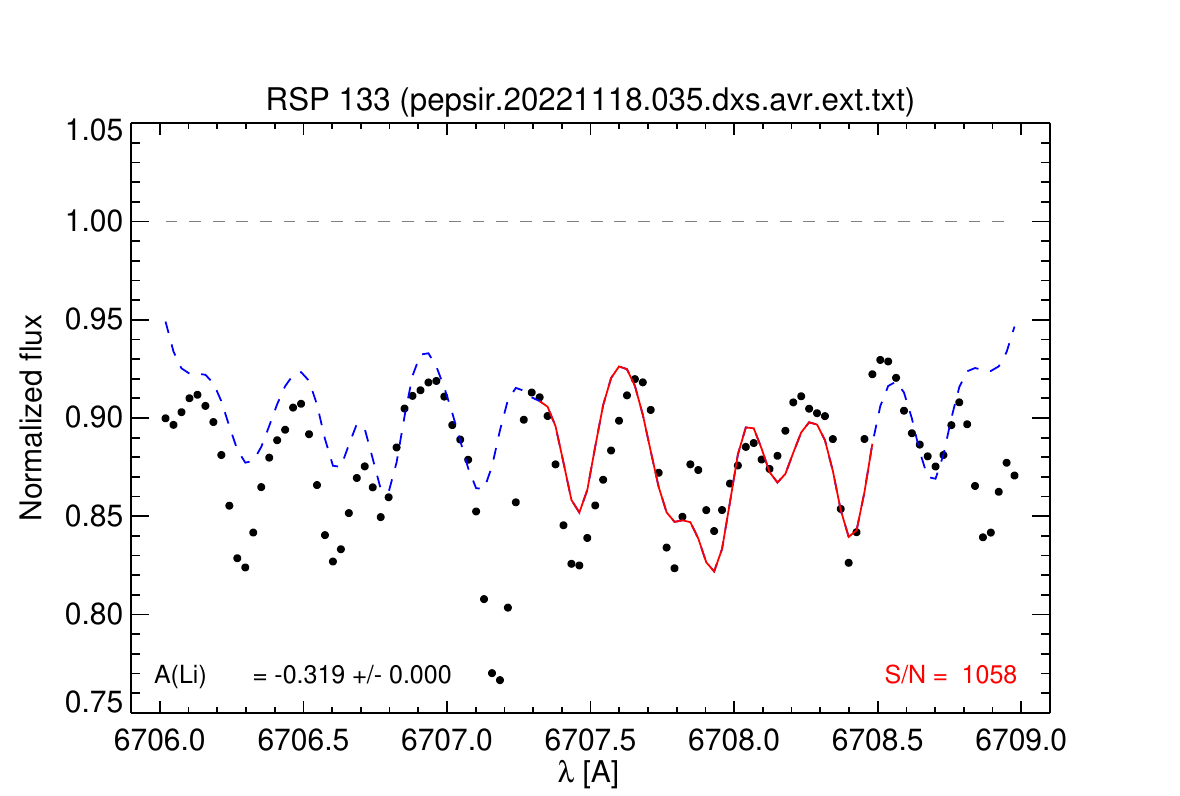}
\includegraphics[angle=0,width=5.cm,clip]{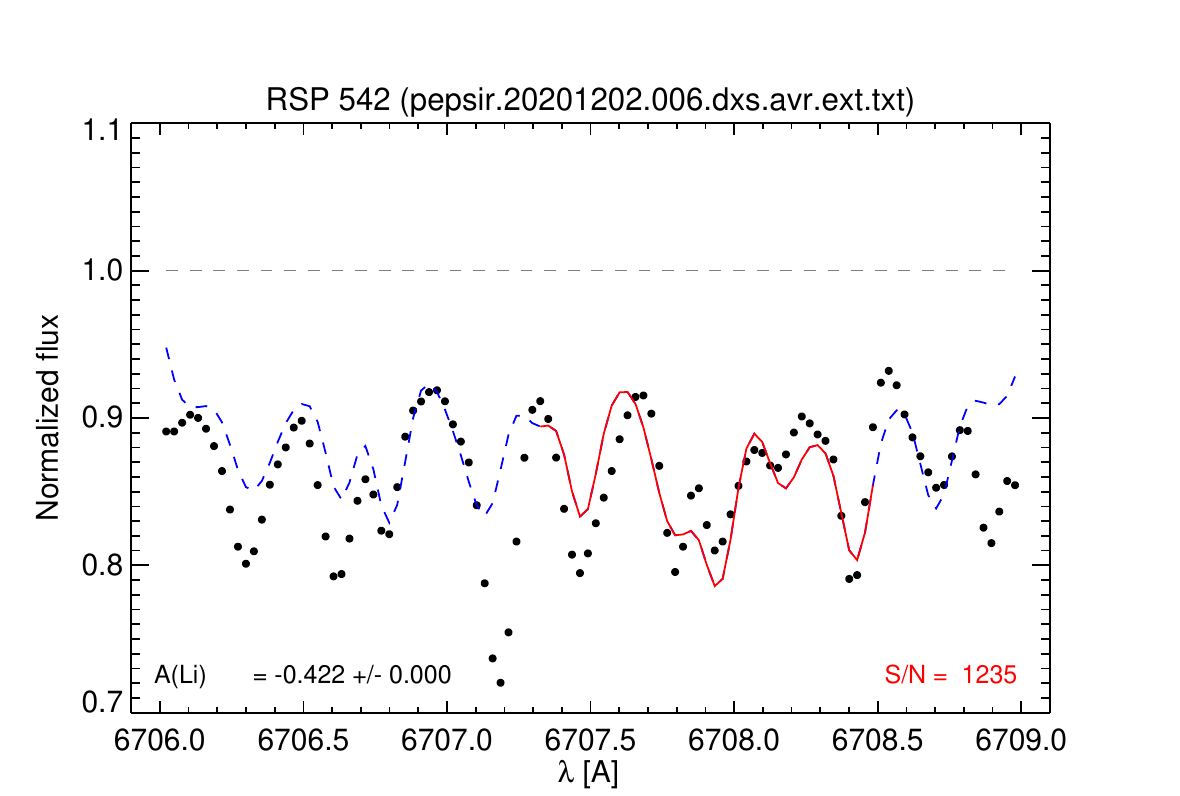}
\includegraphics[angle=0,width=5.cm,clip]{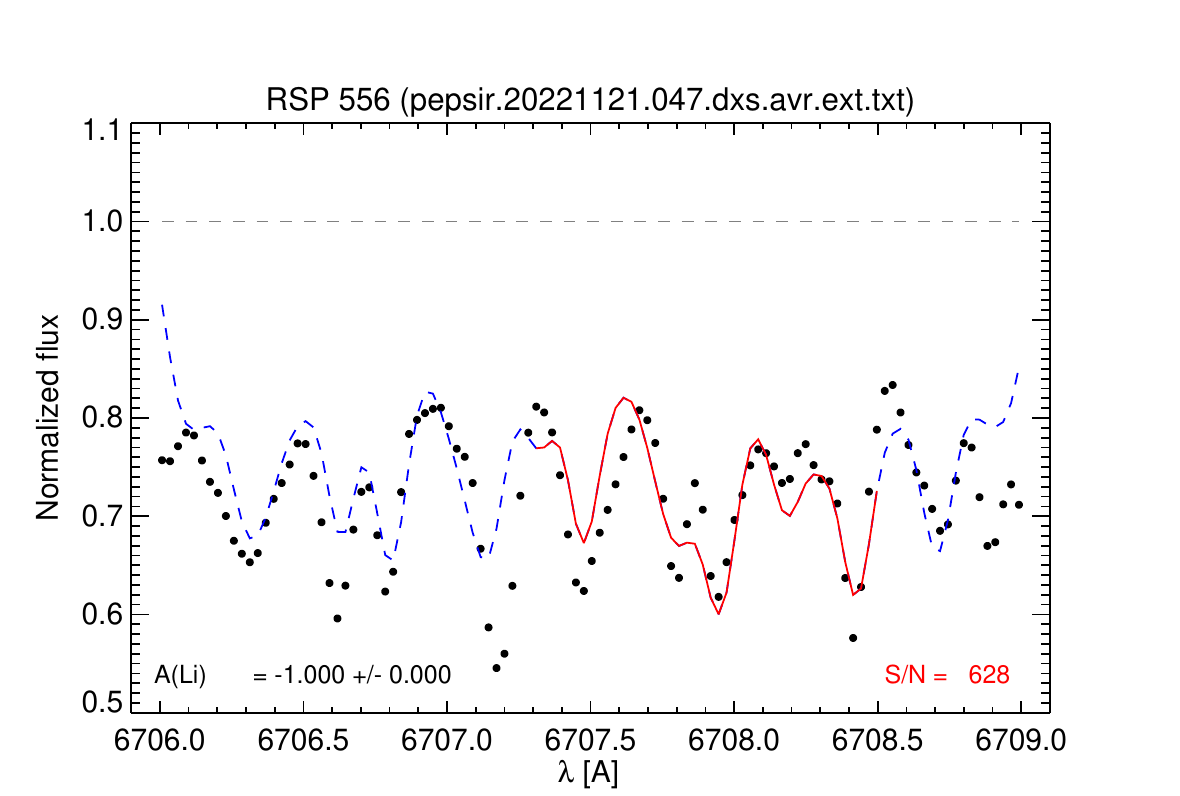}
 \caption{TurboMPfit of the lithium region (lines are the fit, dots are the observations). Fixed parameters were $T_{\rm eff}$, $\log g$, $\xi_t$, and $v\sin i$. Free parameters were A(Li), its isotope ratio, [Fe/H], a global Gaussian line broadening, and the continuum level. A total of 44 pixels were fitted (indicated by the red line). }
 \label{F_App1}
\end{figure*}

\begin{figure*}
 \centering
\includegraphics[angle=0,width=6.0cm,clip]{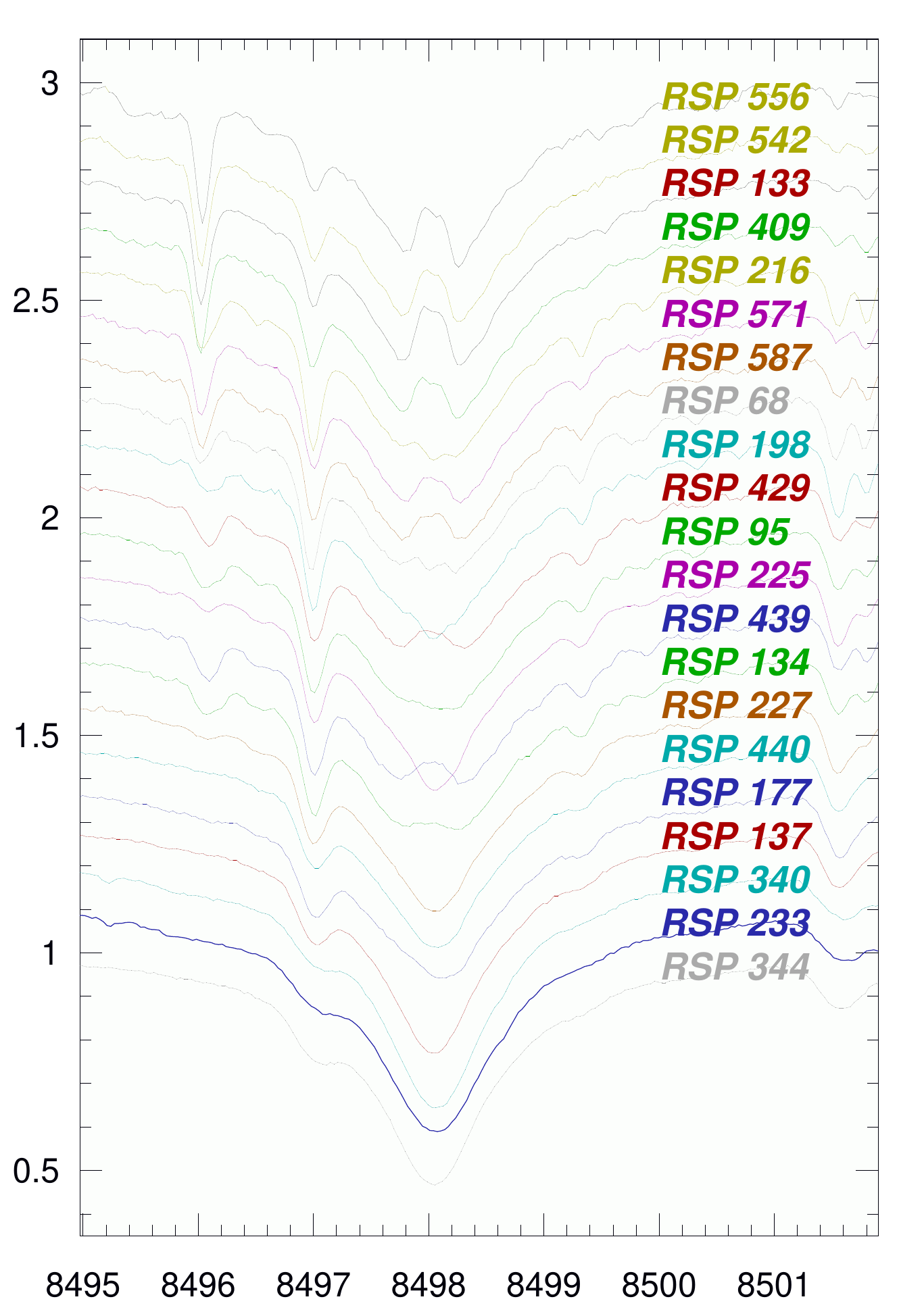}
\includegraphics[angle=0,width=6.0cm,clip]{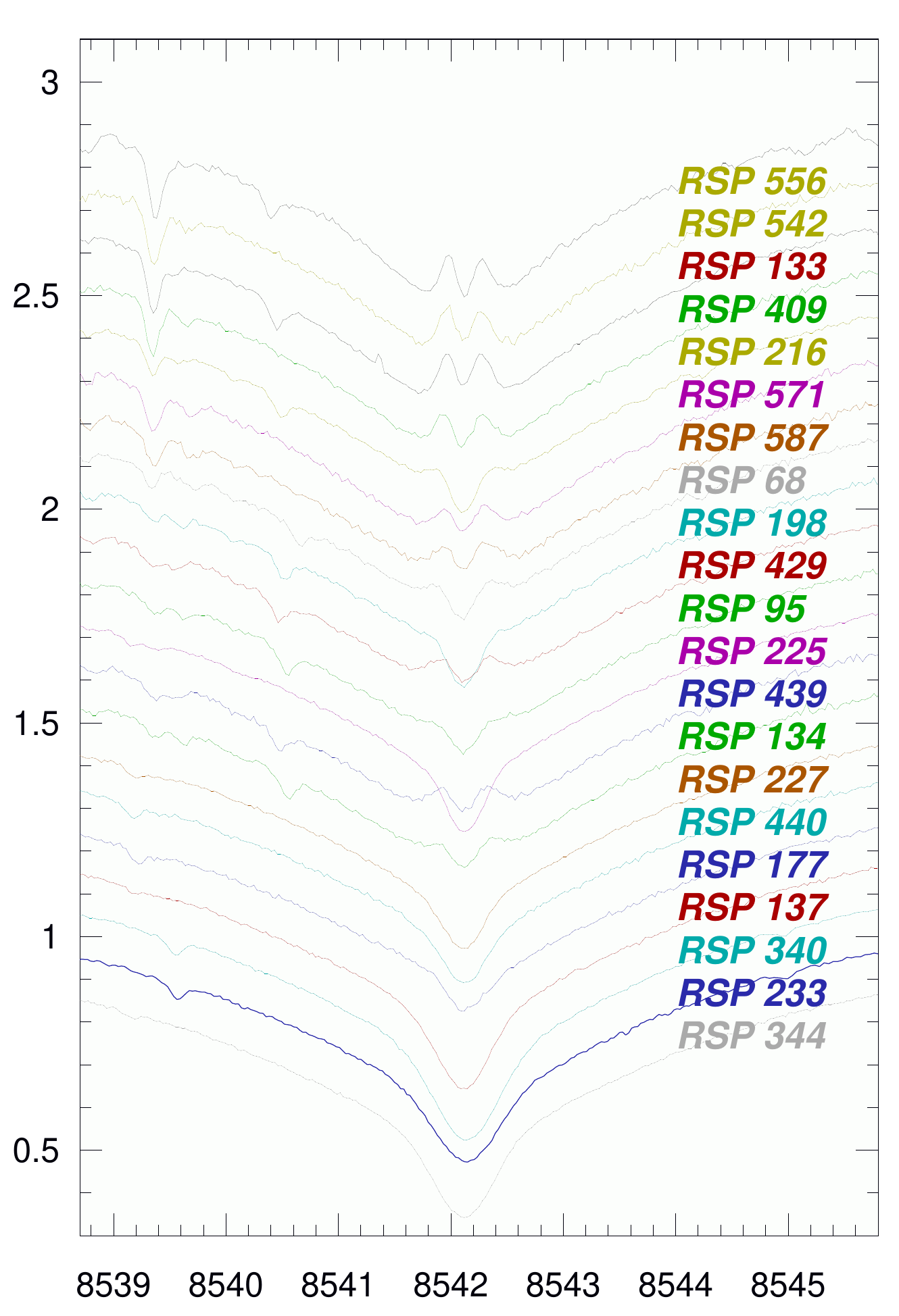}
\includegraphics[angle=0,width=6.0cm,clip]{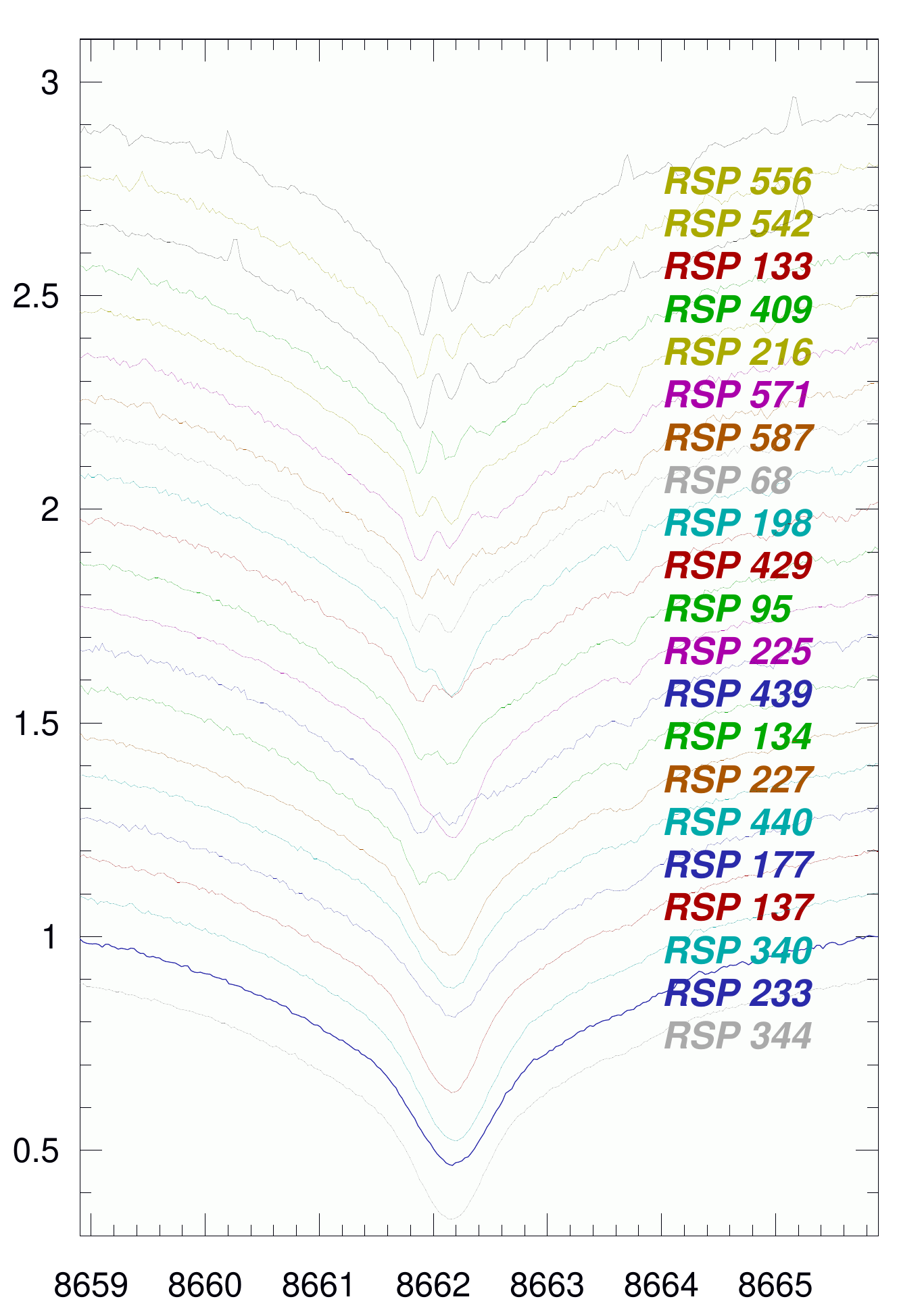}
\caption{Ca\,{\sc ii} IRT line profiles. From left to right: IRT-1 (8498\,\AA), IRT-2 (8542\,\AA), and IRT-3 (8662\,\AA). Targets are arranged with increasing rotation period from bottom to top; each spectrum is consecutively shifted in intensity by 0.1 for better visibility. Targets are identified on the right side of each plot. The x-axis is wavelength in \AA ngstroem. }
 \label{F_App2}
\end{figure*}

\begin{figure*}
 \centering
\includegraphics[angle=0,width=4.35cm,clip]{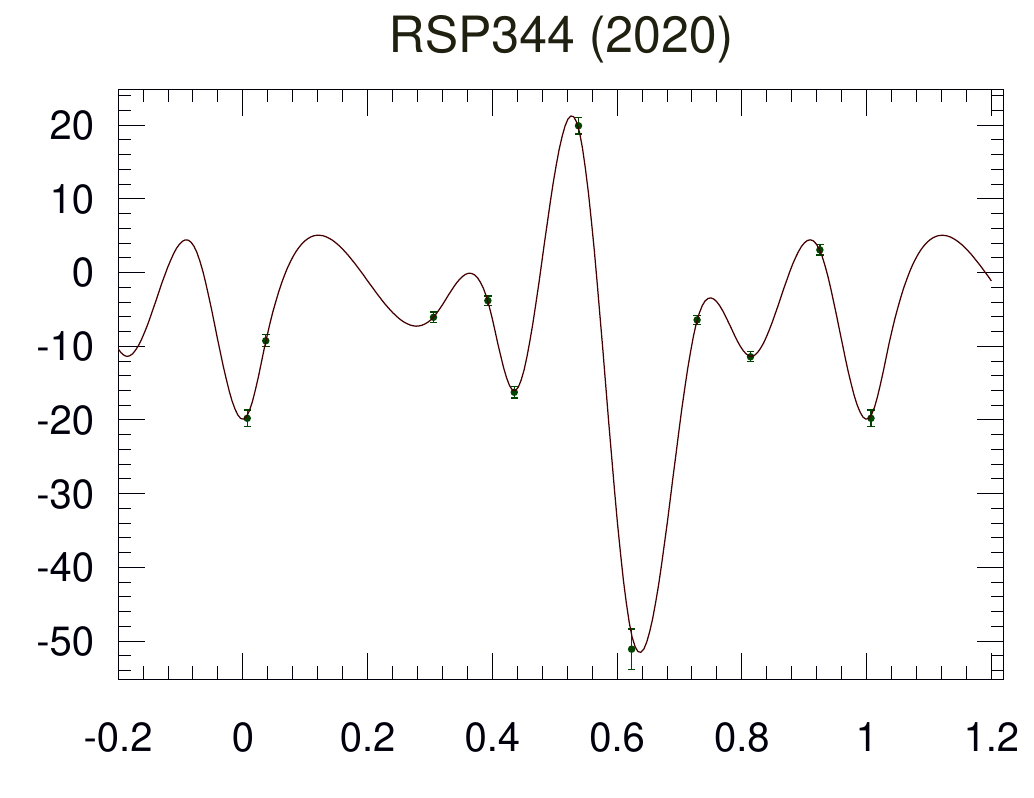}
\includegraphics[angle=0,width=4.35cm,clip]{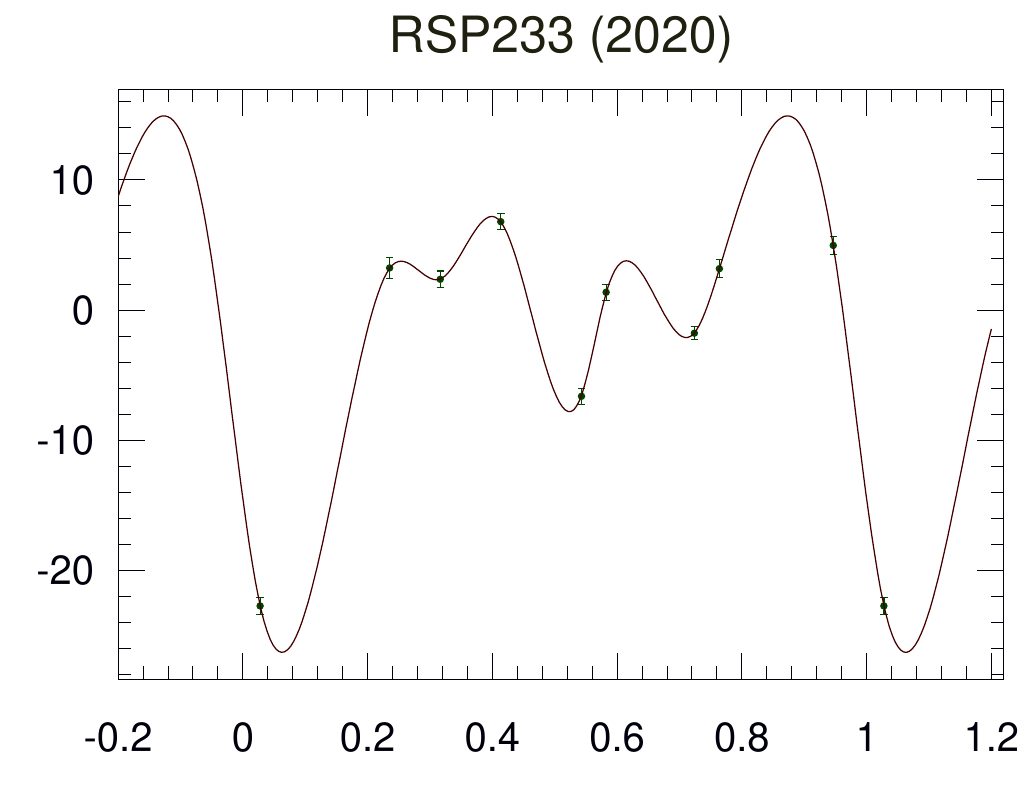}
\includegraphics[angle=0,width=4.35cm,clip]{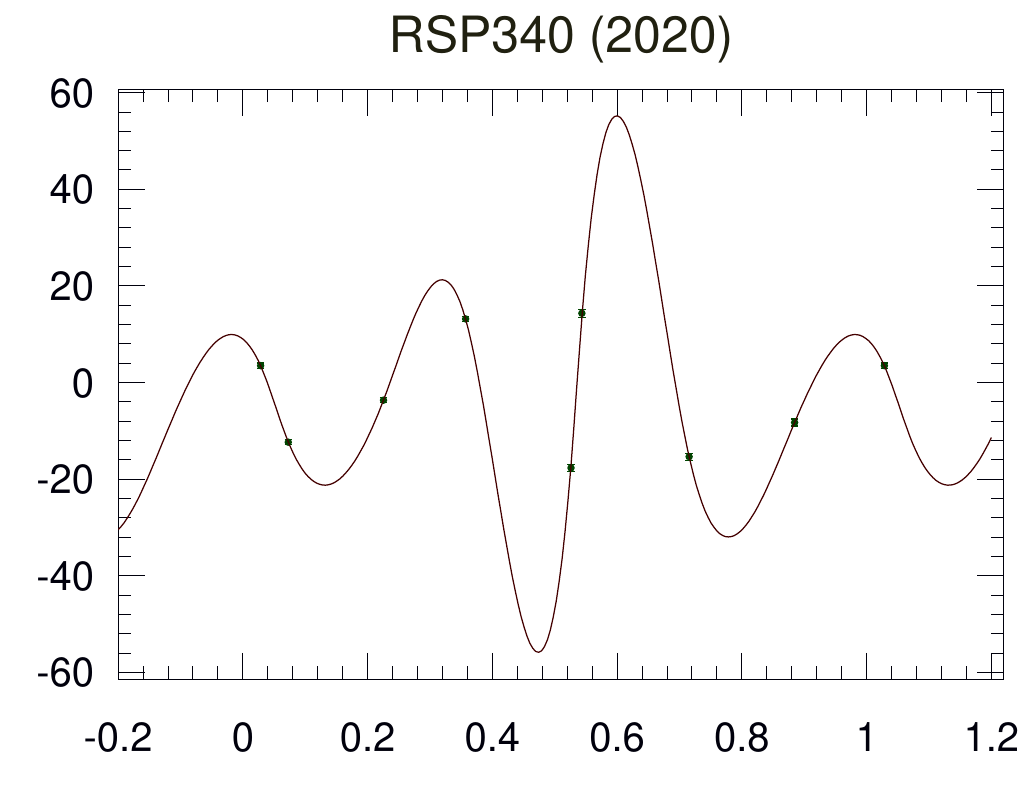}
\includegraphics[angle=0,width=4.35cm,clip]{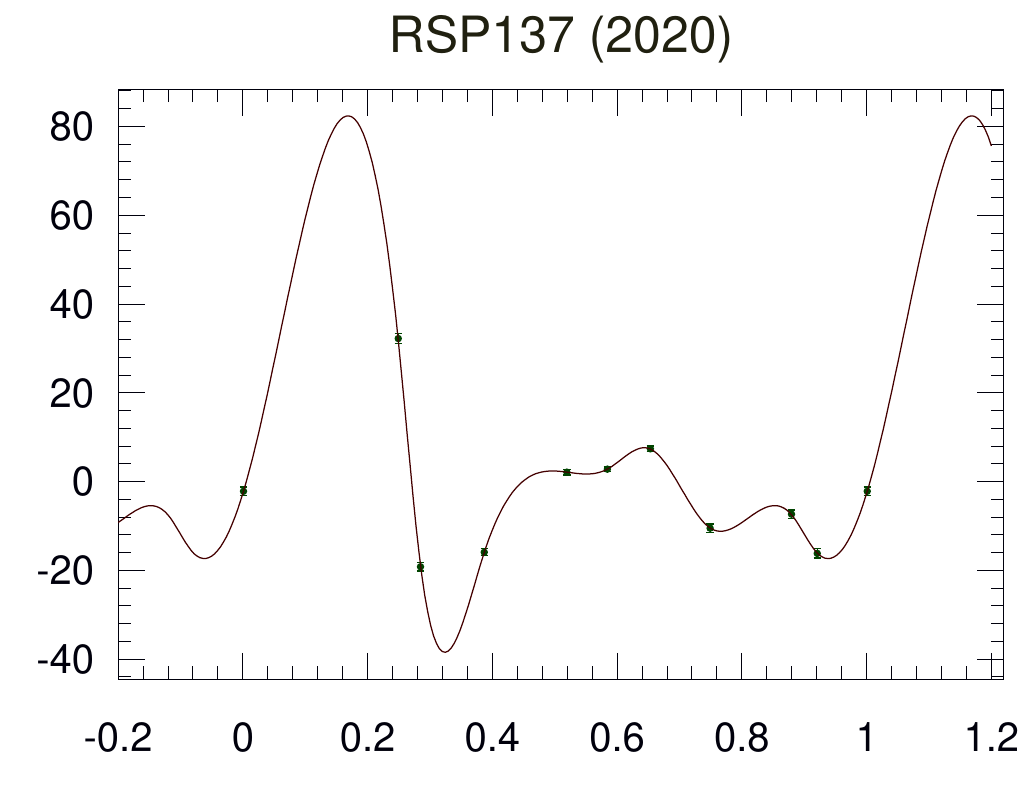}
\includegraphics[angle=0,width=4.35cm,clip]{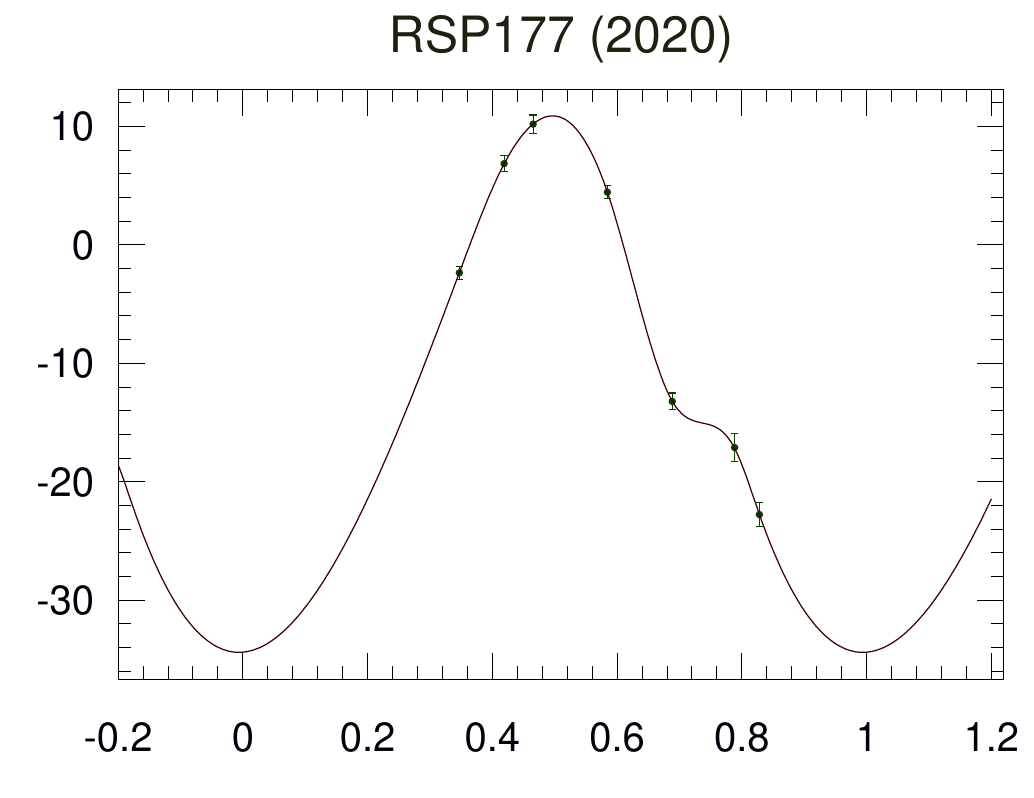}
\includegraphics[angle=0,width=4.35cm,clip]{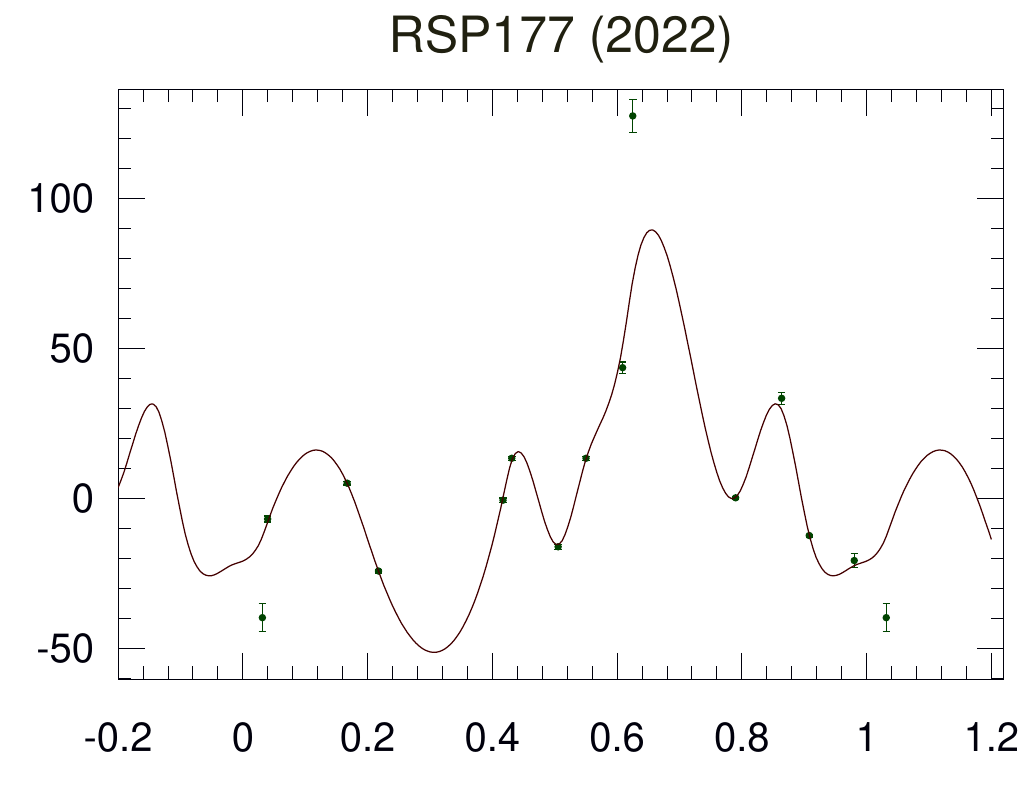}
\includegraphics[angle=0,width=4.35cm,clip]{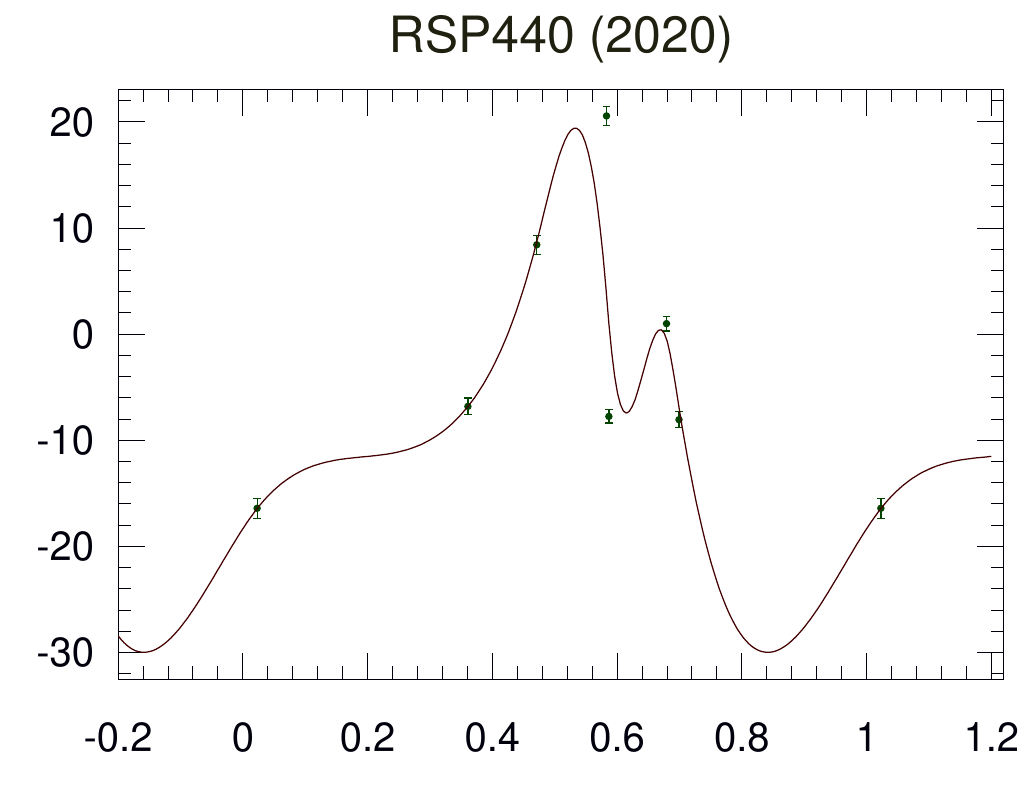}
\includegraphics[angle=0,width=4.35cm,clip]{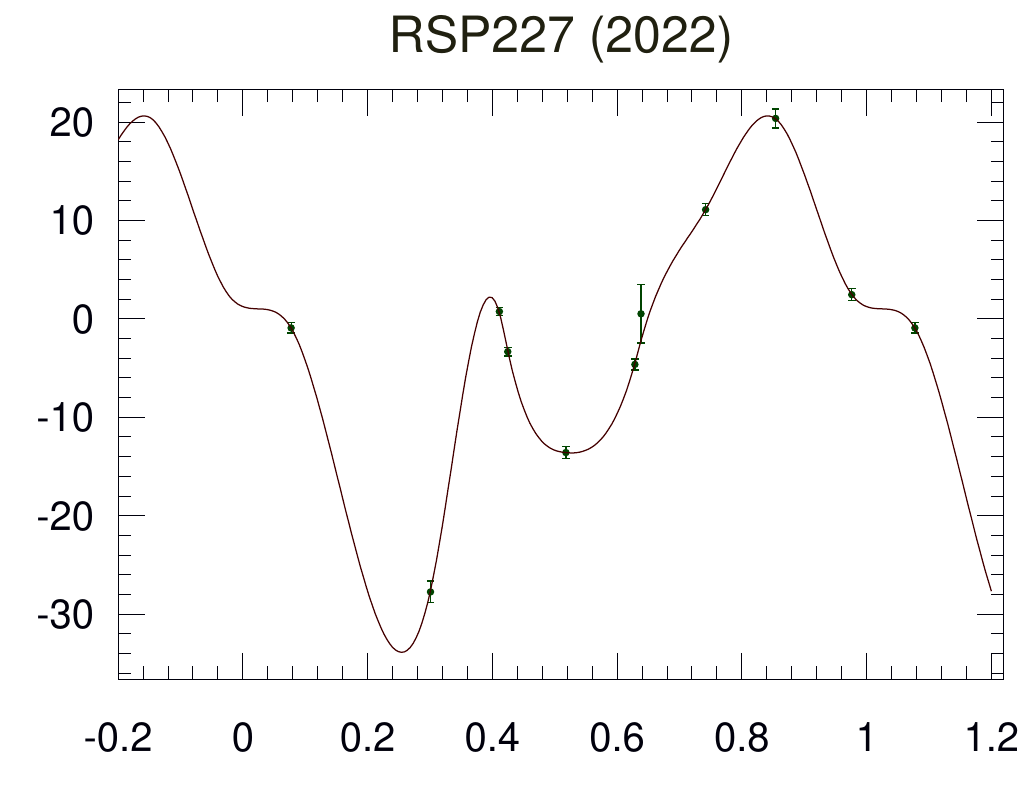}
\includegraphics[angle=0,width=4.35cm,clip]{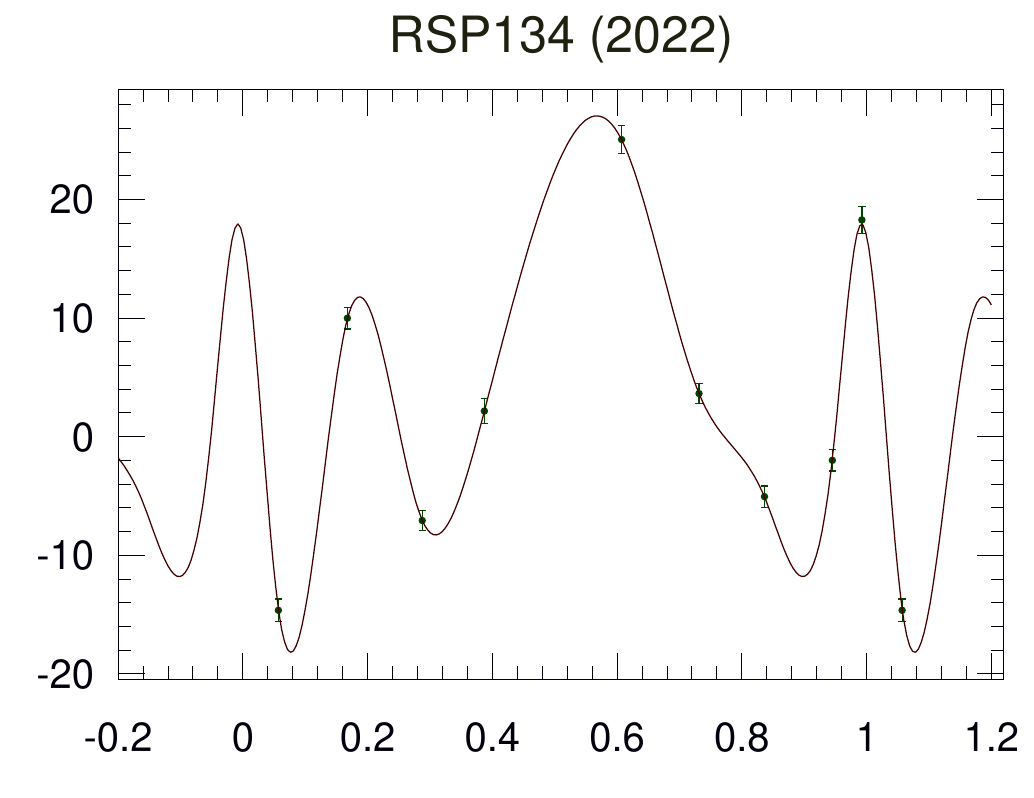}
\includegraphics[angle=0,width=4.35cm,clip]{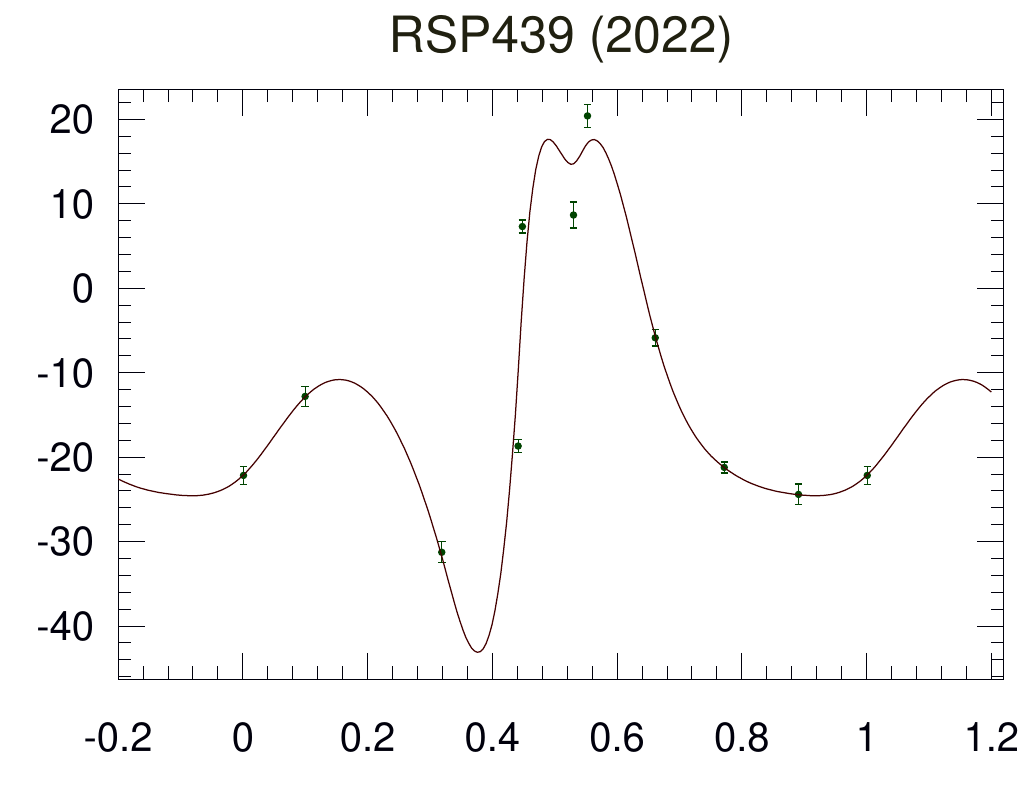}
\includegraphics[angle=0,width=4.35cm,clip]{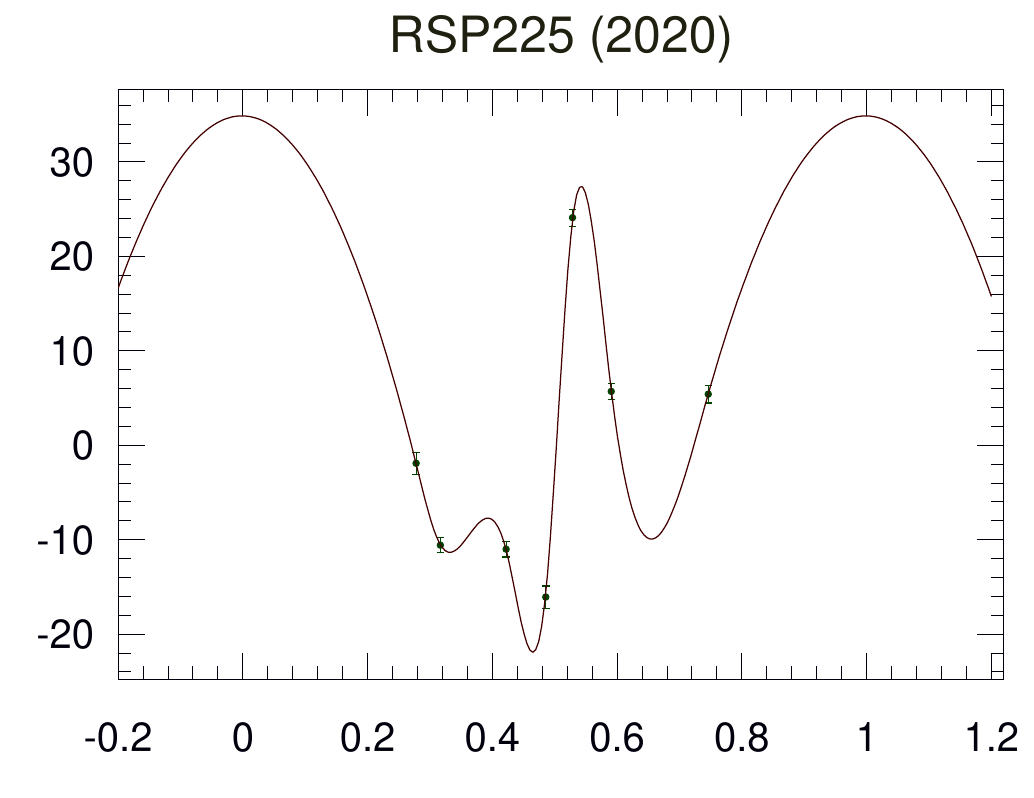}
\includegraphics[angle=0,width=4.35cm,clip]{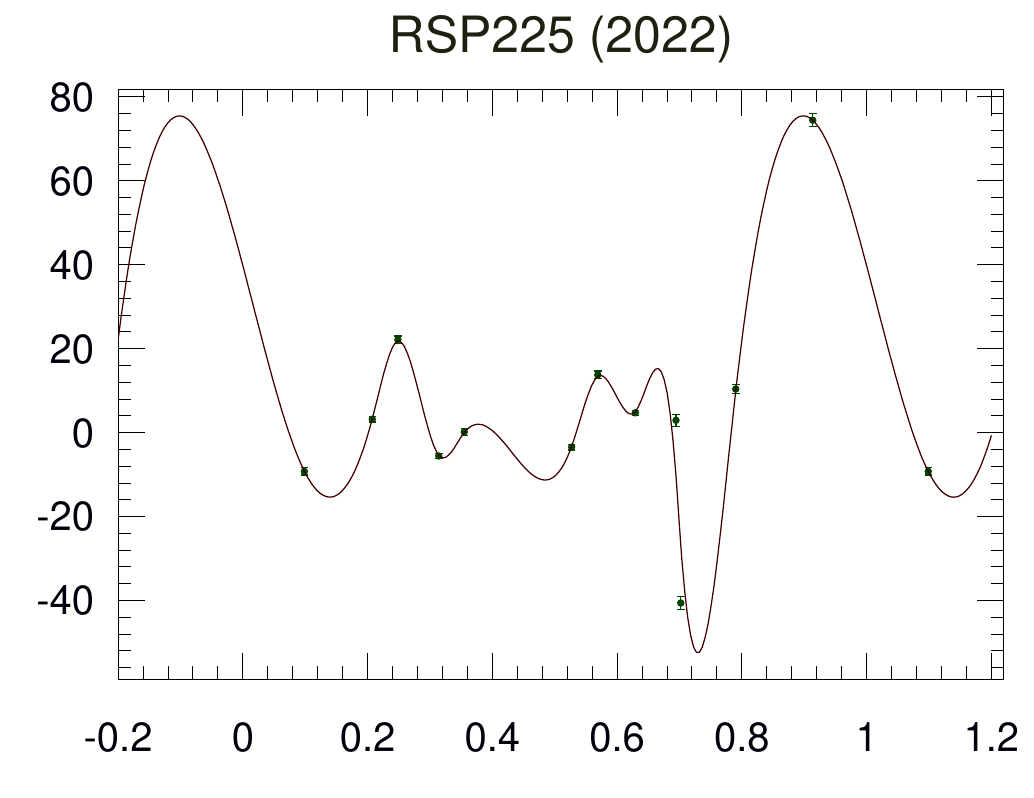}
\caption{Longitudinal magnetic-field measurements in Gauss versus rotational phase. Shown are the data (dots with error bars) and a bicubic spline as a line to guide the eye. Each panel is for a particular target identified in the panel header (the sequence follows Table~\ref{T1}). Targets with data from more than one observing season are shown in separate panels with the year indicated in the header. }
 \label{F_App3}
\end{figure*}
\setcounter{figure}{2}
\begin{figure*}
 \centering
\includegraphics[angle=0,width=4.35cm,clip]{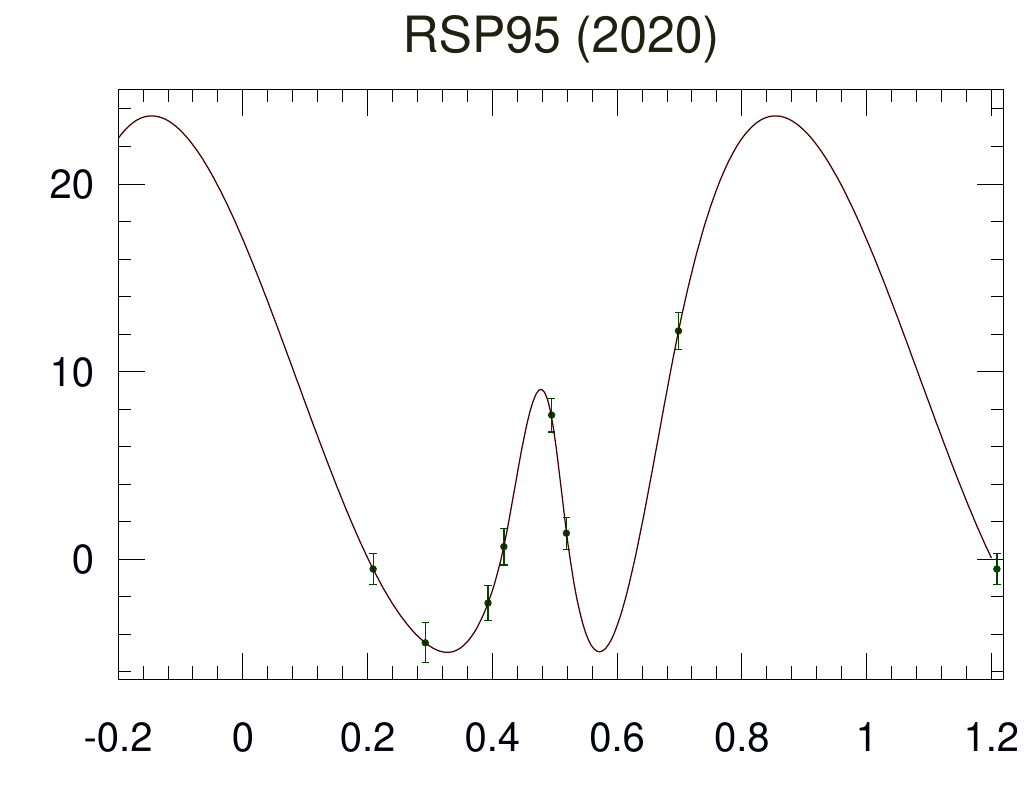}
\includegraphics[angle=0,width=4.35cm,clip]{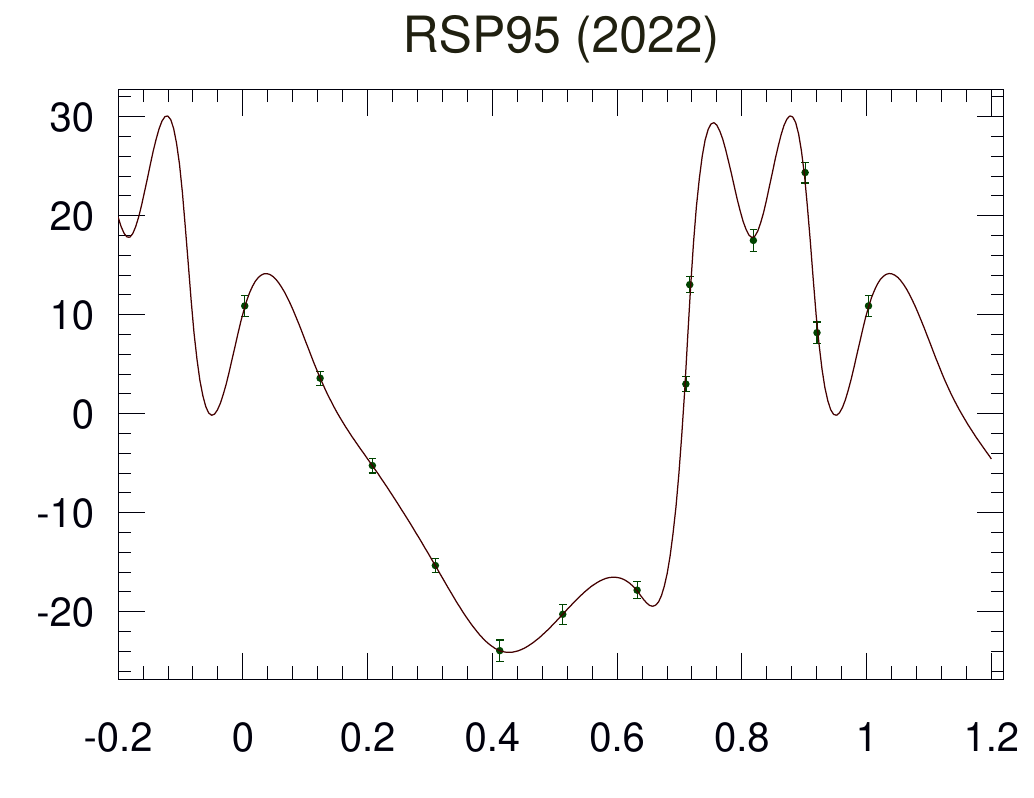}
\includegraphics[angle=0,width=4.35cm,clip]{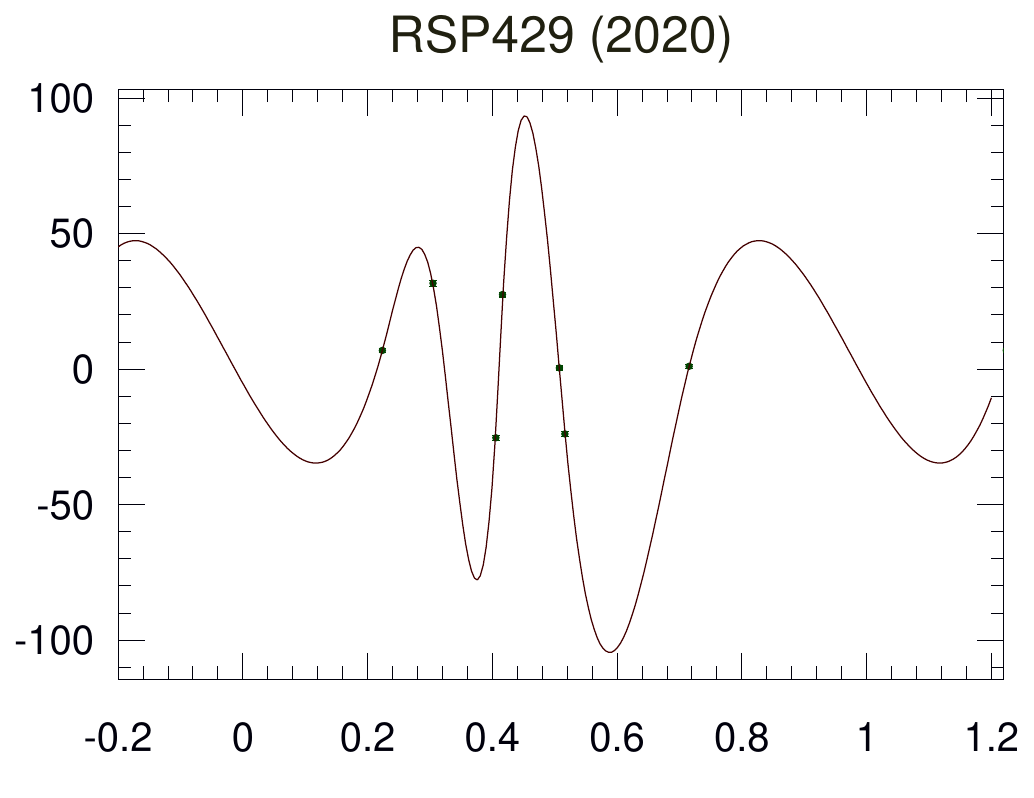}
\includegraphics[angle=0,width=4.35cm,clip]{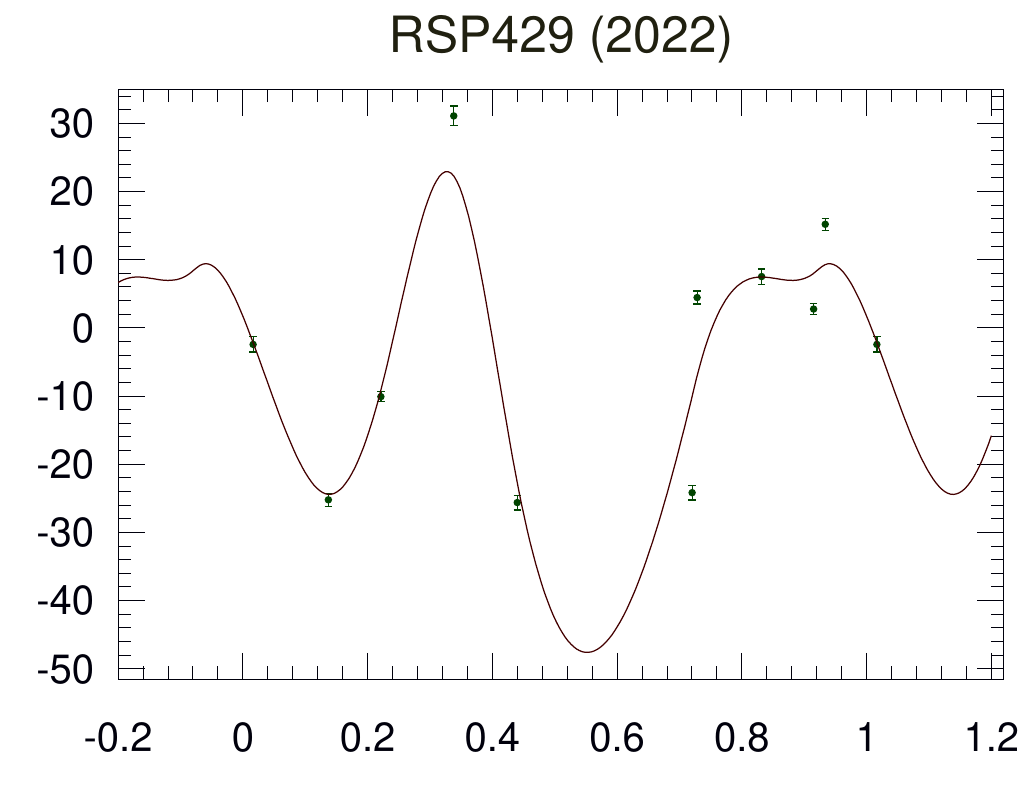}
\includegraphics[angle=0,width=4.35cm,clip]{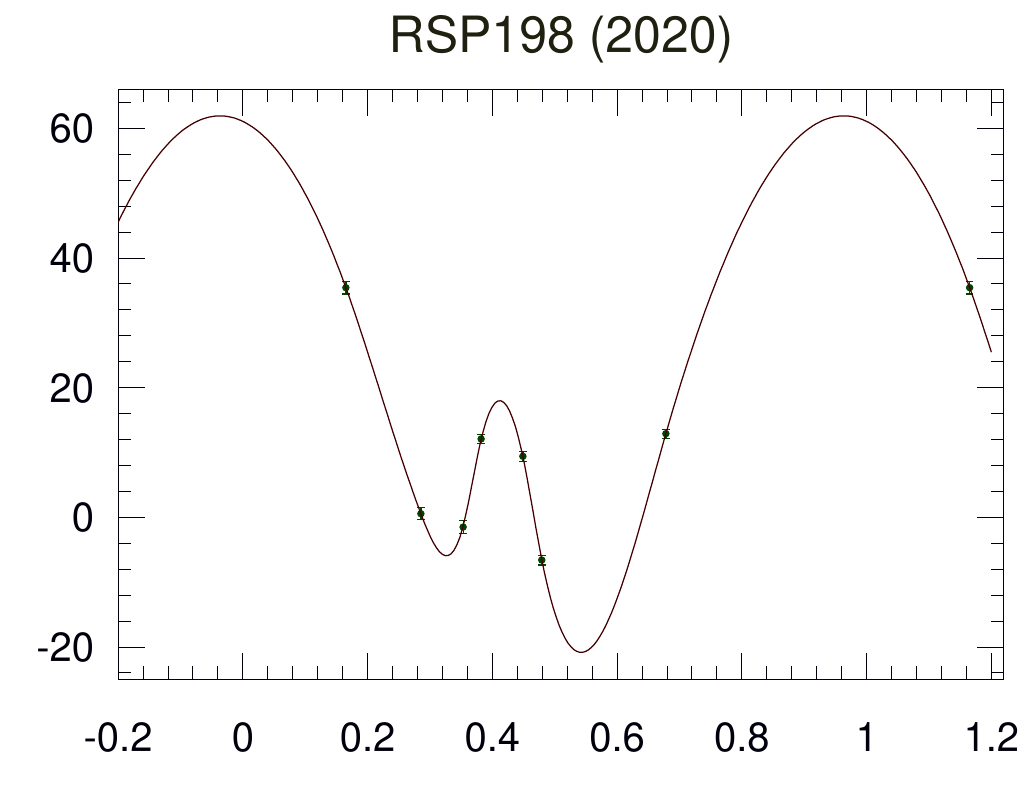}
\includegraphics[angle=0,width=4.35cm,clip]{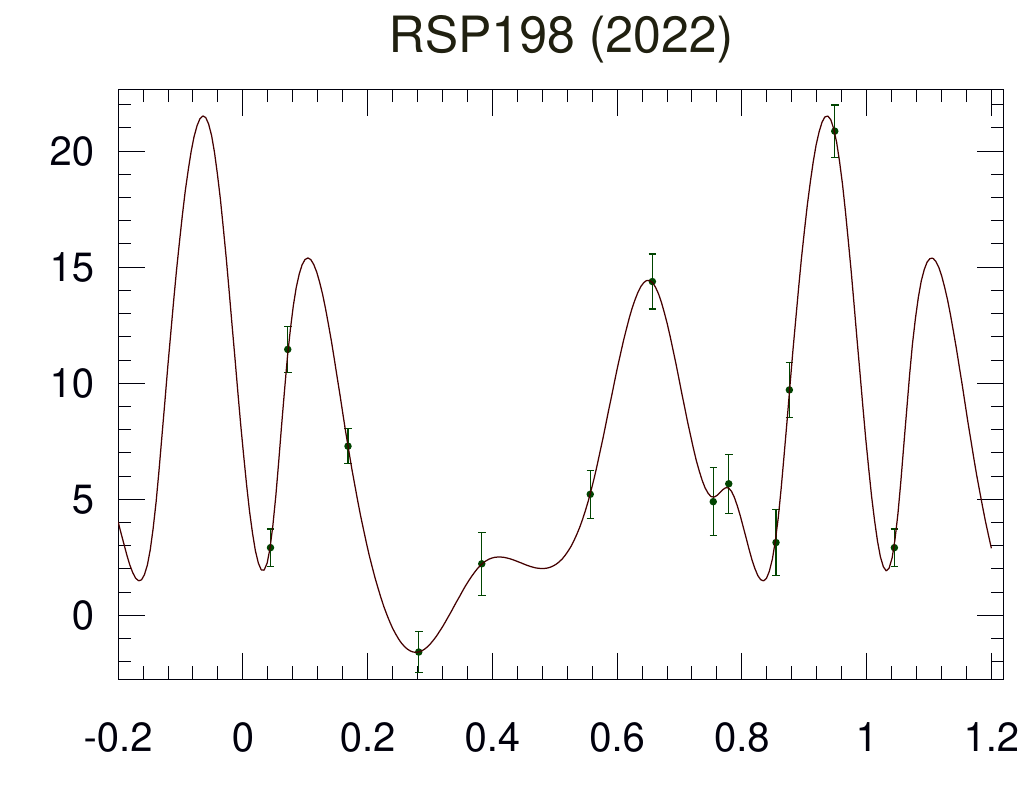}
\includegraphics[angle=0,width=4.35cm,clip]{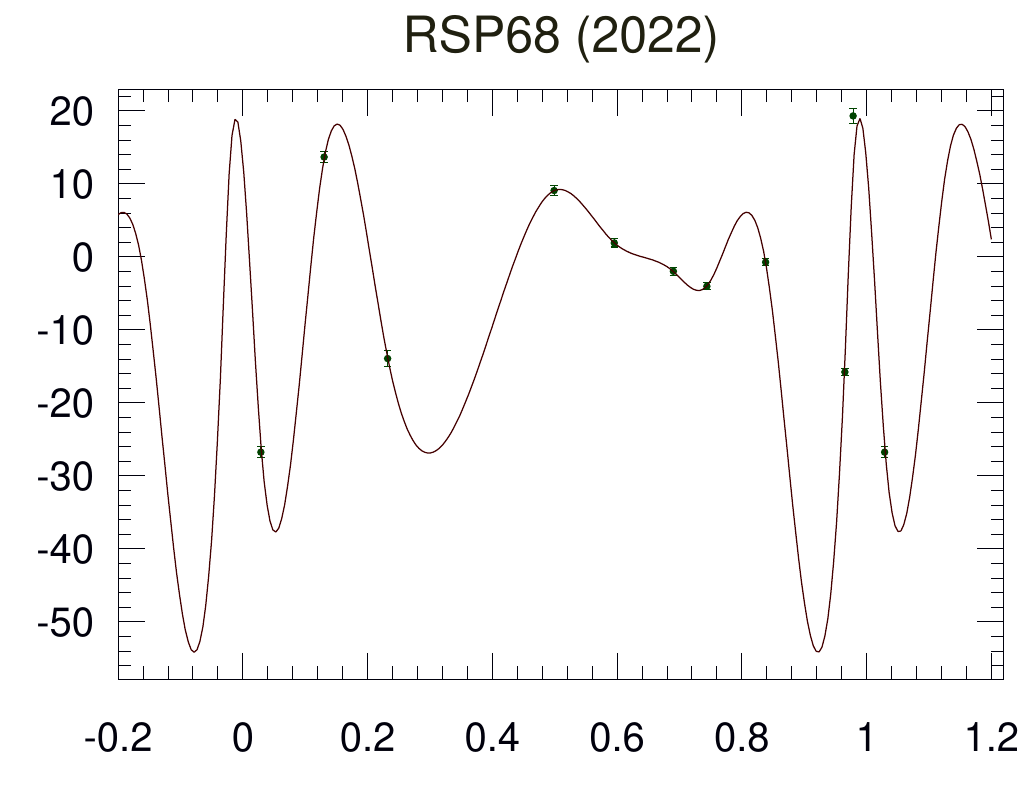}
\includegraphics[angle=0,width=4.35cm,clip]{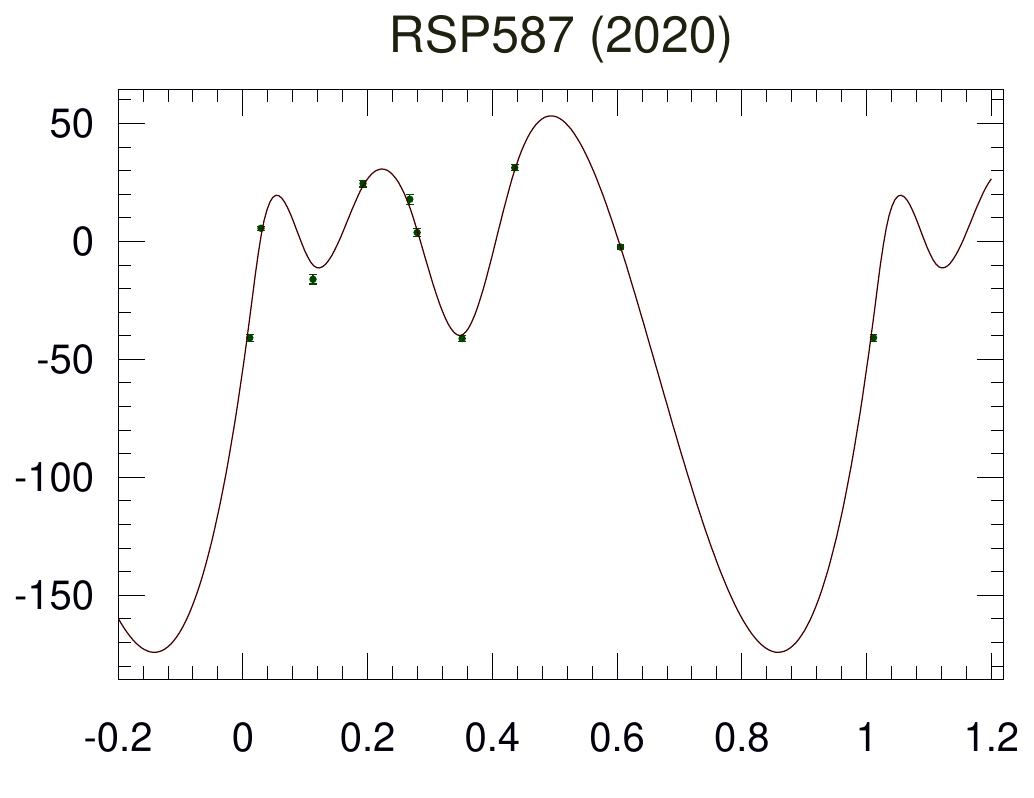}
\includegraphics[angle=0,width=4.35cm,clip]{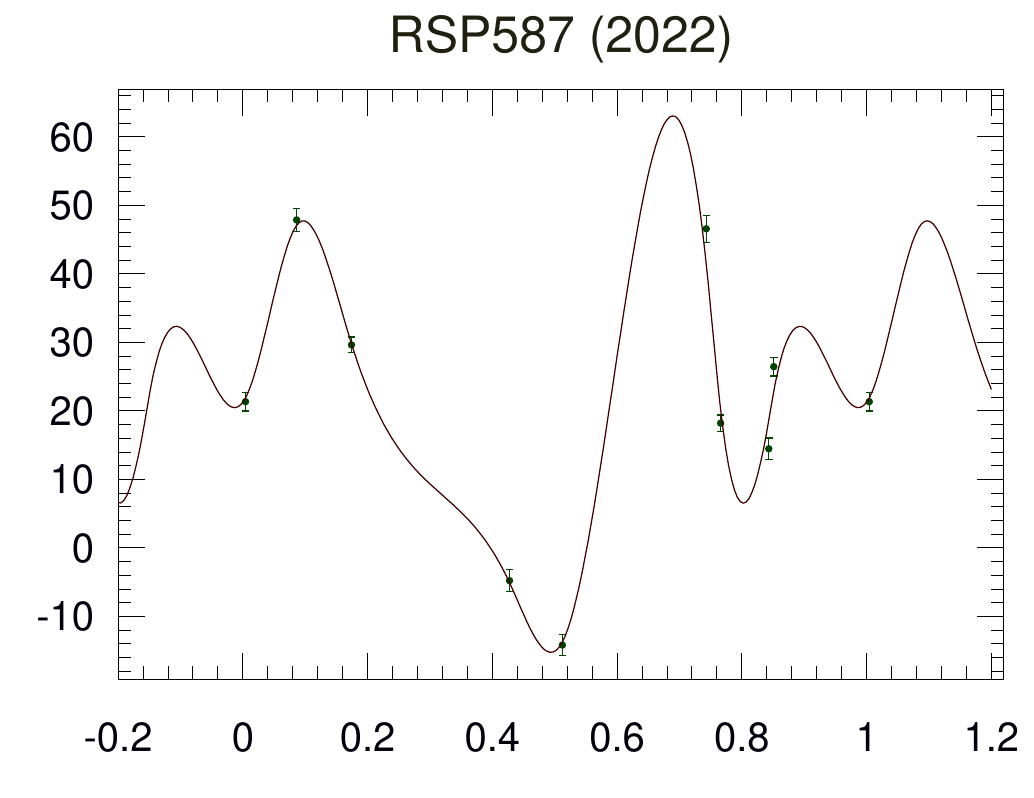}
\includegraphics[angle=0,width=4.35cm,clip]{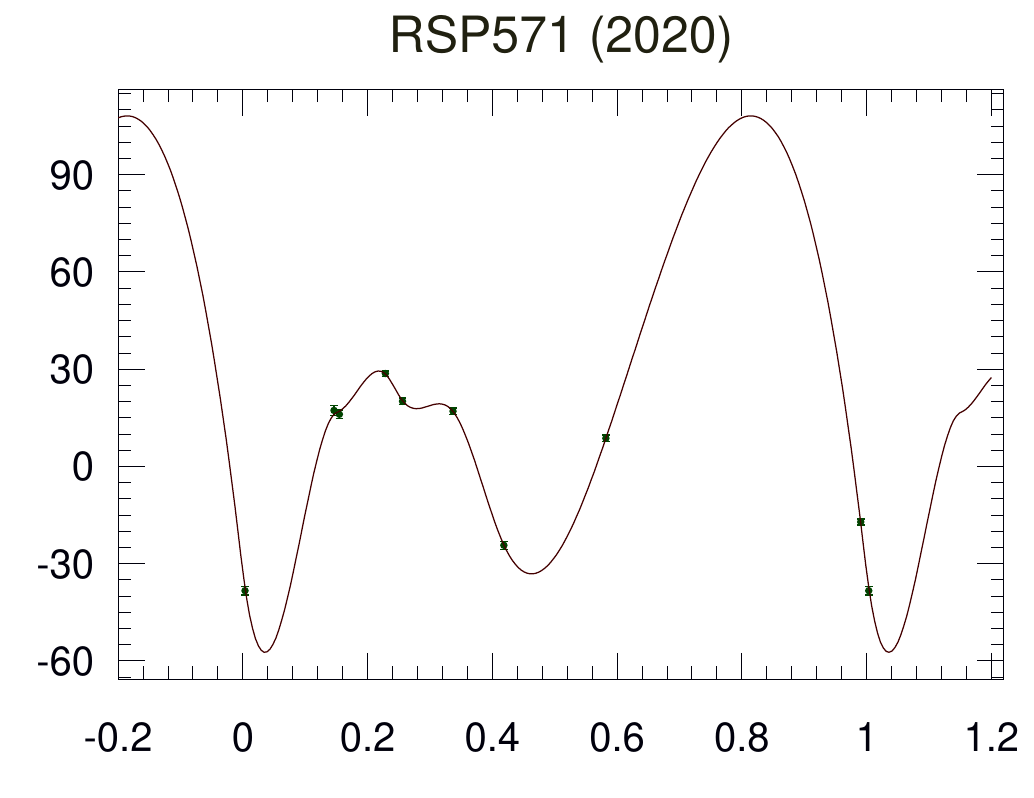}
\includegraphics[angle=0,width=4.35cm,clip]{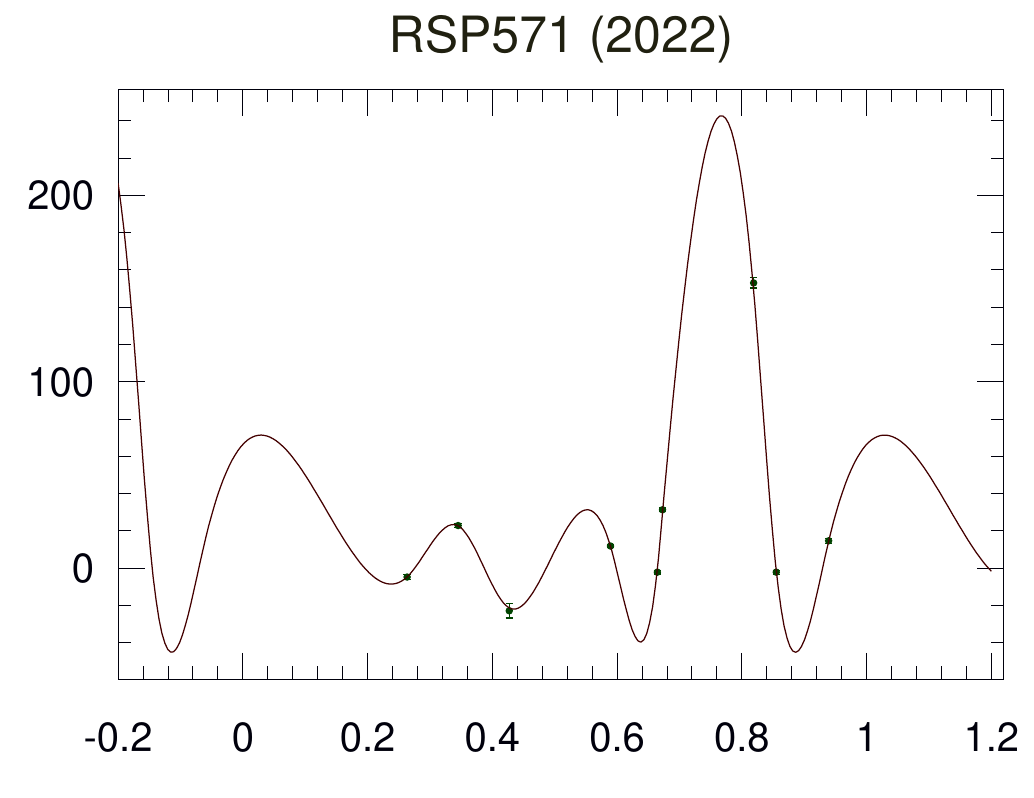}
\includegraphics[angle=0,width=4.35cm,clip]{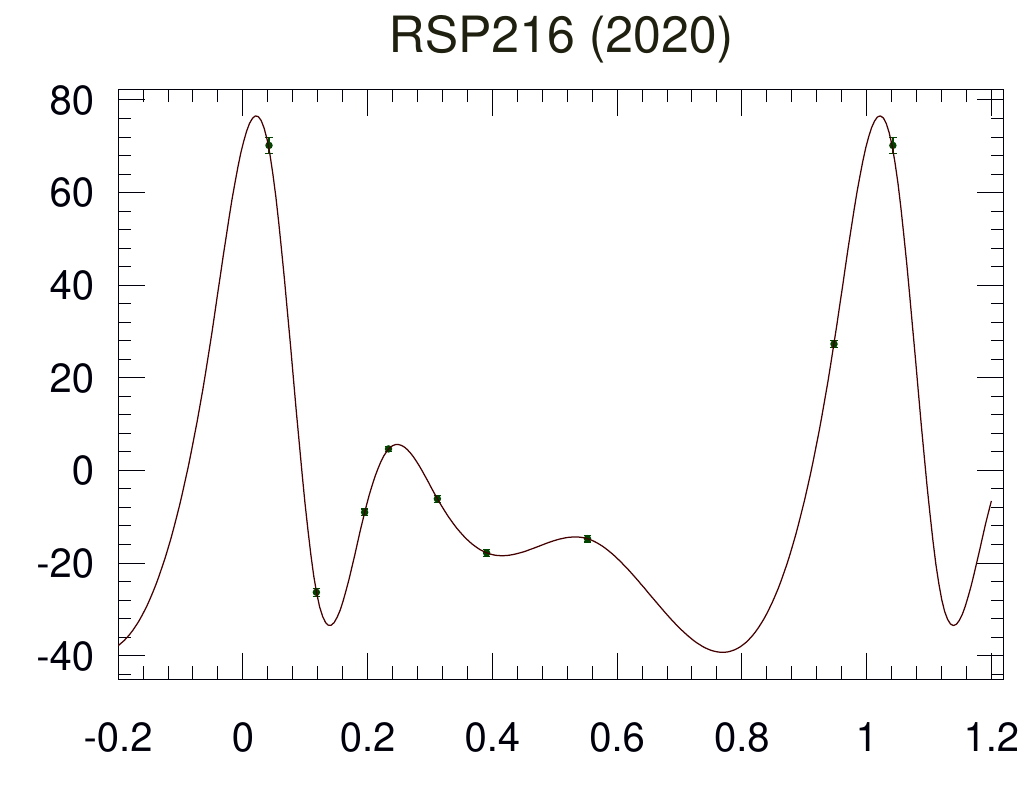}
\includegraphics[angle=0,width=4.35cm,clip]{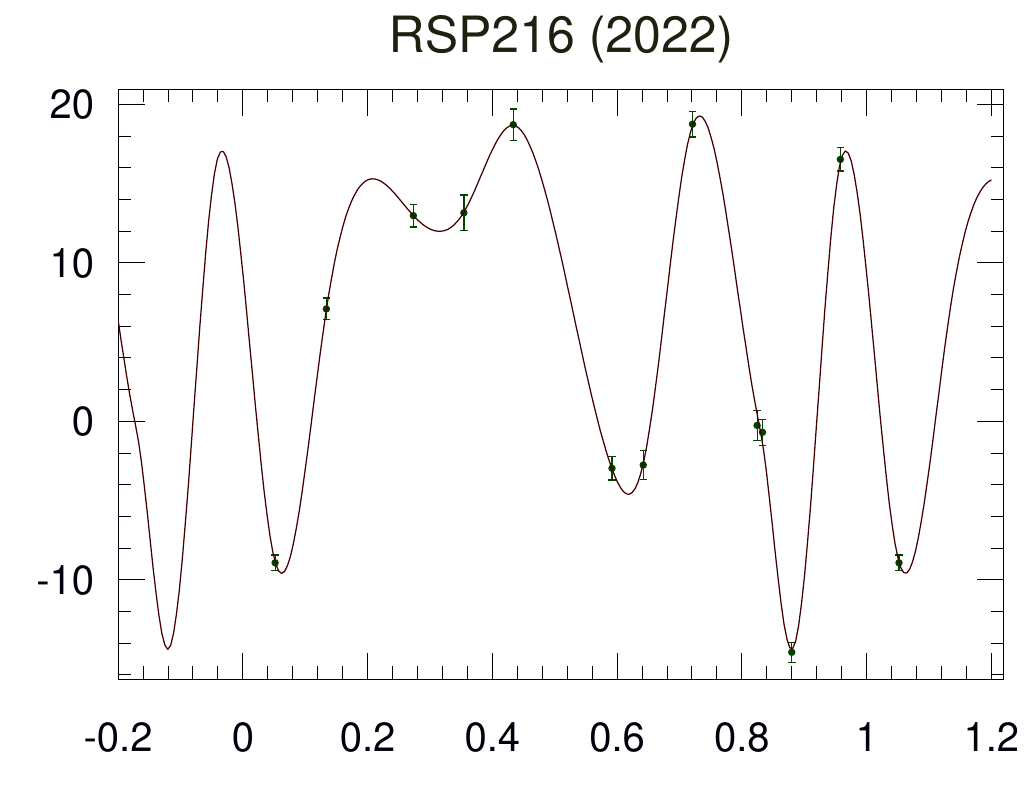}
\includegraphics[angle=0,width=4.35cm,clip]{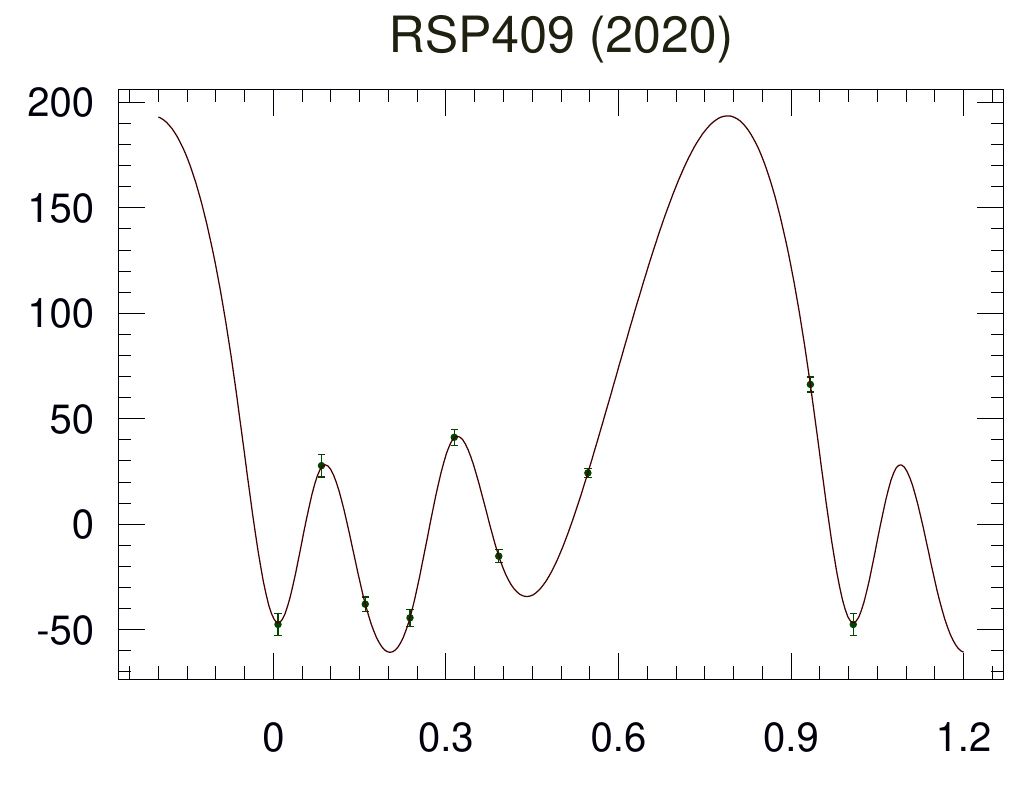}
\includegraphics[angle=0,width=4.35cm,clip]{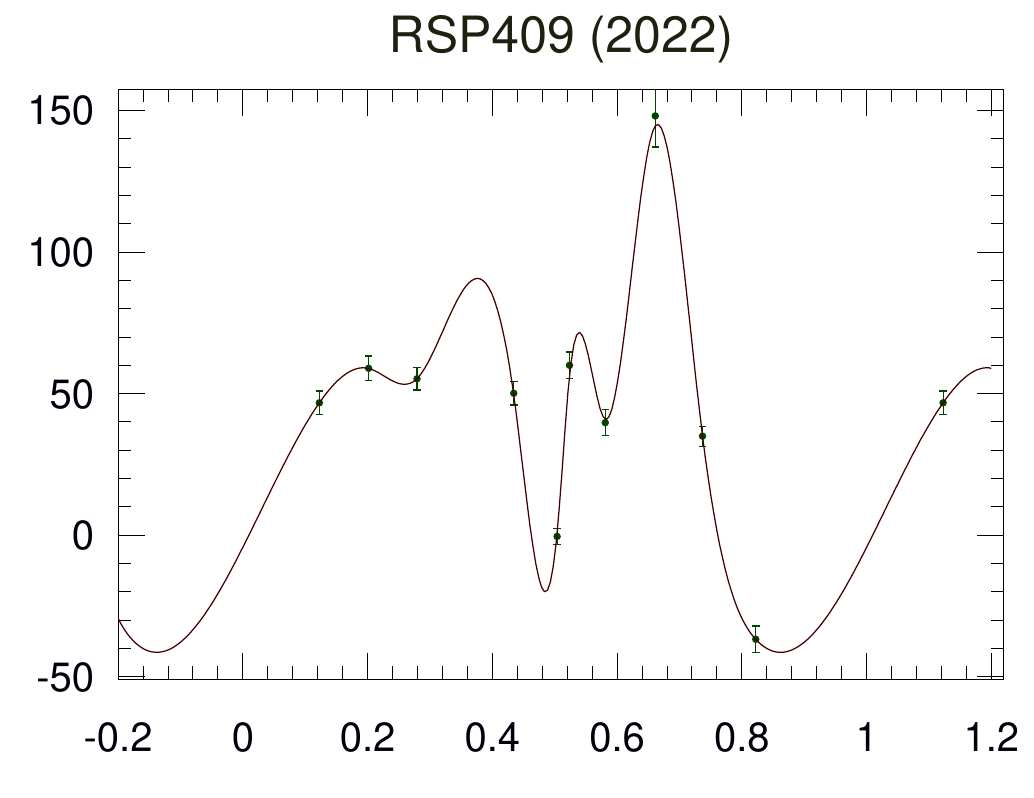}
\includegraphics[angle=0,width=4.35cm,clip]{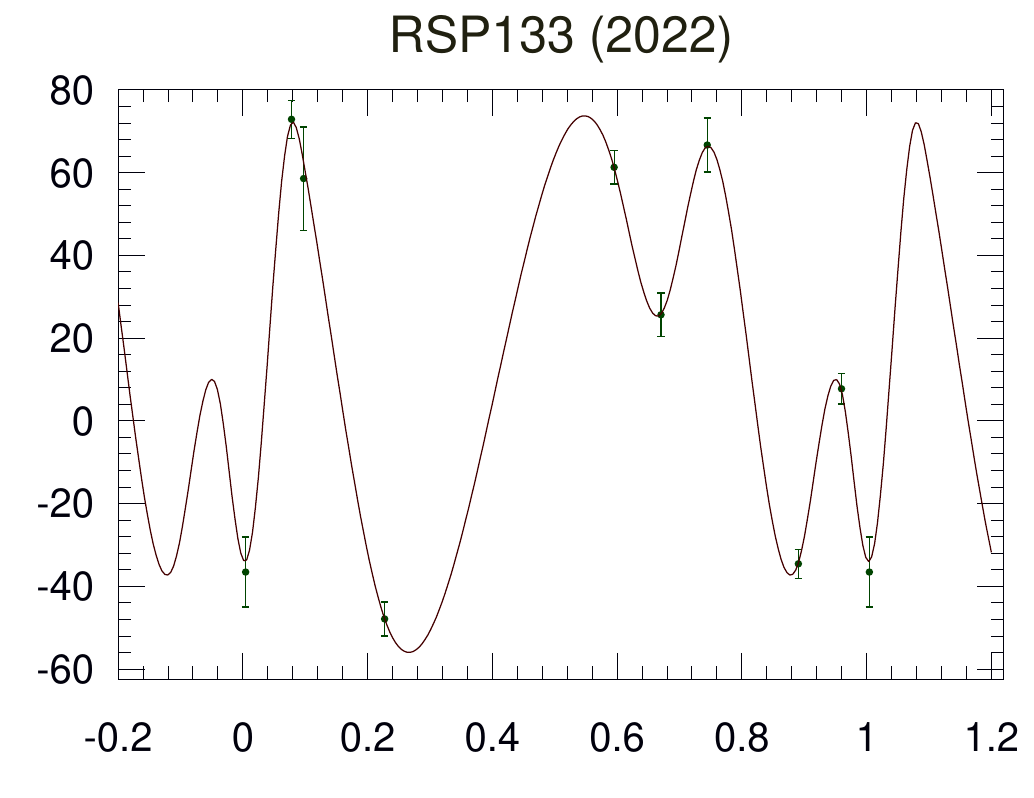}
\includegraphics[angle=0,width=4.35cm,clip]{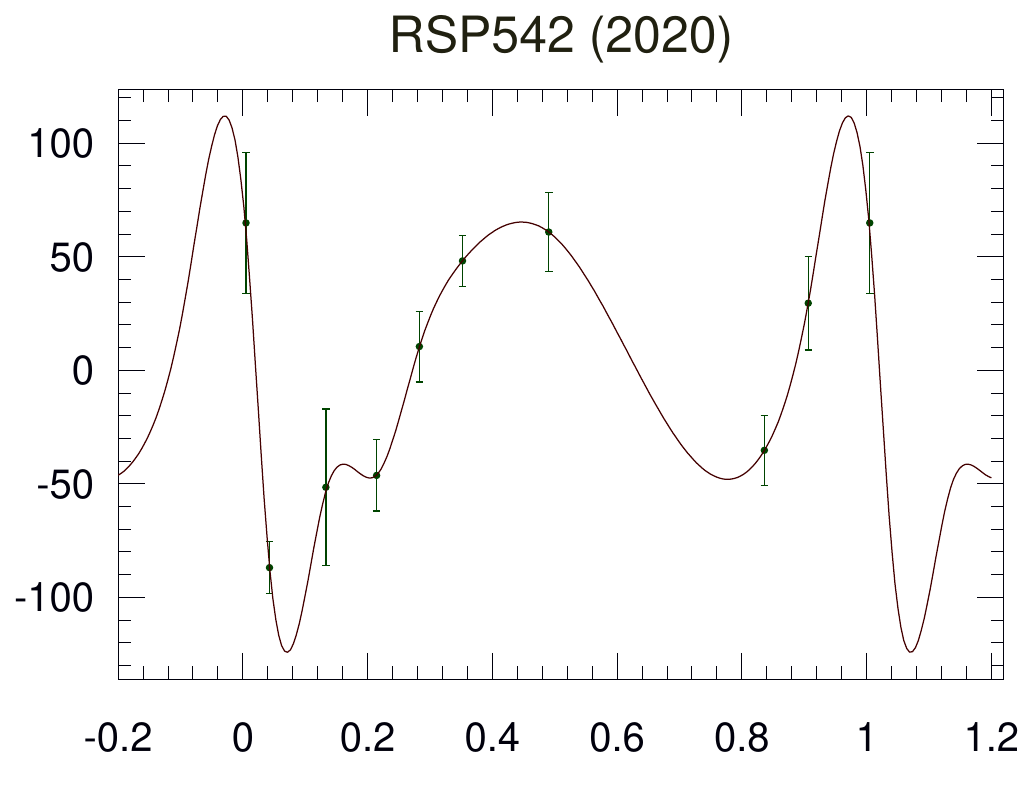}
\includegraphics[angle=0,width=4.35cm,clip]{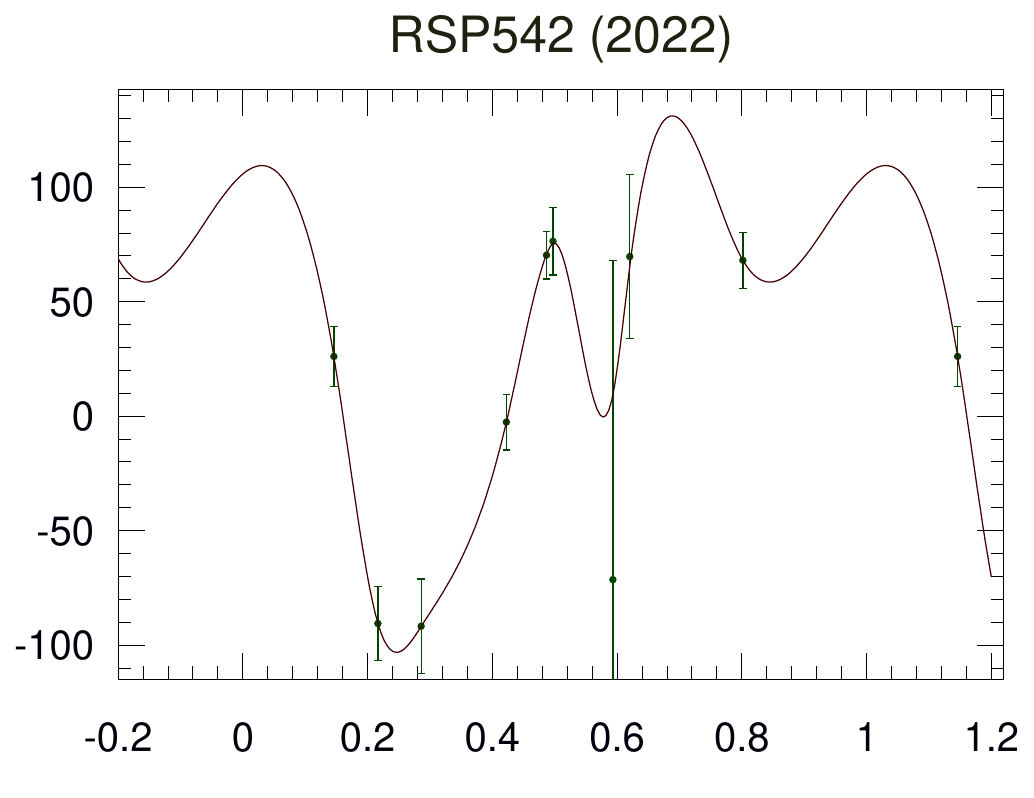}
\includegraphics[angle=0,width=4.35cm,clip]{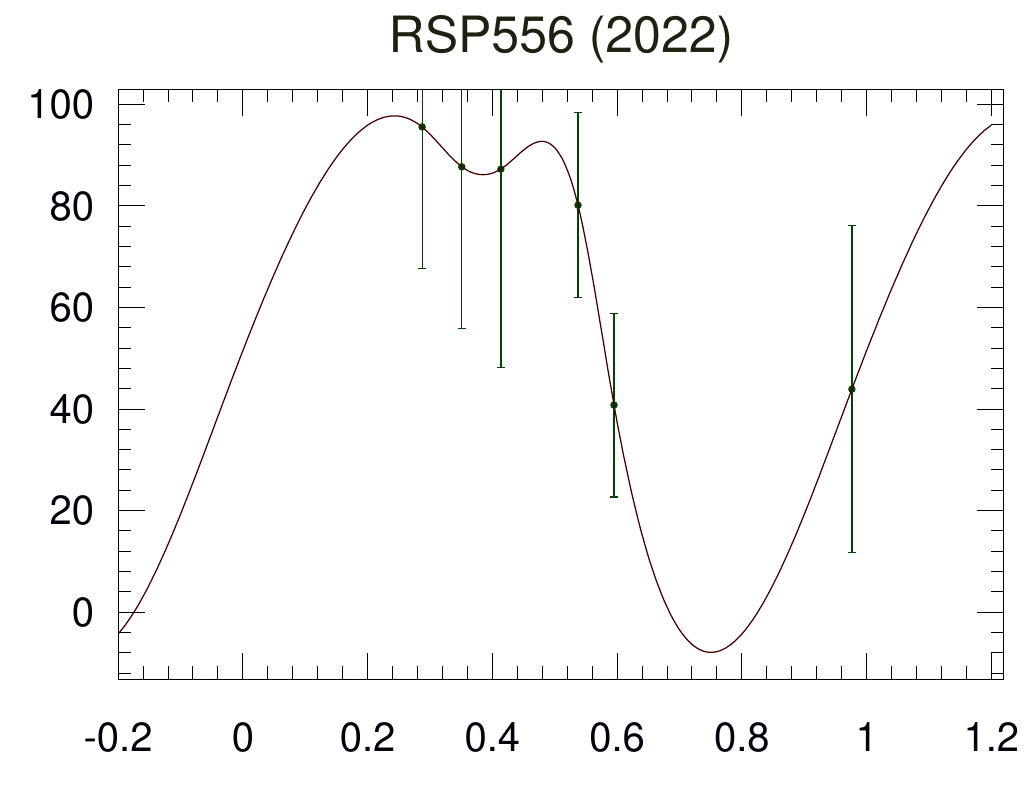}
\caption{(Continued.)}
\end{figure*}

\section{Extra tables}

\begin{table*}[!tbh]
\caption{Target sample and observational phase coverage.}\label{T1-App}
\begin{flushleft}
\begin{tabular}{llllllllllll}
\hline\hline
\noalign{\smallskip}
RSP   & HIP  &2MASS  & Other    & $V$     &$P_{\rm rot}$ & Ref.               & Single?$^1$ & Sp.T. & \multicolumn{3}{c}{Phase coverage}  \\
        &         &            & name    & (mag) &(d)                 & $P_{\rm rot}$ &                   &          & S20 & S21 & S22 \\
\noalign{\smallskip}\hline\noalign{\smallskip}
348   &         &04275919+1845327 & Mel25-R251&14.22   &3.30 & D19 & ?\,NY\,? &M5: & P/7 &  \dots &  \dots \\
344   &HIP20826 &04274607+1144110 & HD 28237  &7.482   &5.14 & K12 & YYYY  &F8 & F/9 & \dots  & \dots  \\
233   &HIP20237 &04201294+1914005 & V986 Tau  &7.444   &5.45 & R87 & YYYY  &F9 & F/9 & \dots  & \dots  \\
340   &HIP20815 &04273587+1535211 & V993 Tau  &7.404   &5.87 & R87 & YYYY  &G0 & F/8 & \dots  & \dots \\
137   &HIP19148 &04061614+1541532 & HD 25825  &7.811   &7.47 & K12 & YYYY  &G0 & F/9 & \dots  & \dots   \\
177   &HIP19793 &04143228+2334298 & V1310 Tau &8.047   &8.38 & K12 & YYYY  &G5 & P/7 & \dots  & F/12 \\
440   &HIP21317 &04343529+1530166 & V938 Tau  &7.898   &8.93 & K12 & YYYY  &F8  & P/7 & \dots  & \dots\\
227   &HIP20146 &04190798+1731291 & HD 27282  &8.427   &8.96 & K12 & YYYY  &G8 & \dots  & F/10 & \dots  \\
134   &HIP19098 &04053965+1756156 & HD 285367 &9.291   &9.10 & K12 & YYYY  &K0 & \dots  & F/8 & \dots  \\
439   &         &04343216+1549392 & HD 28977  &9.644   &9.16 & D19 & YYYY  &K0  & \dots  & F/9 & \dots\\
225   &HIP20130 &04185795+1954241 & V893 Tau  &8.597   &9.39 & R87 & YYYY  &G9 & P/8 & \dots  & F/11 \\
95    &HIP18327 &03550647+1659545 & HD 285252 &8.964   &9.82 & K12 & YYYY  &K0 & P/7 & \dots  & F/11 \\
429   &         &04333795+1645450 & HD 28878  &9.357   &9.82 & D19 & YYYY  &G5 & P/7 & \dots  & F/9 \\
198$^2$&HIP19934&04163346+2154269 & V984 Tau  &9.125   &10.26& D11 & YYYY  &G5 & P/7 & \dots  & F/11\\
      &         &                 &           &        &9.7  & F18 &       &   &   &  &  \\
68    &HIP16908 &03373495+2120355 & Mel25-5   &9.362   &10.57& D11 & YYYY  &K0 & \dots & \dots & F/9\\
587   &HIP22654 &04522352+1859489 & HD 284930 &10.28   &11.79& D19 & YYYY  &K0 & P/9 & \dots  & P/7 \\
      &         &                 &           &        &11.34& D11 &       &   &  &  &  \\
571   &         &04500069+1624436 & HD 286085 &10.69   &12.22& D19 & YYYY  &K6 & P/9 & \dots  & P/8  \\
      &         &                 &           &        &11.98& D11 &       &   &   &  &  \\
216   &HIP20082 &04181926+1605181 & V985 Tau  &9.548   &12.64& R87 & YYYY  &K0  & P/8 & \dots  & F/11\\
409   &HIP21138 &04315244+1529585 & HD 285876 &11.00   &12.90& D19 & YYYY  &K7  & P/8 & \dots  & P/9 \\
      &         &                 &           &        &13.13& D11 &       &   &   &  & \\
133   &HIP19082 &04052565+1926316 & Mel25-226 &11.40   &13.85& D16 & YYYY  &K7 & \dots & \dots & P/9  \\
      &         &                 &           &        &13.51& D11 &       &   &       &       &  \\
542   &         &04471851+0627113 & StKM 1-514&11.35   &14.44& D11 & YYYY  &M0 & P/9 & \dots  & P/7  \\
556   &         &04483062+1623187 & LP 416-570&12.43   &15.99& D19 & YYYY  &M2  & \dots & \dots & P/6\\
      &         &                 &           &        &15.69& D11 &       &   &       &       &  \\
\noalign{\smallskip}
\hline
\end{tabular}
\tablefoot{$^1$Single-star status from four criteria following Douglas et al. (\cite{doug19}), for explanations see text. $^2$Mel\,25-21 in Folsom et al. (\cite{fol}). References for $P_{\rm rot}$: D19 Douglas et al. (\cite{doug19}), K12 Kundert et al. (\cite{kund}), R87 Radick et al. (\cite{rad87}), D11 Delorme et al. (\cite{delo}), D16 Douglas et al. (\cite{doug16}), F18 Folsom et al. (\cite{fol}). Spectral type (Sp.T.) is from various sources in CDS/Simbad. The last three columns are the phase coverage in the three observing runs in season 2020/21 (S20), season 2021/22 (S21), and season 2022/23 (S22); P = Partial, F = Full (the number is the number of spectra; the detailed observing log is given in Table~\ref{T2-App}).}
\end{flushleft}
\end{table*}

\begin{table*}[!tbh]
\caption{Observing log of individual spectra. }\label{T2-App}
\begin{tabular}{lllllllll}
\hline\hline\noalign{\smallskip}
 Target & UT date & UT start    & $t_{\rm exp}$ & CD:\,$\Delta\lambda$ & S/N & CD:\,$\Delta\lambda$ & S/N & phase\\
        & (d/m/y) & (h:m:s)     & (m:s)         & \ \ \ (\AA )         &     & \ \ \ (\AA ) &  &  \\
\noalign{\smallskip}\hline\noalign{\smallskip}
 RSP 348 & 09/12/2020 &  03:37:21.6 & 01:00:00  & 3:4800-5441 &    44  & 5:6278-7419 &   148 & 0.092 \\
         & 06/12/2020 &  06:12:44.1 & 01:00:00  & 3:4800-5441 &    52  & 5:6278-7419 &   163 & 0.215 \\
         & 16/12/2020 &  08:06:08.6 & 01:00:00  & 3:4800-5441 &    50  & 5:6278-7419 &   154 & 0.269 \\
         & 07/12/2020 &  06:11:33.0 & 01:00:00  & 3:4800-5441 &    54  & 5:6278-7419 &   166 & 0.518 \\
         & 17/12/2020 &  09:12:20.8 & 01:00:00  & 3:4800-5441 &    44  & 5:6278-7419 &   138 & 0.586 \\
         & 14/12/2020 &  03:55:06.0 & 01:00:00  & 3:4800-5441 &    46  & 5:6278-7419 &   154 & 0.611 \\
         & 05/12/2020 &  06:28:00.4 & 01:00:00  & 3:4800-5441 &    49  & 5:6278-7419 &   148 & 0.915 \\
 RSP 344 & 07/12/2020 &  10:22:30.0 & 00:20:00  & 3:4800-5441 &   368  & 5:6278-7419 &   542 & 0.006 \\
         & 02/12/2020 &  10:40:08.1 & 00:20:00  & 3:4800-5441 &   464  & 5:6278-7419 &   646 & 0.035 \\
         & 14/12/2020 &  02:34:51.1 & 00:20:00  & 3:4800-5441 &   762  & 5:6278-7419 &   941 & 0.305 \\
         & \vdots &  \vdots & \vdots  & \vdots &  \vdots & \vdots &  \vdots & \vdots \\
\noalign{\smallskip}
\hline
\end{tabular}
\tablefoot{The full table is available at the CDS. UT start is the UT start time of the first of the two CP integrations. $t_{\rm exp}$ is the sum for both CP exposures. S/N is given for the sum (Stokes-I) and are always per pixel for the 95\%\ quantile. Phase is given for the mid time of the two individual CP exposures. }
\end{table*}

\end{document}